\documentclass[preprint,review,12pt,numbers,sort&compress]{elsarticle}

\usepackage{amssymb}
\usepackage{amsmath}
\usepackage{makecell}
\usepackage{ulem}
\usepackage{hyperref} 
\usepackage{graphicx}

\usepackage{lineno}
\usepackage{booktabs}
\usepackage[percent]{overpic}
\usepackage{subfig}
\journal{ }

\begin{document}

\begin{frontmatter}



\title{Battery Recycling: Mechanistic Modelling of LiCoO$_2$ Leaching with Coupled Diffusion–Reaction Kinetics and Film Passivation} 


\author{Uddipta Sarma and Ganesh Madabattula\corref{cor1}} 

\affiliation{organization={Department of Chemical Engineering and Technology, Indian Institute of Technology (BHU) Varanasi},
            city={Varanasi},
            postcode={221005}, 
            state={Uttar Pradesh},
            country={India} }
\cortext[cor1]{Corresponding author: ganesh.che@iitbhu.ac.in}

\begin{abstract}
We developed a new mechanistic model for lithium and cobalt recovery during acidic-reductive leaching from LiCoO$_2$. The system considers HCl as an acid and H$_2$O$_2$ as a reducing agent. The model tracks conversion of LCO, the particle radius, the acid, H$_2$O$_2$, formation-dissolution dynamics of Co$_3$O$_4$ film, the film thickness, recovery of lithium and cobalt, and moles of O$_2$. We introduced the film passivation effects for reduced recovery in the absence of H$_2$O$_2$. We validated the model against the experimental data of recovered Li and Co reported in literature at three acid concentrations (0.5 M, 1.5 M, and 2.5 M) and four H$_2$O$_2$ concentrations (0\%, 0.2\%, 0.4\%, and 0.6\% (v/v)). The model works reasonably well at 0.5 M and 1.5 M HCl for the 0\% and 0.6\% concentrations of H$_2$O$_2$. At 2.5 M HCl, the model over-predicts the data in the presence of H$_2$O$_2$, while it works well at 0\% H$_2$O$_2$. We suggest pathways for further improvements in the model. The comprehensive mechanistic modelling framework for the leaching with a full list of the equations, the parameters, and the variables, reported for the first time, can be extended to other cathode chemistries and acid-reductive leaching systems. 
\end{abstract}




\end{frontmatter}



\section{Introduction}
Who will cry when lithium-ion batteries die if you have efficient, sustainable, and cost-effective technologies for recycling? The rapid global adoption of lithium-ion batteries (LIBs), driven by their use in portable electronics, electric vehicles, and grid-scale energy storage, has created a high demand for lithium, cobalt, nickel, and manganese\cite{IEA2024, Li2024}. At the same time, the limited mineral resources, geopolitics around the metal supply chains, increasing raw-material costs, and growing volume of end-of-life LIB waste have led battery recycling crucial for import dependent countries\cite{IEA2024recycling,IEA2024critical, Li2024}.  

For battery recycling, hydro-metallurgy has been a preferred choice due to its comparatively lower energy consumption, higher metal selectivity, and potential for closed-loop recovery of battery-grade materials \cite{Wu2022, Biswal2024, jung2021review}. In the process, the leaching step is the most critical and rate-determining stage, as it governs the dissolution of valuable metals from complex cathode matrices into the liquid phase \cite{Wu2022,Biswal2024,jung2021review}. 

Acidic leaching, often assisted by reducing agents, is effective for layered oxide cathodes such as LCO, NMC, and NCA \cite{Wu2024, Davis2023}. The extraction efficiency is controlled by breakdown of the crystal lattice and the reduction of high-valence transition metals. Mechanistic understanding and optimization of the leaching stage will help efficiency, selectivity, and sustainability of the recycling process.

Leaching behaviour in LIB recycling is governed by a set of interdependent physicochemical parameters such as acid type, concentration, temperature, solid-to-liquid ratio, particle size, agitation speed, reaction time, redox environment, and the presence of additives or reducing agents \cite{Yu2019, Xu2021}. Furthermore, intrinsic material properties such as cathode composition, crystallinity, and surface area significantly influence dissolution kinetics \cite{Wang2018, Wang2021}. The simultaneous interaction of chemical reaction, mass transfer, phase transformation, and structural evolution during leaching makes the process highly non-linear and system-specific; these interactions make empirical methods insufficient for predictive process design and scale-up.

Mechanistic modelling enables quantitative interpretation of rate-controlling steps and rational evaluation of the effects of operating parameters \cite{Raschman2019,cerrillo2022acid, Cerrillo-Gonzalez2020}. Beyond fitting experimental data, the models provide predictive capability, facilitate reactor design, support process optimization, and reduce experimental redundancy. 

Despite extensive experimental research \cite{Wongnaree2024, Meshram2015LeachingKinetics,Meshram2015ReducingAgent,Pinna2020LCO,Guzolu2017LiCoRecovery, Yu2024SelectiveLiLeaching, Xuan2021NMCLeaching, Tzanetakis2004NiMH, MolinaMontesdeOca2024Electrochemical, Faraji2022LeachingReview}, the field still lacks a universally accepted, mechanistically comprehensive kinetic models for LIB leaching. Most reported studies rely on empirical rate expressions, such as Avrami model\cite{Wongnaree2024,wongnaree2024leaching,Avrami1939PhaseChange, Khawam2006SolidStateKinetics, Grenman2011SolidLiquidKinetics, Muzayanha2020NCAKinetics, Faraji2022LeachingReview}. Another set of studies focussed on identifying the controlling regimes, such as boundary layer mass-transfer control, surface reaction control or residue layer diffusion control \cite{gao2018selective, zhang2015closed, xuan2021new, natarajan2018recovery}. These studies plot experimental data of recovery against a predefined equation to identify the mechanisms. However, we need mechanistic models beyond identifying the controlling regime to capture multicomponent evolution, dynamic structural changes, acidic-reductive interactions, evolving reactive interfaces, thermal influence and multiphase transformations characteristic of real cathode materials \cite{Cerrillo-Gonzalez2020,Wongnaree2024, Faraji2022LeachingReview, Khawam2006SolidStateKinetics, Grenman2011SolidLiquidKinetics, Muzayanha2020NCAKinetics, Chen2017ImprovedSCM}. 

In this work, we develop a mechanistic modelling framework for leaching of Co and Li from LCO cathodes, using hydrochloric acid and H$_2$O$_2$ as a reducing agent. \citet{cerrillo2022acid} presented a mechanistic model for the same system and forms a foundation for our work; they presented the reaction pathways for LCO leaching with and without the reducing agent. Their model traced the species balance of acid, H$_2$O$_2$, Li and Co. Building on their work, we develop an improved mechanistic model from a fresh perspective on solid-liquid acidic-reductive leaching kinetics and provide detailed analysis with additional variables. 

Along with the revised reaction rate expressions for heterogeneous reactions,  our model adds the equations to track LCO, particle radius, Co$_3$O$_4$ film, and oxygen evolution, which are necessary for holistic mechanistic modelling for scale-up. In our model, the LCO particle radius decreases with conversion. We also discuss formation-dissolution kinetics for Co$_3$O$_4$ film thickness in response to different concentrations of the acid and the reducing agent. We introduce the film passivation effects for reduced recovery in the absence of H$_2$O$_2$. Later, we parameterize the model and compare it with the experimental data published by \citet{cerrillo2022acid} at three acid concentrations (0.5 M, 1.5 M, and 2.5 M) and four H$_2$O$_2$ concentrations (0\%, 0.2\%, 0.4\%, and 0.6\% (v/v)). 
We believe that this work provides a foundation for well-defined mechanistic models for the recycling of more complex cathodes, such as NMC and NCA, in a cost-effective manner to increase throughput, scale-up, and support environmental sustainability and circular economy goals; so that nobody will cry when the batteries die..  

\section{Reactions for LiCoO$_2$ leaching}
\paragraph{Without H$_2$O$_2$ \cite{cerrillo2022acid}}
\begin{equation}
\mathrm{LiCoO_2 (s)} + 2\mathrm{H^+} \rightarrow \mathrm{Li^+} + \tfrac{1}{2}\mathrm{Co^{2+}} + \tfrac{1}{6}\mathrm{Co_3O_4 (s)} + \mathrm{H_2O} + \tfrac{1}{6}\mathrm{O_2} \tag{R1}
\label{eqn:R1}
\end{equation}
\begin{equation}
\mathrm{Co_3O_4 (s)} + 6\mathrm{H^+}
\rightarrow 3\mathrm{Co^{2+}} + 3\mathrm{H_2O} + \tfrac{1}{2}\mathrm{O_2} \tag{R2}
\label{eqn:R2}
\end{equation}
\paragraph{With H$_2$O$_2$: In addition to \ref{eqn:R1} and \ref{eqn:R2} \cite{cerrillo2022acid}}
\begin{equation}
\mathrm{LiCoO_2 (s)} + 3\mathrm{H^+} + \tfrac{1}{2}\mathrm{H_2O_2}
\rightarrow \mathrm{Li^+} + \mathrm{Co^{2+}} + 2\mathrm{H_2O} + \tfrac{1}{2}\mathrm{O_2} \tag{R3}
\label{eqn:R3}
\end{equation}
\begin{equation}
\mathrm{Co_3O_4 (s)} + 6\mathrm{H^+} + \mathrm{H_2O_2}
\rightarrow 3\mathrm{Co^{2+}} + 4\mathrm{H_2O} + \mathrm{O_2} \tag{R4}
\label{eqn:R4}
\end{equation}

\begin{figure}
    \centering
    \includegraphics[width=1\linewidth]{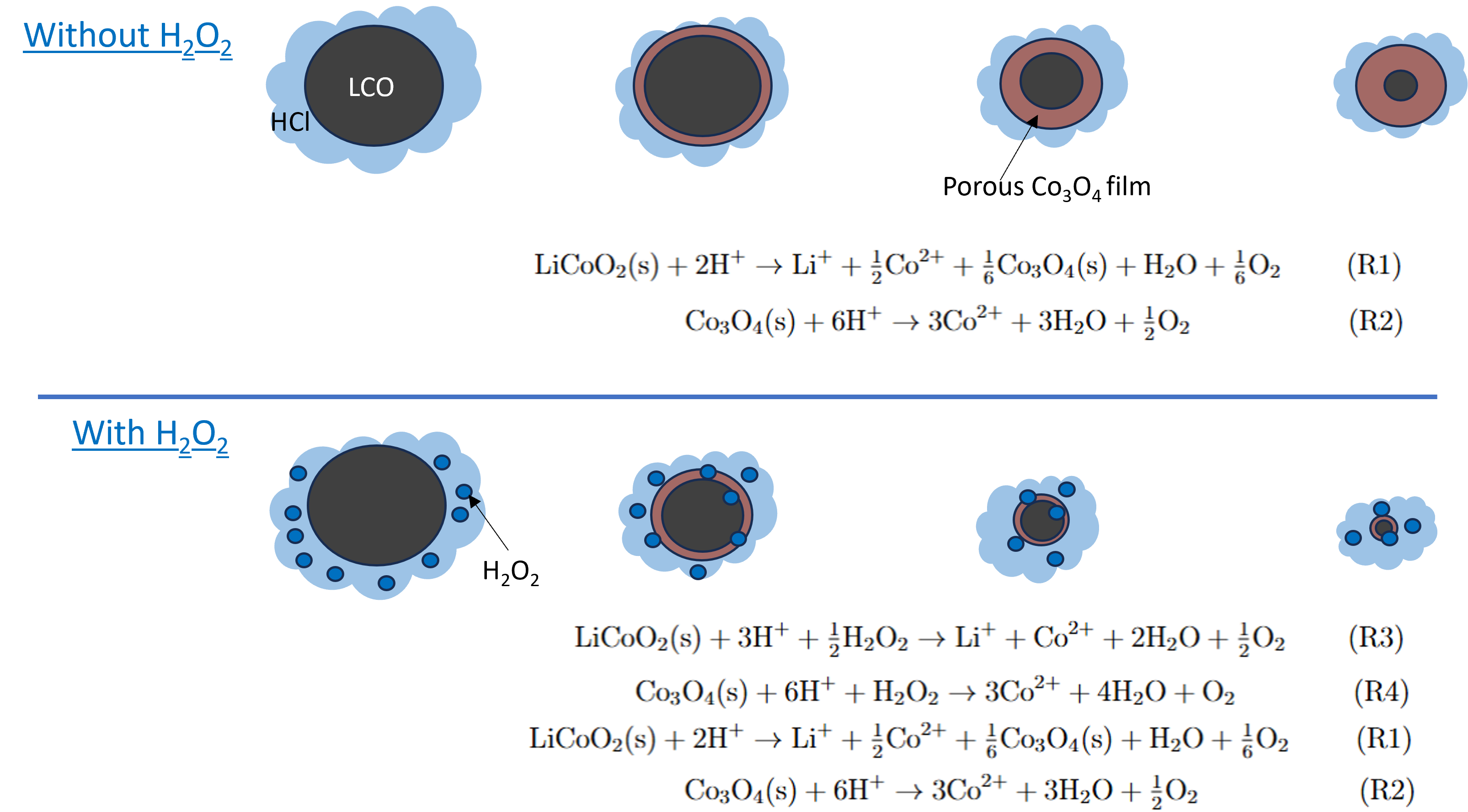}
    \caption{A schematic of LCO leaching process with HCl acid and H$_2$O$_2$, a reducing agent. Co$_3$O$_4$ film forms on LCO interface without H$_2$O$_2$ and leads to incomplete leaching of Co$^{2+}$ \citep{cerrillo2022acid}.}
    \label{fig:schemeLCO}
\end{figure}

\citet{cerrillo2022acid} reported the reactions for acidic-reductive leaching of LiCoO$_2$ cathode, using hydrochloric acid and H$_2$O$_2$ as a reducing agent. As represented in Figure~\ref{fig:schemeLCO}, in the absence of H$_2$O$_2$, LCO would react according to Eqn.~\ref{eqn:R1} and \ref{eqn:R2}. When reacted with acid, Co$^{2+}$ partially leaches out into the acid and some part forms an insoluble porous layer of Co$_3$O$_4$ on the LCO surface, hindering further leaching. The Co$_3$O$_4$ film formation was detected using XPS measurements experimentally in \cite{cerrillo2020recovery}. Later, the  Co$_3$O$_4$ layer slowly gets attacked by acid to recover Co$^{2+}$. Whereas, Li$^+$ leaches directly into the acid, faster than Co$^{2+}$. 
Their experimental observations show that, in the absence of the reducing agent, the ratio of leaching rates of Li:Co is around 2:1. 

On the other hand, the authors also presented two additional reactions for the leaching in the presence of H$_2$O$_2$, given in Eqn.~\ref{eqn:R3} and \ref{eqn:R4}. In this case, H$_2$O$_2$, along with acid, facilitates direct leaching of Co$^{2+}$ and do not let the Co$_3$O$_4$ layer form (\ref{eqn:R3}). However, as acid still reacts with LCO without H$_2$O$_2$ (\ref{eqn:R1}) at a reduced rate, the formed Co$_3$O$_4$ layer will then be dissolved by H$_2$O$_2$ (\ref{eqn:R4}). Reactions \ref{eqn:R3} and \ref{eqn:R4} occur simultaneously with \ref{eqn:R1} and \ref{eqn:R2}. Following their work, we use these four reactions as a basis for the mechanistic model.

\section{Model Equations}

To help with understanding the model equations, Table~\ref{tab:variables} and Table~\ref{tab:parameters}, in \ref{sec:appendix}, present a description of the variables and parameters involved for the mechanistic model of acidic-reductive leaching of LiCoO$_2$ particles. 
We assume the model is 0-dimensional/lumped and the variables evolve only with time. 

\subsection{Mixed regime (diffusion-reaction) controlled reactions}
\begin{figure}
    \centering
    \includegraphics[width=0.5\linewidth]{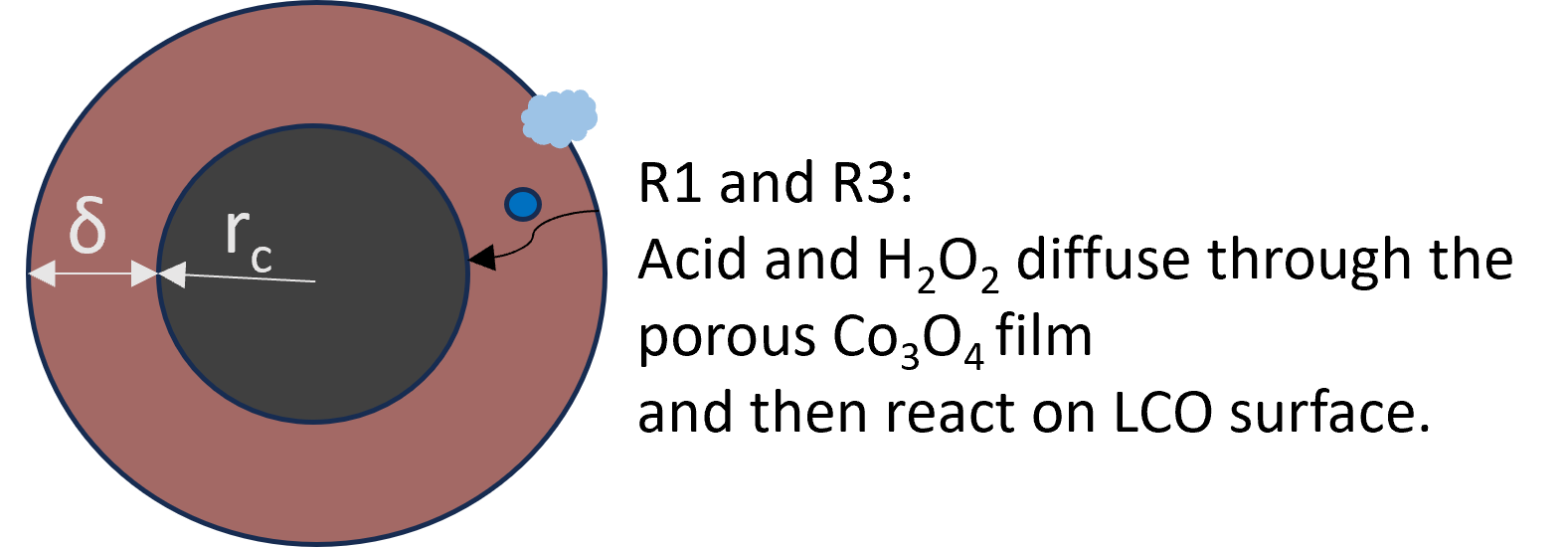}
    \caption{A schematic of reaction-diffusion mechanism for the R1 and R3. The acid and H$_2$O$_2$ will have to diffuse through the Co$_3$O$_4$ film for leaching of Li and Co from the LCO core.}
    \label{fig:schemeR1nR2}
\end{figure}

We consider that reactions \ref{eqn:R1} and \ref{eqn:R3} are controlled by mixed regime (diffusion-reaction). Acid and reducing agent diffuse through the porous Co$_3$O$_4$  layer to react with LCO at the interface, as represented in Figure~\ref{fig:schemeR1nR2}. 

Therefore, surface concentration of species $i$ at the LCO-Co$_3$O$_4$ interface, governed by reaction-diffusion control regime, is:
\begin{align}
C_{i,s} &= \frac{C_i}{1 + \dfrac{k_{s,i}\delta}{D_{eff,i}}},
\end{align}
where $C_i$ is a concentration on the Co$_3$O$_4$ film surface of the particle. $\delta$ is the film thickness of Co$_3$O$_4$ layer on a LiCoO$_2$ particle. $D_{eff,i}$ is the effective diffusivity of species $i$ through the porous Co$_3$O$_4$ film, and is given by:
\begin{align}
D_{eff,i} &= \frac{\varepsilon}{\tau} D_i.
\end{align}
$\varepsilon$ and $\tau$ are porosity and tortuosity of the Co$_3$O$_4$ film. 

\subsection{Governing equations}
\subsubsection{Reaction rates}
The surface rates for the four solid-liquid interface reactions (R1-R4) are defined respectively as follows. Each rate expression is related with its surface rate constant (per unit area of the solid surface) and concentration of species in the liquid at the surface, following the heterogenous reaction engineering practices \cite{fogler1999elements, levenspiel1999chemical}. Reactions R1 and R3 occur at the interface of LCO-Co$_3$O$_4$ film at radius $r_c$, governed by both diffusion and kinetics. The subscript, $s$, indicates the local variable at the interface, $r_c$. Reactions R2 and R4 occur on the Co$_3$O$_4$ film surface, governed by kinetics. 
\begin{align}
r_1 &= k_{s,H} C_{H,s} \\
r_2 &= k_2 C_H \\
r_3 &= k_3 C_{H,s} C_{H_2O_2,s} \\
r_4 &= k_4 C_H C_{H_2O_2}
\end{align}
 
\subsubsection{LiCoO$_2$ core volume balance}
Total core volume of LiCoO$_2$ available for leaching via reactions R1 and R3 is given as
\begin{equation}
\frac{dV_{LCO}}{dt} = -\frac{M_{LCO}}{\rho_{LCO}} (r_1 + r_3)\,A_{eff},
\end{equation}
from which we can calculate the core radius of the particle as
\begin{align}
r_c &= \left(\frac{3V_{LCO}}{4\pi N_p}\right)^{1/3}.
\end{align}
The leaching reaction stops when $A_{eff}$ becomes zero. 
Total number of LCO particles, $N_p$, is equal to $\frac{m_{LCO}/\rho_{LCO}}{\tfrac{4}{3}\pi R_0^3}$. At t = 0, $r_c$ = $R_0$.
\subsubsection{Co$_3$O$_4$ film volume balance and the particle radius}
The total solid volume of Co$_3$O$_4$ film that forms through R1 and dissolves via R2 and R4 is given as
\begin{equation}
\frac{dV_{s}}{dt} = \frac{M_{Co_3O_4}}{\rho_{Co_3O_4}} \left[\frac{1}{6} r_1 A_{eff} - (r_2 + r_4)\, a_s V_{s}\right],
\end{equation}
where $a_s$ is interfacial surface area ($m^2/m^3$) per unit solid volume of the film. The film formation stops when $A_{eff}$ becomes zero, and the dissolution stops when $V_{s}$ becomes zero.

From the solid volume, we can calculate porous volume of the film ($V_f$), with porosity $\varepsilon$, using
\begin{align}
V_f &= \frac{V_{s}}{1-\varepsilon}.
\end{align}
Then, the evolving particle radius ($R$) and the film thickness ($\delta$) as a function of time can be calculated as
\begin{align}
R &= \left(\frac{3(V_{LCO} + V_f)}{4\pi N_p}\right)^{1/3} \\
\delta &= R - r_c.
\end{align}

\subsubsection{Effective surface area}
$A_{eff}$ is the effective surface area of all the LCO particles (core) available for reactions R1 and R2, which is a summation over surface area ($A_c$) of individual particles of $N_p$ in number. All particles are assumed to have uniform radius, $r_c$, which evolves with time. 
\begin{align}
A_c &= 4\pi r_c^2
\end{align}
\begin{equation}
   A_{\mathrm{eff}} = \alpha N_p A_c f_{\mathrm{acc}}
\label{eqn:aeff} 
\end{equation}

The effective reactive area was corrected using an accessibility factor,
$f_{\mathrm{acc}}$, to account for the progressive passivation caused by
Co$_3$O$_4$ film accumulation on the LCO particle, and $\alpha$ is effective area factor due to binder in a black mass. 

The $f_{\mathrm{acc}}$ is dependent on the film volume. It varies between 0 and 1. Lower the film volume, higher the $f_{\mathrm{acc}}$ value (lower the inhibition).

\begin{equation}
f_{\mathrm{acc}}=\exp\left(-\lambda \frac{V_s}{V_{LCO} + V_s}\right)
\label{eqn:fcc}
\end{equation}

\begin{equation}
A_{\mathrm{eff}} = \alpha N_p A_c\exp\left[-\lambda \frac{V_s}{V_{LCO} + V_s}\right]
\label{eqn:accfinal}
\end{equation}

\subsubsection{Species balances}
The surface reactions proceed as long as LCO and Co$_3$O$_4$ film are available for the reactions via $A_{eff}$ and $V_{s}$, along with local concentrations. The change in concentrations of acid/proton and the reducing agent are given as follows. 
\paragraph{Acid}
\begin{equation}
\frac{dC_H}{dt} = -\frac{1}{V} \left[A_{eff}(2r_1 + 3r_3) + a_s V_{s} (6r_2 + 6r_4)\right],
\end{equation}

\paragraph{Hydrogen peroxide}
\begin{equation}
\frac{dC_{H_2O_2}}{dt} = -\frac{1}{V} \left[A_{eff}(0.5 r_3) + a_s V_{s} (r_4)\right].
\end{equation}

The extracted lithium and cobalt concentrations in the acid are given as follows.
\paragraph{Lithium}
\begin{equation}
\frac{dC_{Li}}{dt} = \frac{1}{V} \left[A_{eff}(r_1 + r_3)\right],
\end{equation}

\paragraph{Cobalt}
\begin{equation}
\frac{dC_{Co^{2+}}}{dt} = \frac{1}{V} \left[A_{eff}(0.5 r_1 + r_3) + a_s V_{s} (3r_2 + 3r_4) \right].
\end{equation}
Then, the conversions, $X_{Li}$ and $X_{Co}$ can be calculated as $\frac{V\ C_i\, M_i}{x_{i,m}\, m_{LCO}}$, where $x_{i,m}$ is mass fraction of species $i$ in the LCO mass.

\subsubsection{Oxygen evolution}
We also track the release of moles of oxygen during leaching using the following equation. 
\begin{equation}
\frac{dn_{O_2}}{dt} =
A_{eff}\left(\frac{1}{6}r_1 + \frac{1}{2}r_3\right) + a_s V_{s,f}\left(\frac{1}{2}r_2 + r_4\right)
\end{equation}

\section{Details of the simulations}
The model was simulated using COMSOL multiphysics (v6.3) on a Windows based laptop with 16 GB RAM and Intel Core Ultra 7 processor. Global ODEs and DAEs (ge) module on COMSOL was used. The initial values for the variables are listed in Table~\ref{tab:initCond}. The parameters values are listed in Table~\ref{tab:paramsValue}. The initial values were taken to reflect the experimental conditions \cite{cerrillo2022acid}. Apart from the material properties, the kinetic rate constants, mass-transfer parameters, and morphological parameters of the active LCO mass are assumed and iteratively tuned to predict the data. 

\begin{table}[h]
\centering
\caption{Initial values for the model variables. Values are taken to reflect the experimental conditions of \citet{cerrillo2022acid}.}
\begin{tabular}{ll}
\toprule
Variable & Initial value \\
\midrule
$V_{LCO}(0)$ & $\dfrac{m_{LCO}}{\rho_{LCO}}$ m$^3$ \\

$V_{s,f}(0)$  & $0$ (or $10^{-12}$) m$^3$  \\

$C_H(0)$  & $1.5 \times 10^{3}$ mol/m$^3$  (varies)\\

$C_{H_2O_2}(0)$ & $178$ mol/m$^3$ (varies)  \\

$C_{Li}(0)$  & $0$ mol/m$^3$ \\

$C_{Co^{2+}}(0)$ & $0$ mol/m$^3$  \\

$n_{O_2}(0)$  & $0$ mol \\

$r_c(0)$  & $R_0$ m \\

$R(0)$  & $R_0$  m \\

$\delta(0)$  & $0$ m \\
\bottomrule
\end{tabular}
\label{tab:initCond}
\end{table}

\begin{table}[h]
\centering
\caption{Values of the model parameters. Apart from the material properties, the kinetic rate constants, mass-transfer parameters, and morphological parameters of the active LCO mass are assumed.}
\resizebox{0.4\textwidth}{!}{
\begin{tabular}{ll}
\toprule
Parameter  & Value \\
\midrule
$R_0$  & $5 \times 10^{-6}$ m \cite{cerrillo2022acid}\\
$V$  & $5 \times 10^{-5}$ m$^3$ \cite{cerrillo2022acid}\\
$N_p$ & $\dfrac{m_{LCO}/\rho_{LCO}}{\frac{4}{3}\pi R_0^3}$ \\
$\alpha$  & $1$ \\
$\varepsilon$ & $0.4$ \\
$\tau$ & $3$ \\
$a_s$  & $S_{BET}\,\rho_{Co_3O_4}$ m$^2$/m$^3$ \\
$S_{BET}$  & 50 m$^2$/g \\
$D_{H^+}$  & $5 \times 10^{-12}$ m$^2$/s \\
$D_{H_2O_2}$ & $1.0 \times 10^{-13}$ m$^2$/s \\
$k_{s,H}$  & $2.5 \times 10^{-8}$ m/s \\
$k_{s,H_2O_2}$  & $1 \times 10^{-11}$ m/s \\
$k_2$ & $5 \times 10^{-14}$ m/s \\
$k_3$  & $7 \times 10^{-11}$ m/s \\
$k_4$ & $2 \times 10^{-13}$ m/s \\
$\lambda$ & 8\\
$M_{LCO}$ & $97.87 \times 10^{-3}$ kg/mol \\
$M_{Co_3O_4}$ & $240.8 \times 10^{-3}$ kg/mol \\
$M_{Li}$ & $6.94 \times 10^{-3}$ kg/mol \\
$M_{Co}$ & $58.93 \times 10^{-3}$ kg/mol \\
$\rho_{LCO}$  & $5100$ kg/m$^3$ \\
$\rho_{Co_3O_4}$ & $6000$ kg/m$^3$ \\
$x_{Li,m}$ & 0.0709 \\
$x_{Co,m}$ & 0.602 \\
$m_{LCO}$ & 2.5 g \cite{cerrillo2022acid}\\

\bottomrule
\end{tabular}
}
\label{tab:paramsValue}
\end{table}
\clearpage
\section{Results and Discussion}
For the model comparison, we use  \citet{cerrillo2022acid}'s LCO leaching experimental data at different concentrations  of HCl (0.5 M, 1.5 M and 2.5 M) and H$_2$O$_2$ (0\% (v/v) (0 mM); 0.2\% (v/v) (86 mM); 0.4\% (v/v) (173 mM); and 0.6\% (v/v) (260 mM)). The conversion vs. time data clearly indicates that the recovery of Li and Co increases with increase in the acid concentration and the  H$_2$O$_2$ concentration; The data shows Li is easier and faster to recover than Co. 

\subsection{At 1.5 M acid}
\begin{figure}
    \centering
     
    \subfloat[]{\includegraphics[width=0.5\linewidth]{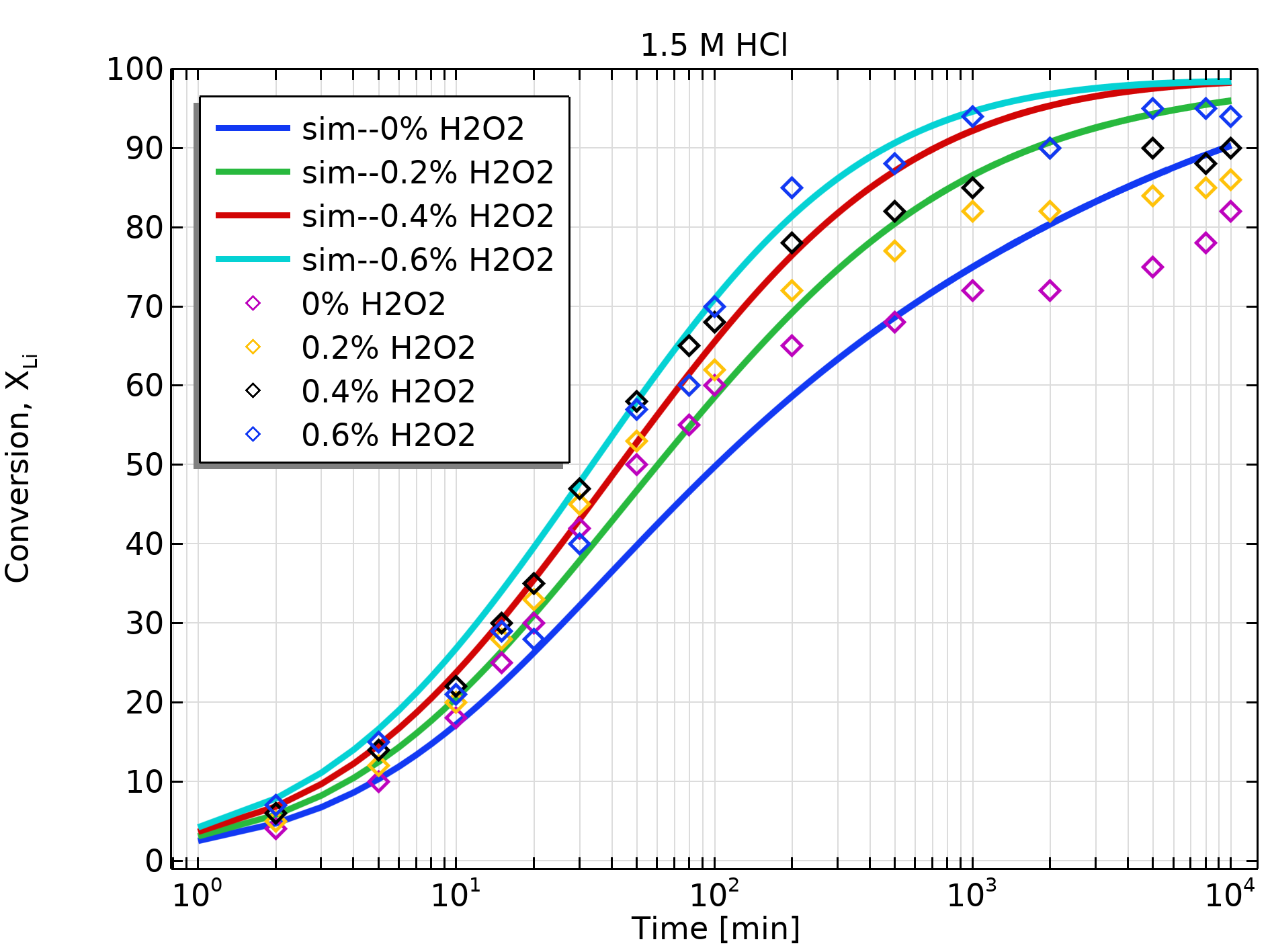}}
    \subfloat[]{\includegraphics[width=0.5\linewidth]{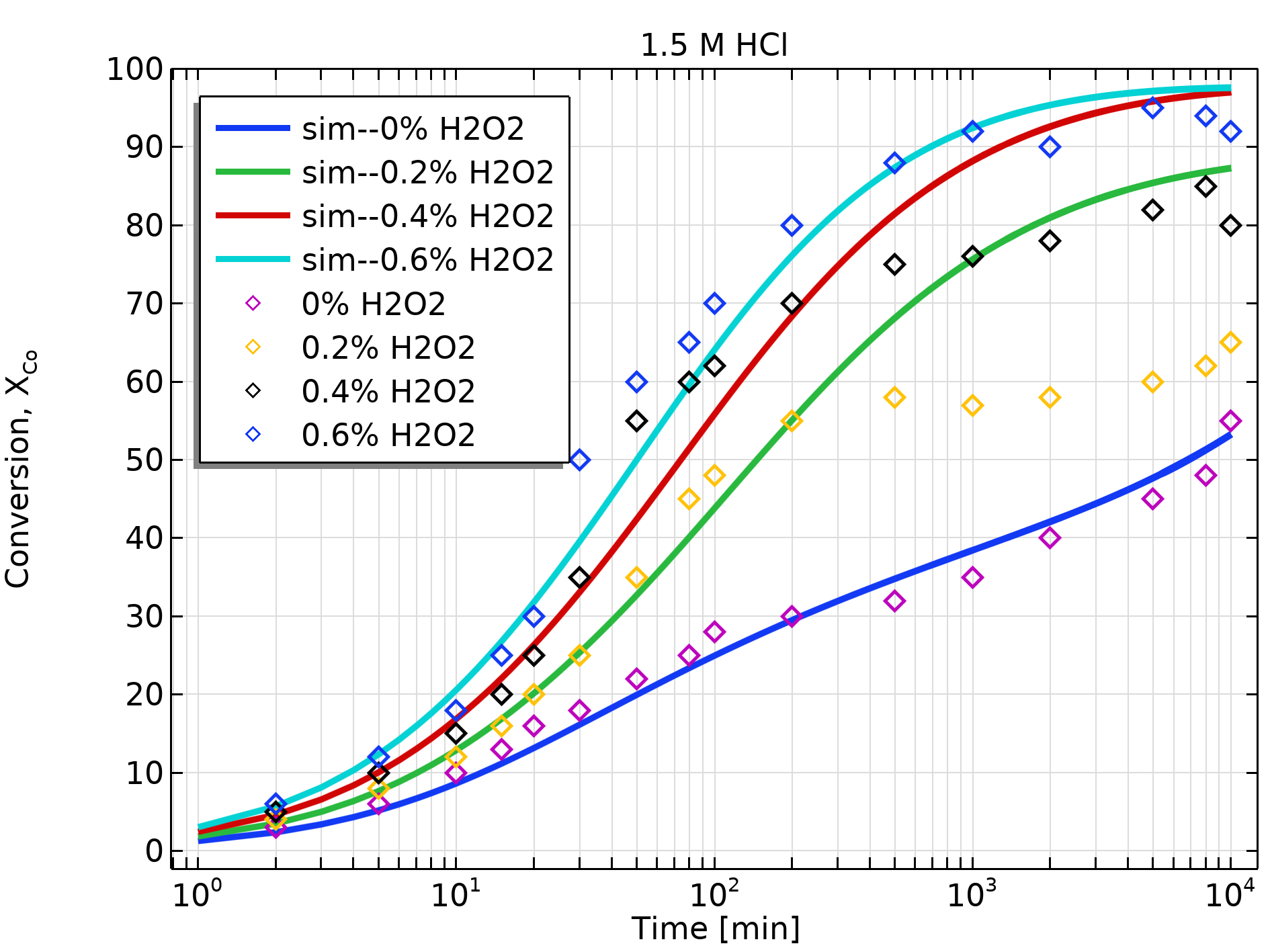}}
    \caption{At 1.5 M Acid: The model predictions with the film passivation compared against the experimental data of \citet{cerrillo2022acid} at four H$_2$O$_2$ concentrations. a) Conversion of Li and b) conversion of Co.}
    \label{fig:1p5M_X}
\end{figure}

Figure~\ref{fig:1p5M_X} shows the model predictions of the conversions of Li and Co at 1.5~M acid with different H$_2$O$_2$  concentrations vs. time compared against the data. The predictions include film passivation (Eqns.~\ref{eqn:aeff}-\ref{eqn:accfinal}) in the model. The model shows reasonable agreement with the data for Co and Li, capturing the overall behaviour while observing some quantitative deviations at intermediate concentrations of H$_2$O$_2$. The results for Co indicate that the film dynamics are well captured in the absence and at the 0.6\%  of H$_2$O$_2$. However, the film transition dynamics need improvement in the model. The model deviates at 0.2 and 0.4\% H$_2$O$_2$ at later stages (2E2 min) of leaching. For Li, the model performs well for all the H$_2$O$_2$ concentrations.   Overall, the mechanistic model demonstrates consistent prediction reasonably well and validates the parameterisation approach.
\begin{figure}
    \centering
    \subfloat[]{\includegraphics[width=0.48\linewidth]{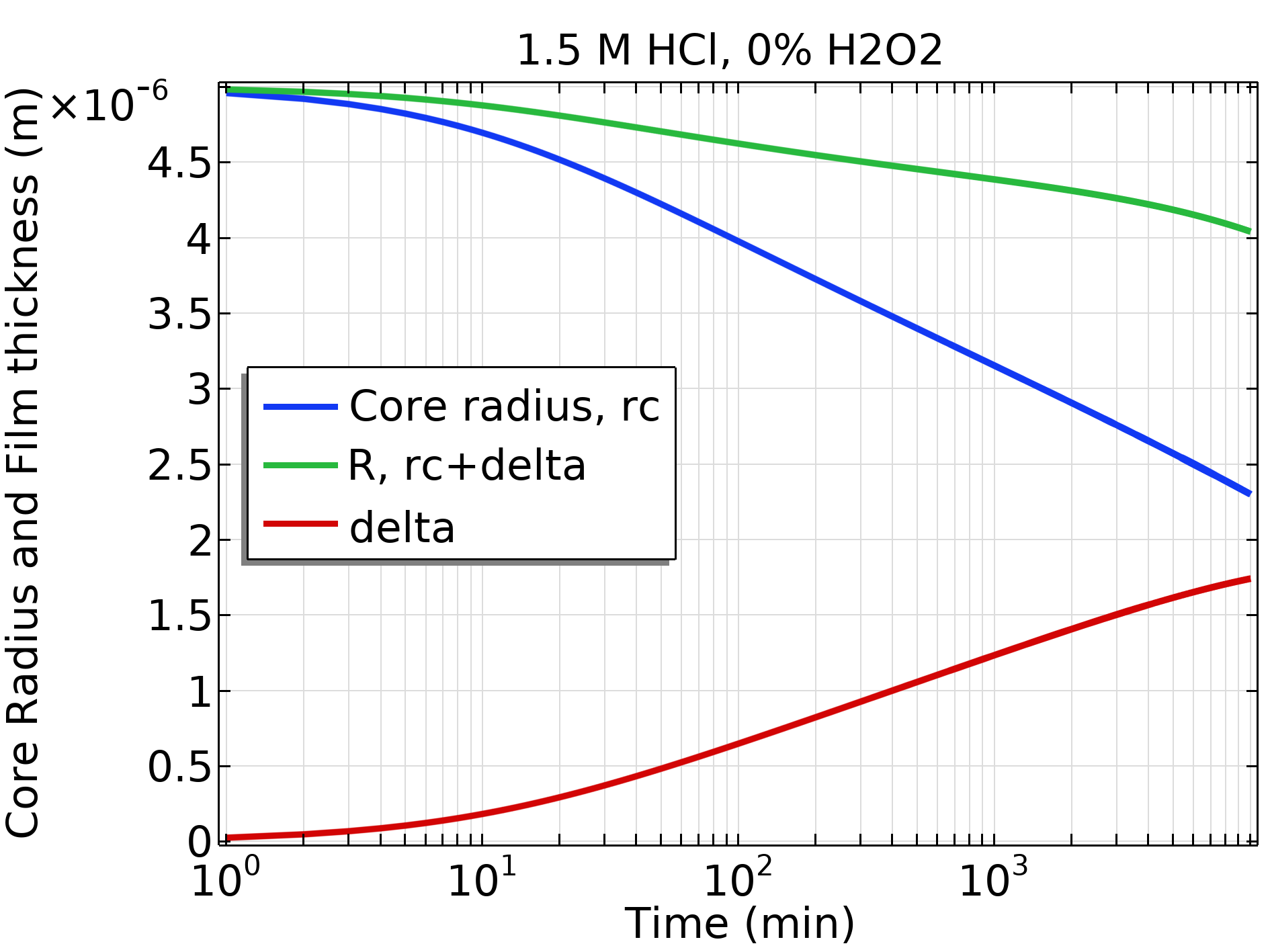}}
    \subfloat[]{\includegraphics[width=0.48\linewidth]{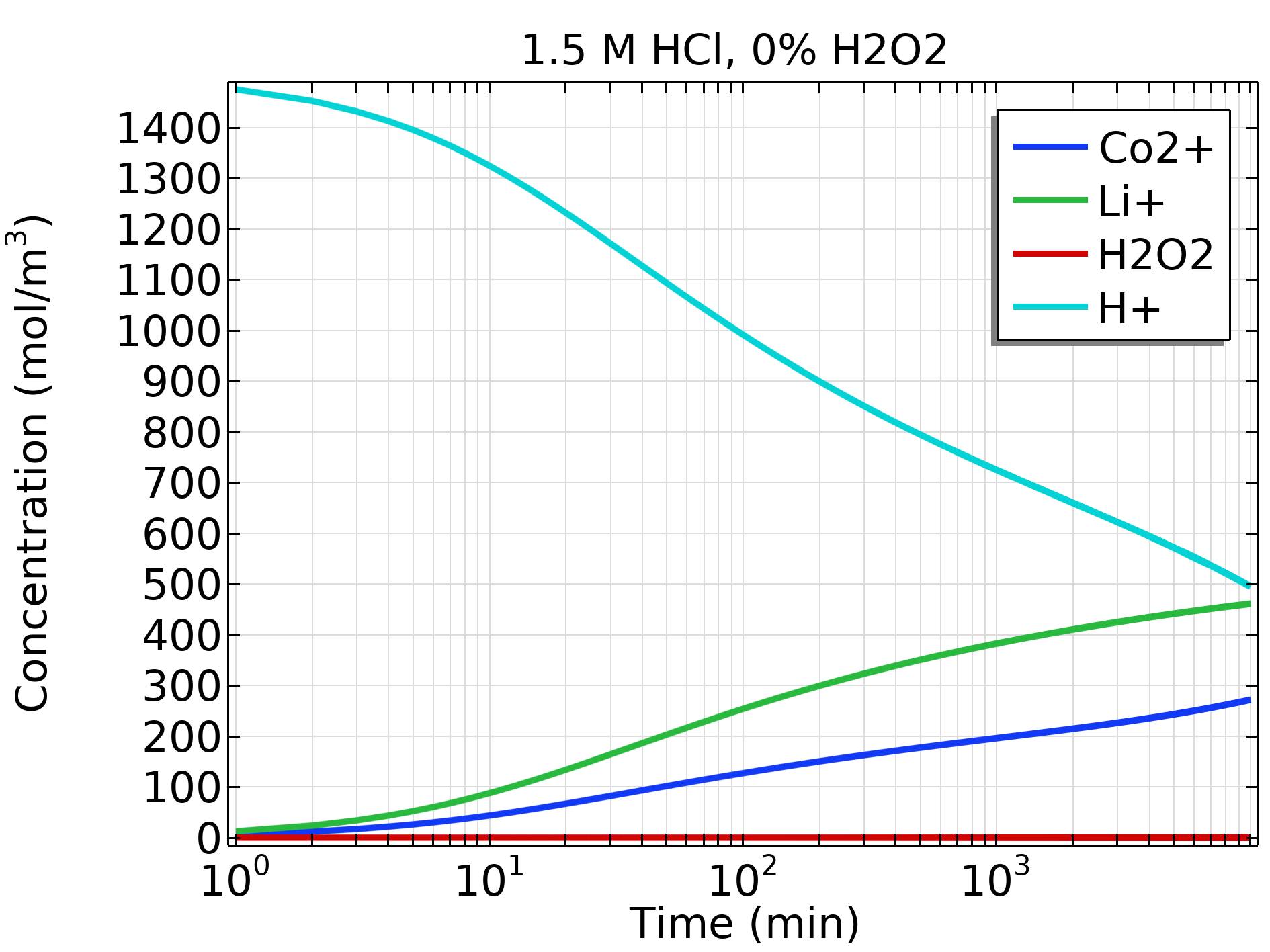}}
    \hfill
     \subfloat[]{\includegraphics[width=0.48\linewidth]{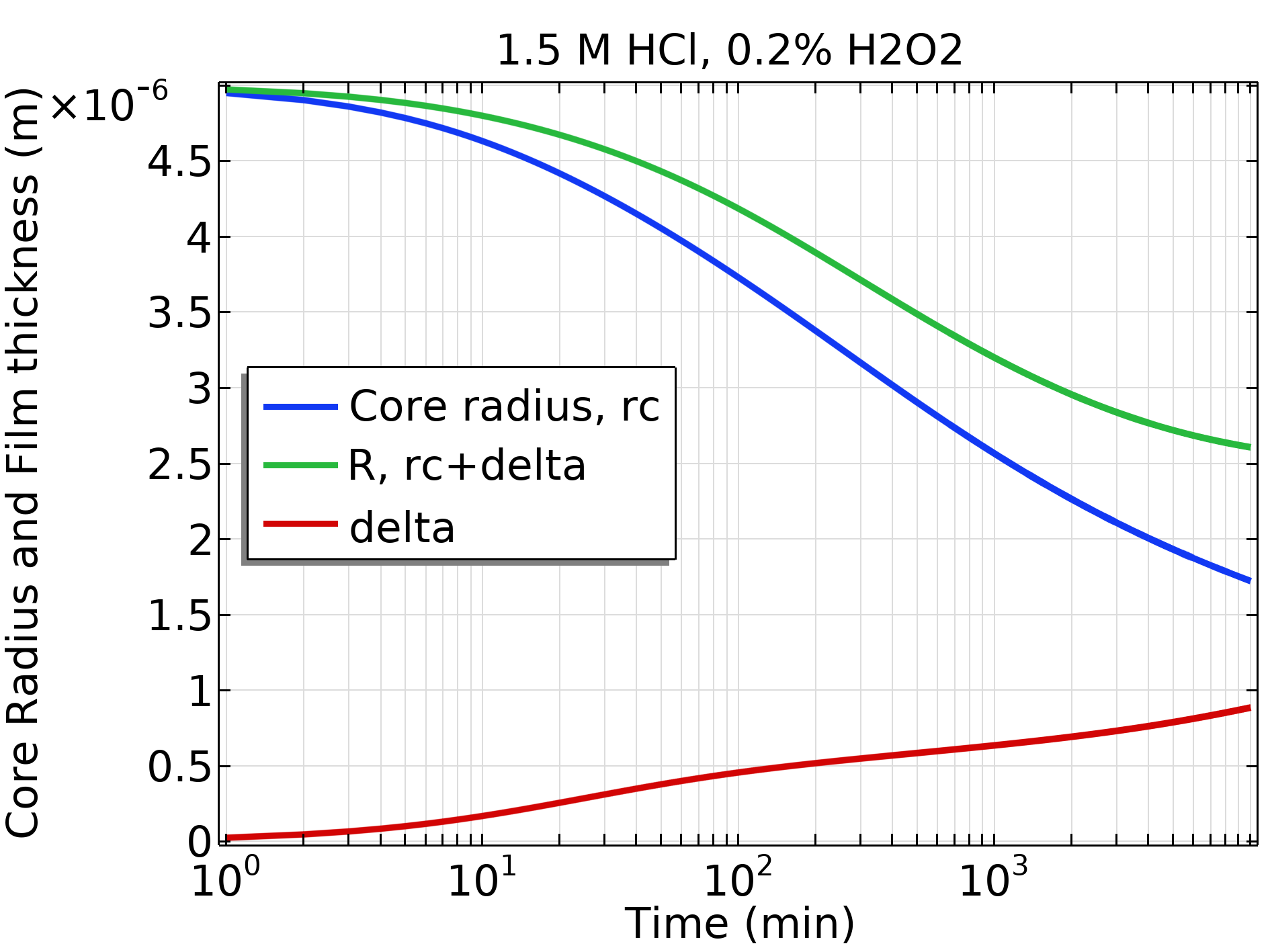}}
    \subfloat[]{\includegraphics[width=0.48\linewidth]{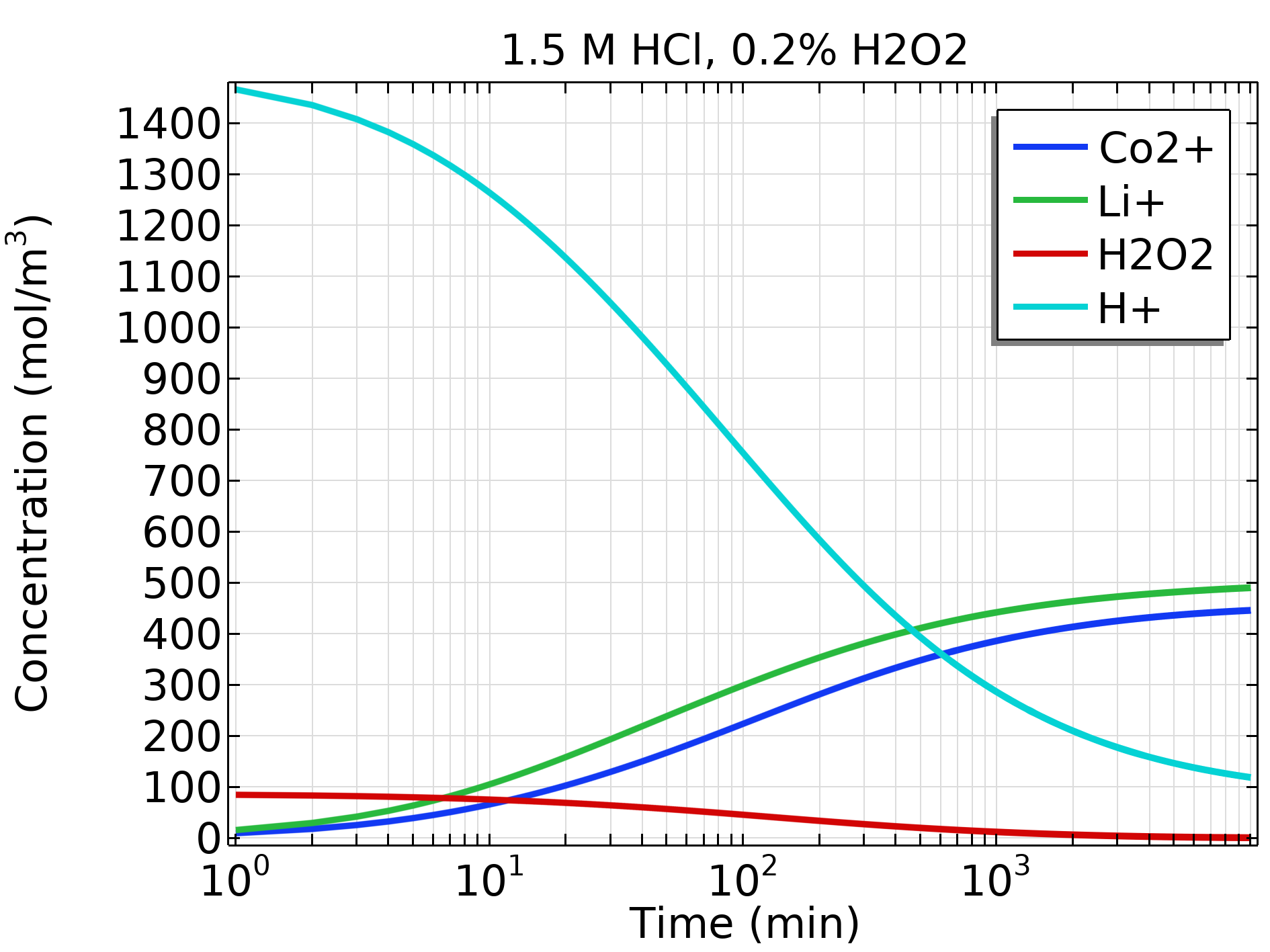}}
    \caption{At 1.5 M Acid: the model predictions with film passivation. The profiles of LCO core particle radius (rc), the film thickness (delta), and total particle radius (rc + delta); and concentrations of Co$^{2+}$, Li$^{+}$, the acid and H$_2$O$_2$. a \& b) H$_2$O$_2$. c \& d) 0.2\% H$_2$O$_2$.}
    \label{fig:1p5M_Xconc1}
\end{figure}

\begin{figure}
    \centering
    \subfloat[]{\includegraphics[width=0.48\linewidth]{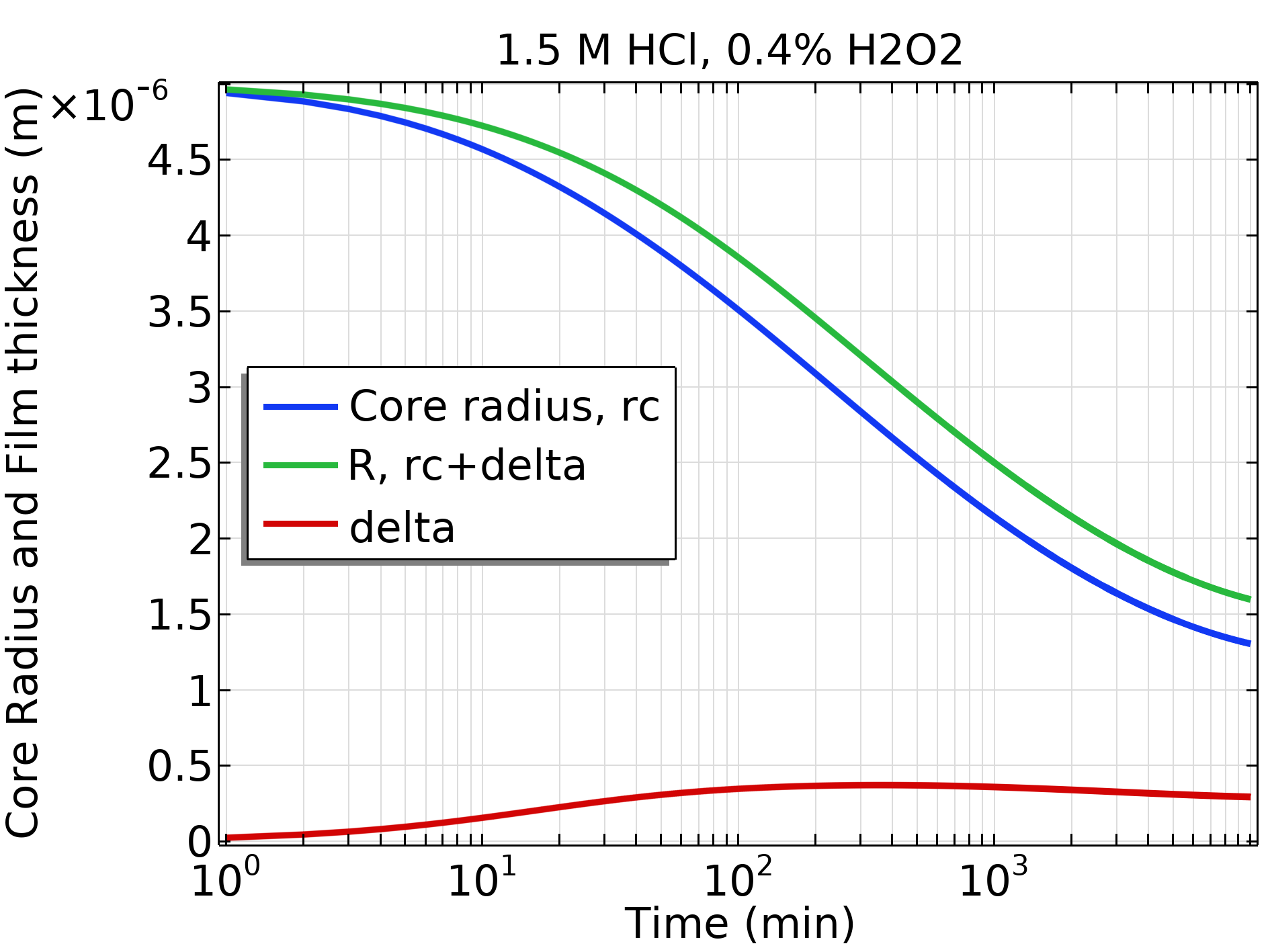}}
    \subfloat[]{\includegraphics[width=0.48\linewidth]{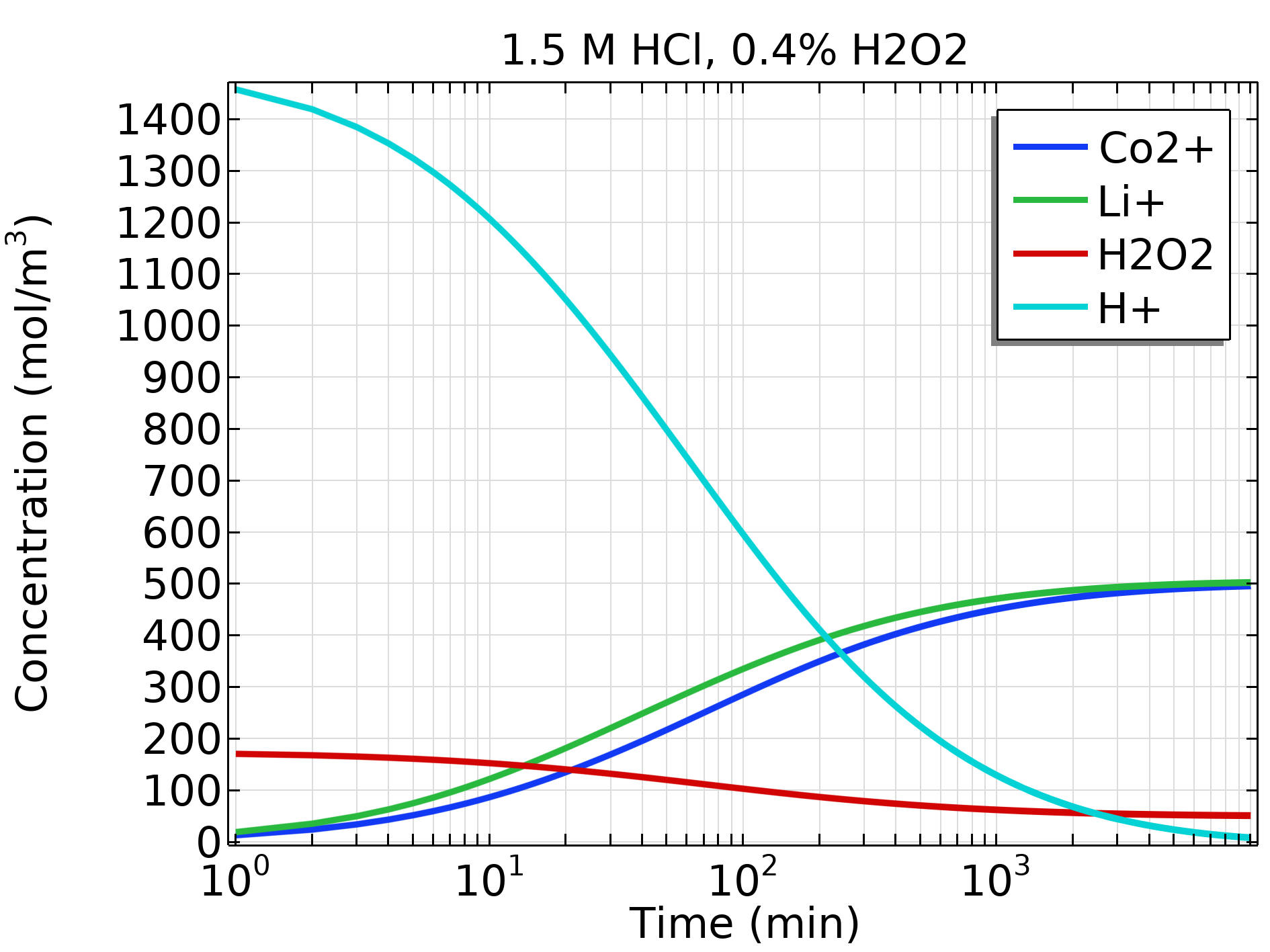}}
    \hfill
    \subfloat[]{\includegraphics[width=0.48\linewidth]{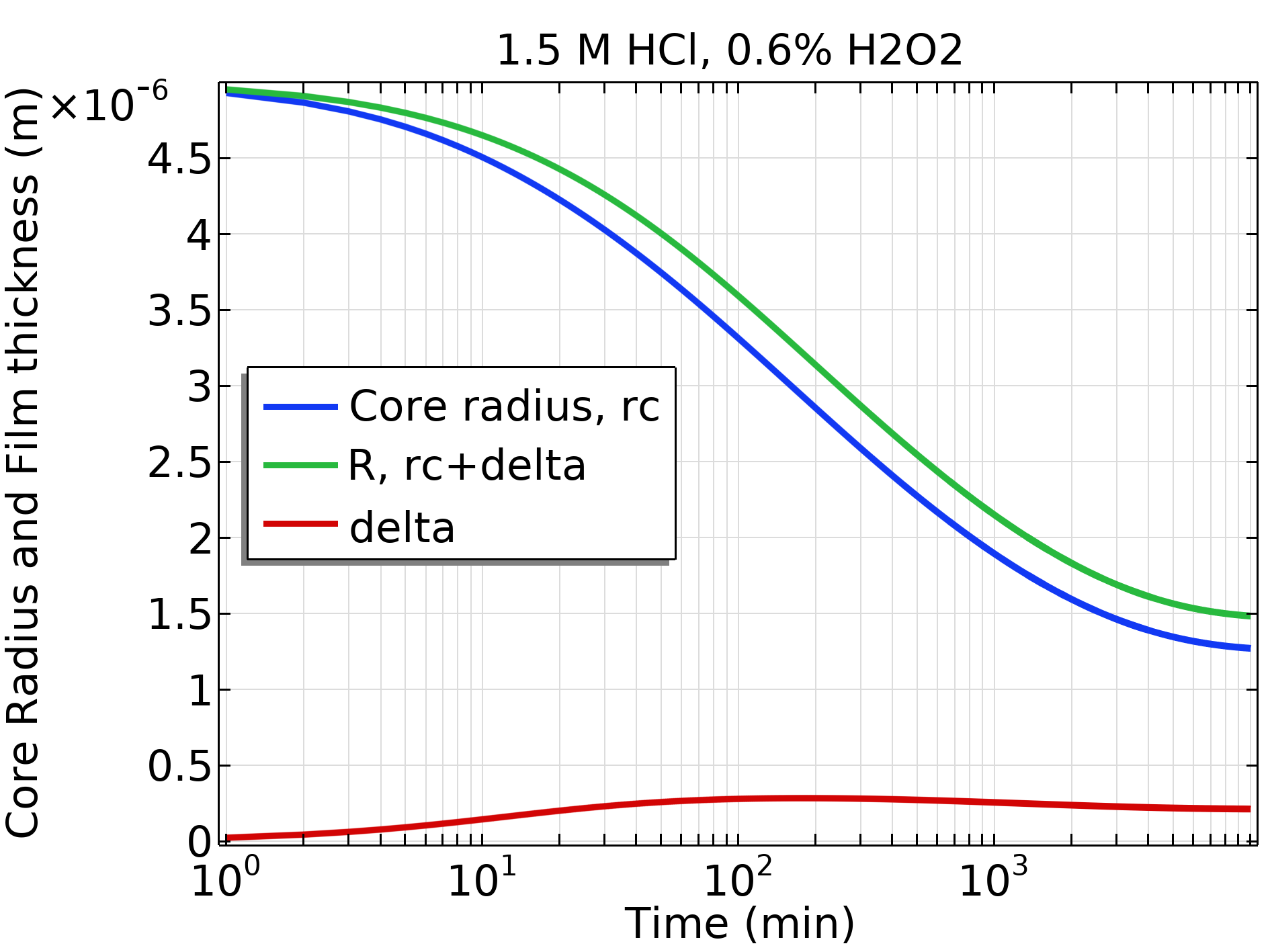}}
    \subfloat[]{\includegraphics[width=0.48\linewidth]{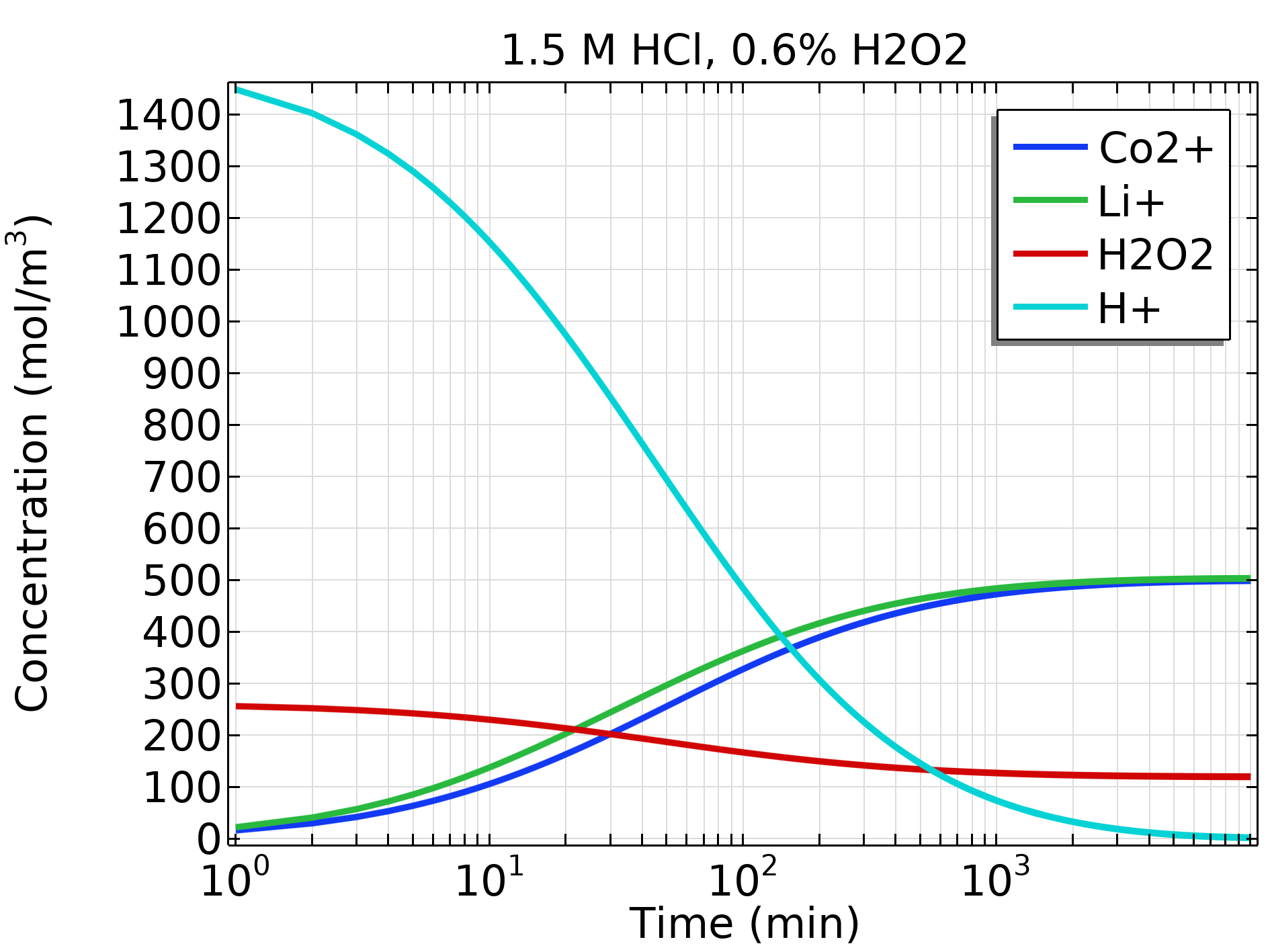}}
    \caption{At 1.5 M Acid: the model predictions with film passivation. The profiles of LCO core particle radius (rc), the film thickness (delta), and total particle radius (rc + delta); and concentrations of Co$^{2+}$, Li$^{+}$, the acid and H$_2$O$_2$. a \& b) at 0.4\% H$_2$O$_2$. c \& d) 0.6\% H$_2$O$_2$. }
    \label{fig:1p5M_Xconc2}
\end{figure}

Figures~\ref{fig:1p5M_Xconc1} and~\ref{fig:1p5M_Xconc2} show change in the LCO core particle radius, evolution of the film thickness, the total particle radius, and evolution of concentrations of Co$^{2+}$, Li$^{+}$, H$_2$O$_2$, and H$^{+}$ over the leaching period, for the four H$_2$O$_2$ concentrations, at 1.5~M HCl. We can observe the following: the film thickness keeps increasing at 0\% and 0.2\% H$_2$O$_2$ at different rates, being slower at 0.2\%; whereas for the 0.4\% and 0.6\% cases, the film thickness reaches to a peak value and then decreases, indicating the interplay between the formation-dissolution kinetics w.r.t H$_2$O$_2$ concentration. At 0\% and 0.2\% H$_2$O$_2$, the partial recovery of the Li and Co is indicated by the non-zero values of the core radius and the film thicknesses. The corresponding changes in particle core radius, \textit{r$_c$}, can be seen. The changes in the concentrations of the four variables are consistent with the recovery of Li and Co and w.r.t the H$_2$O$_2$ concentrations. The concentrations of Li and Co can evolve up to their theoretical maximum values ($\sim$ 510 mol/m$^3$) if conversion is 100\%.

From the 0.6\% case, we can see that complete leaching of LCO is limited by the acid concentration (indicated by non-zero values of \textit{r$_c$},  the film thickness and the H$_2$O$_2$ concentration), even at the 0.6\% H$_2$O$_2$ for the 1.5 M HCl, highlighting the depletion of the acid as the reason for the slowdown of the recovery at later stage. The model suggest that the experiments need higher concentration of the acid than 1.5 M for the possible complete recovery.

\begin{figure}
    \centering
    \subfloat[]{\includegraphics[width=0.48\linewidth]{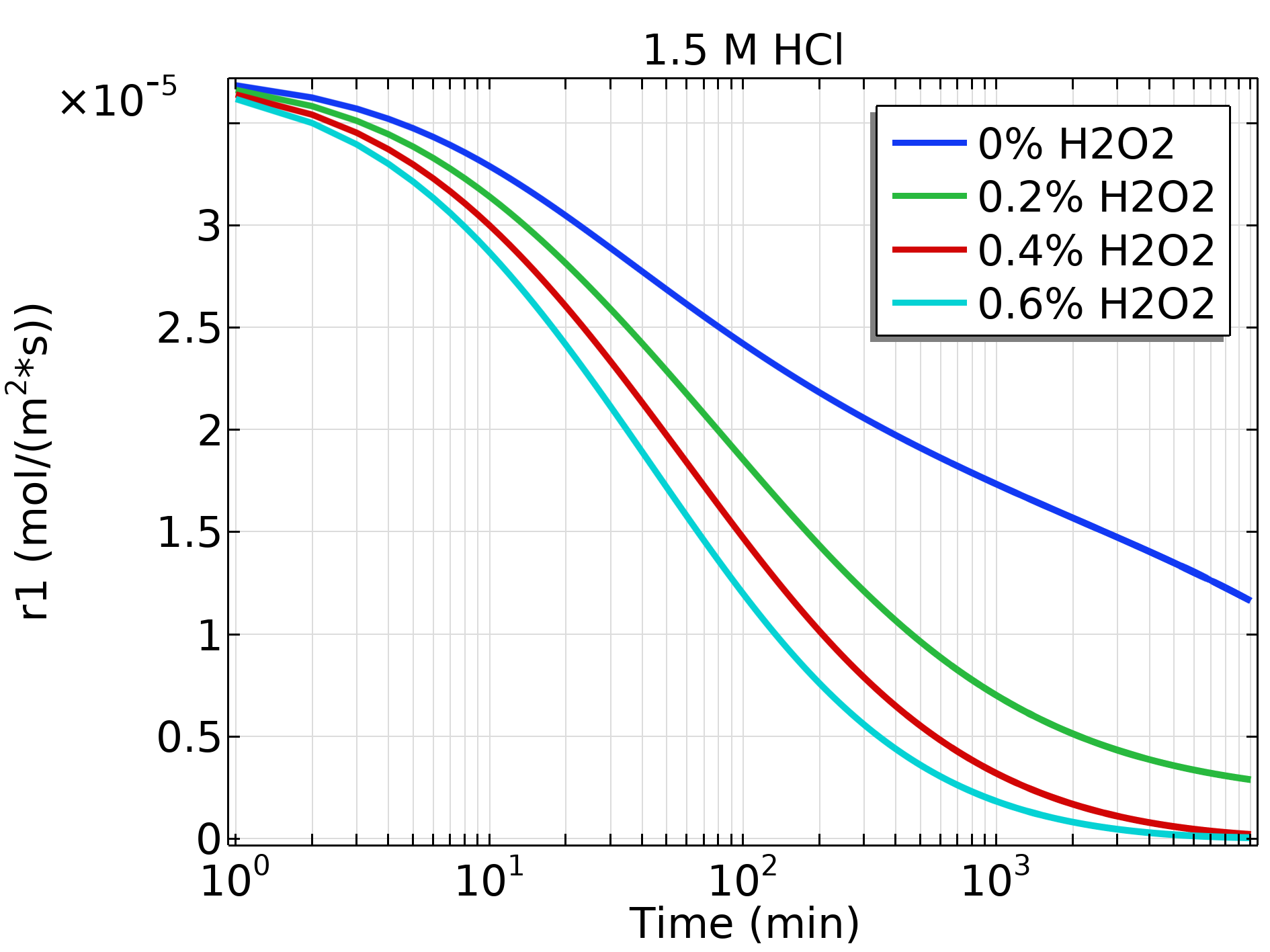}}
    \subfloat[]{\includegraphics[width=0.48\linewidth]{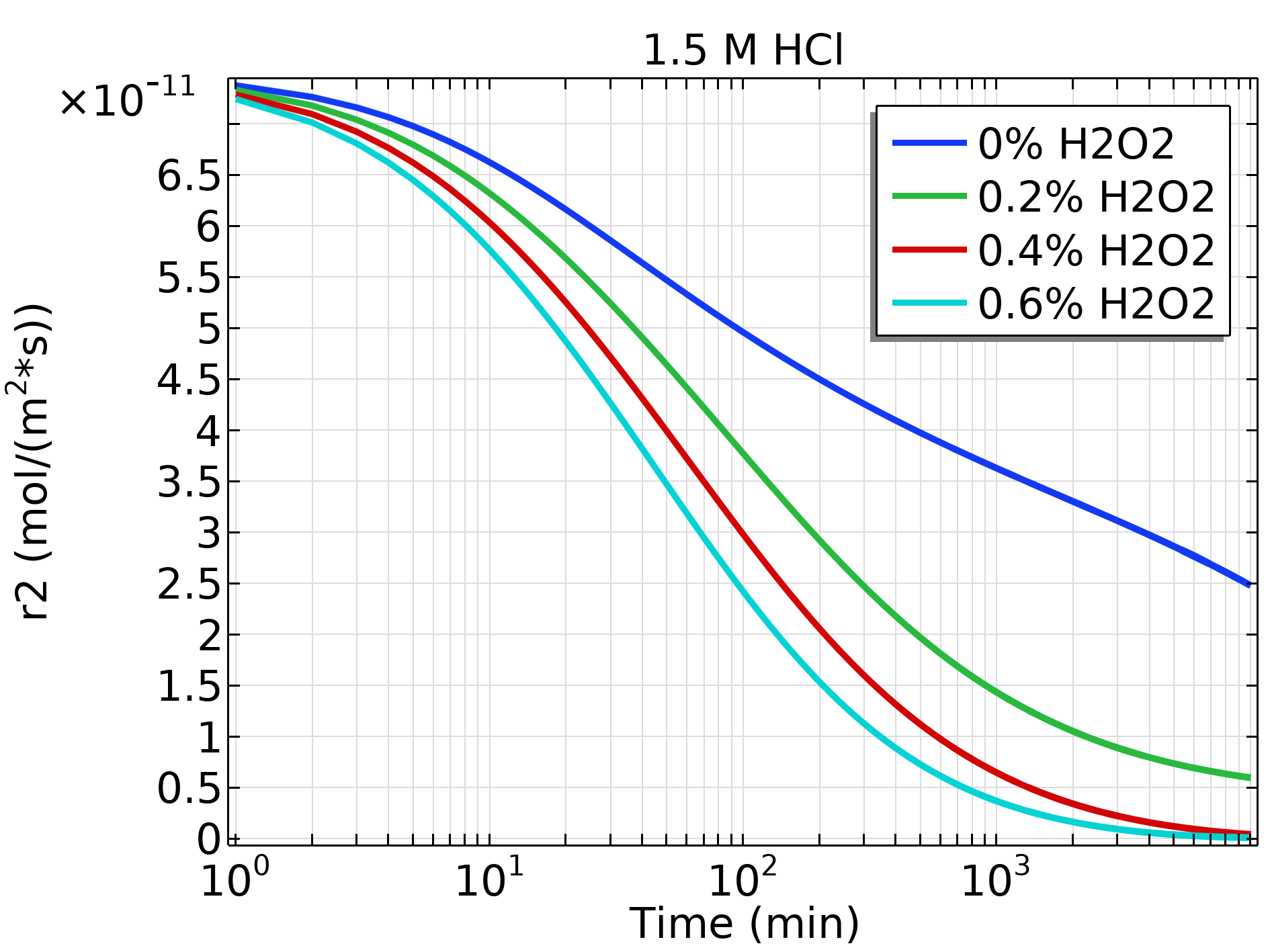}}
    \hfill
    \subfloat[]{\includegraphics[width=0.48\linewidth]{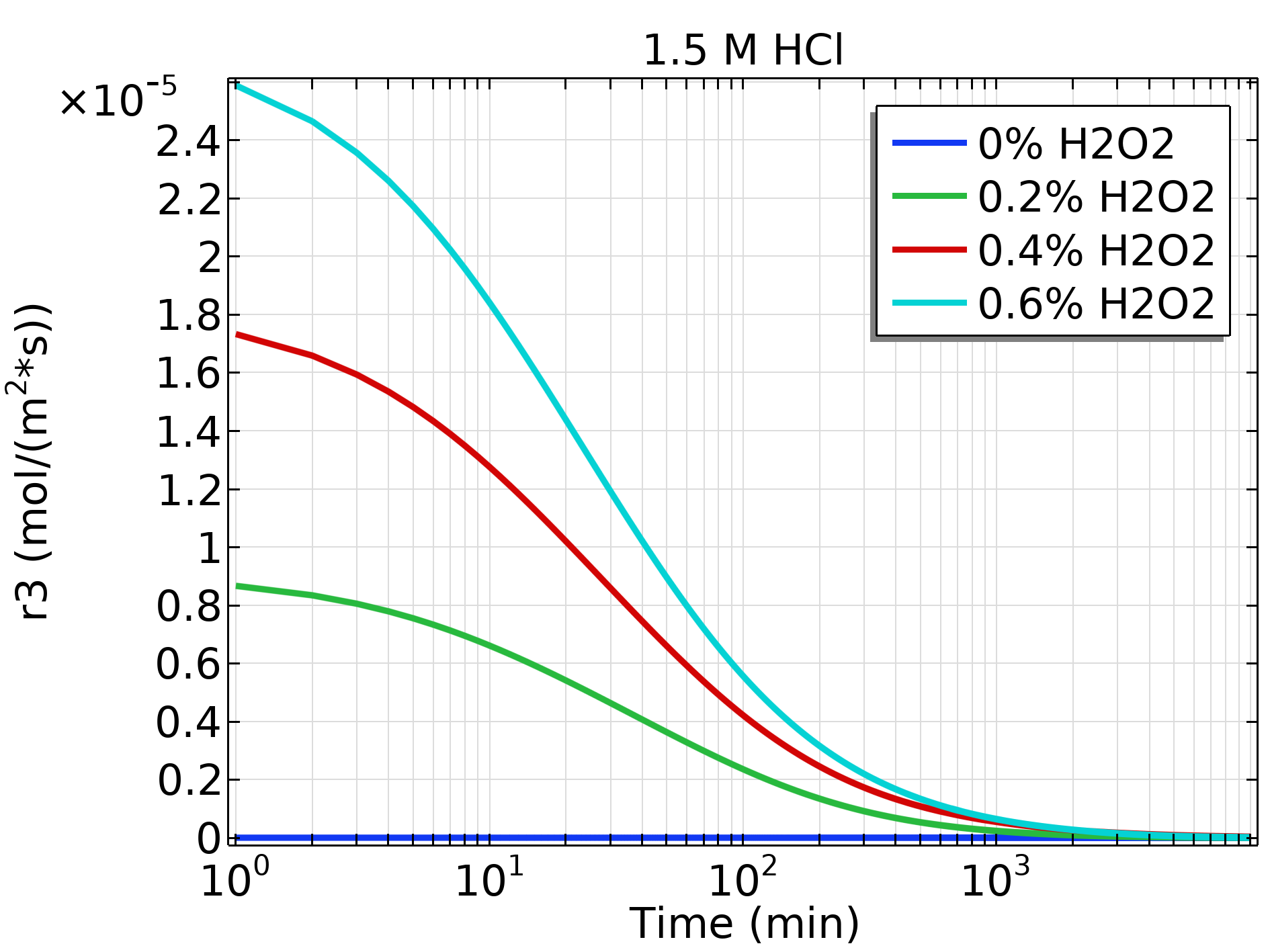}}
    \subfloat[]{\includegraphics[width=0.48\linewidth]{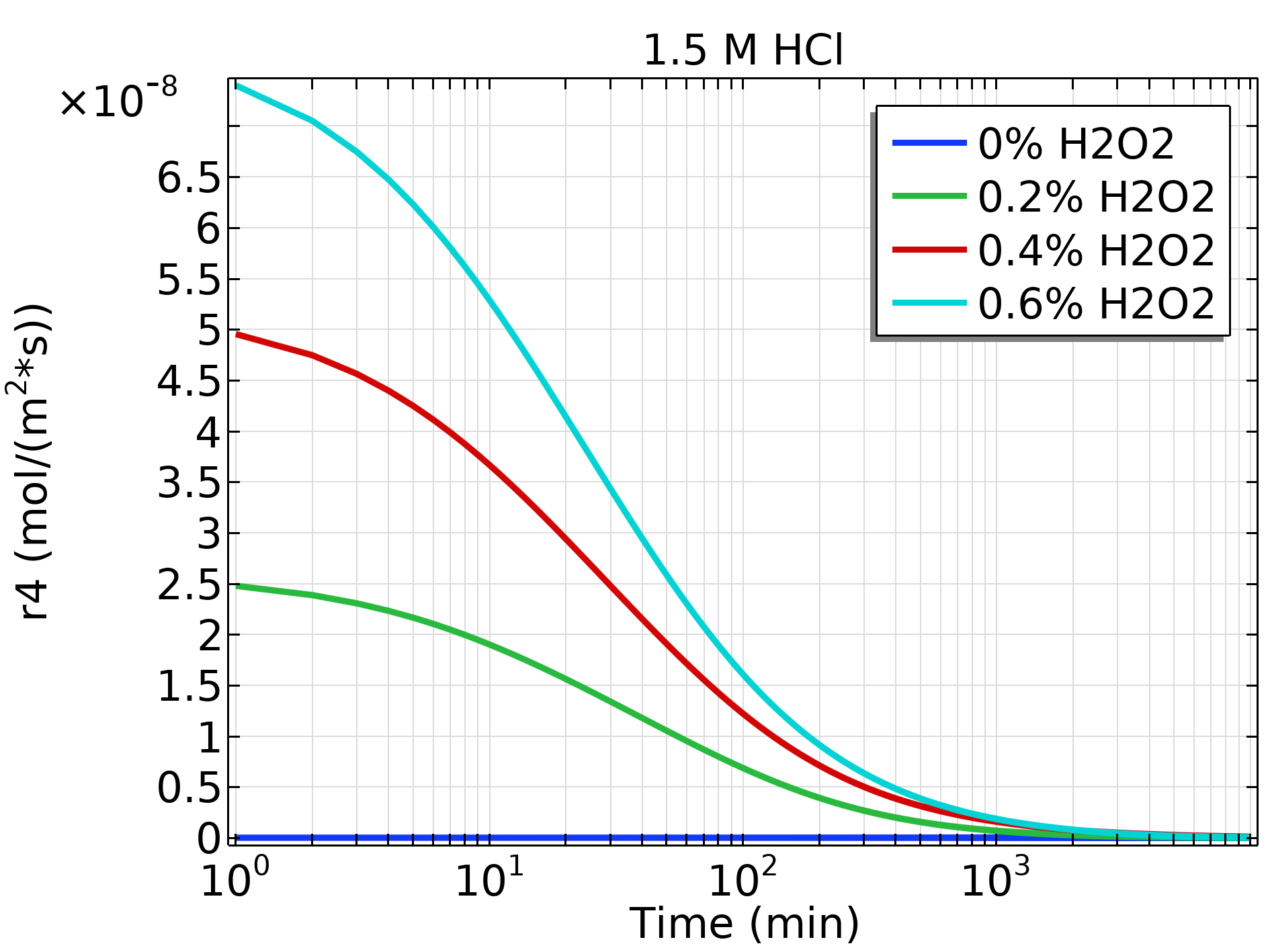}}
    \caption{At 1.5 M Acid: The model predictions of the rates (mol/m$^2$.s) of the four reactions at the four H$_2$O$_2$ concentrations  a) r$_1$, b) r$_2$, c) r$_3$ and d) r$_4$. Order of the rates: R1 ($\sim$10$^{-5}$) > R3 ($\sim$10$^{-5}$) > R4 ($\sim$10$^{-8}$) > R2 ($\sim$10$^{-11}$).}
    \label{fig:1p5M_rates}
\end{figure}

Using the mechanistic model, we can also track the individual rates of the four reactions quantitatively. Figure~\ref{fig:1p5M_rates} shows the rates of the four reactions (\textit{r$_1$} - \textit{r$_4$}) as time progress during the leaching at the four H$_2$O$_2$ concentrations for 1.5 M HCl. The rates are high at beginning of the leaching and keep dropping as time progresses. R1 and R3 occur at higher rates ($\sim$10$^{-5}$). As expected, R2 is the slowest one, to dissolve the film with the acid ($\sim$10$^{-11}$). Interestingly, addition of H$_2$O$_2$ slows down the rates of R1 and R2, with higher concentrations. The reduced availability of acid to R1 and R2 due to the competitive reactions, R3 and R4, is the reason. Whereas, the rates of R3 and R4 are higher in presences of H$_2$O$_2$ and increase with increase in H$_2$O$_2$. 

We can note that the rates of R3 and R4 are zero at 0\% H$_2$O$_2$ as expected. Among the four, fastest is R1 and slowest is R2; the order of rates as: R1 ($\sim$10$^{-5}$) > R3 ($\sim$10$^{-5}$) > R4 ($\sim$10$^{-8}$) > R2 ($\sim$10$^{-11}$). 

\begin{figure}
    \centering
    \includegraphics[width=0.5\linewidth]{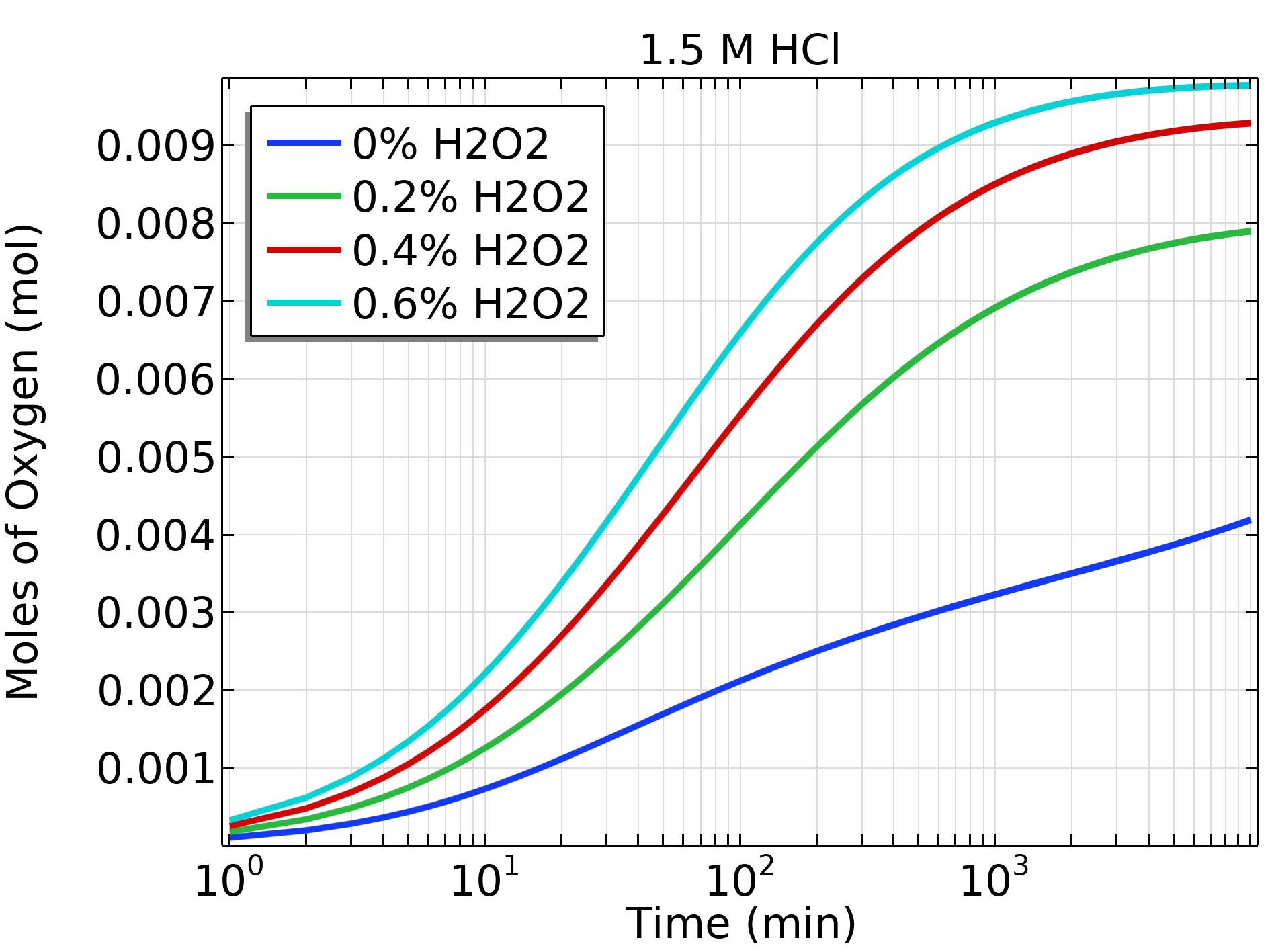}
    \caption{At 1.5 M HCl: The model predictions of the moles of O$_2$ released during leaching at different H$_2$O$_2$ concentrations. Higher the conversion, higher the moles of O$_2$ released.}
    \label{fig:molesO2}
\end{figure}

The moles of oxygen released, another feature of the mechanistic model, during the leaching at the four H$_2$O$_2$ concentrations for 1.5 M HCl are shown in Figure~\ref{fig:molesO2}. The release of the gas is proportional to the conversion of the LCO, which is again proportional to the concentration of H$_2$O$_2$; higher the conversion, higher the moles of the gas released. 

\subsubsection{At 1.5 M acid: without the film passivation}
\begin{figure}
    \centering
    \subfloat[]{\includegraphics[width=0.48\linewidth]{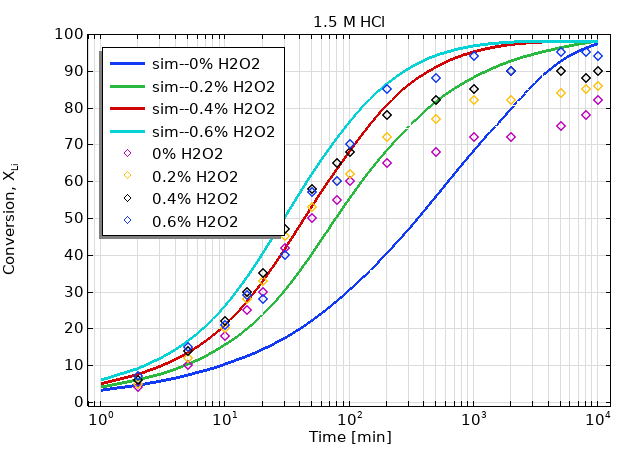}}
    \subfloat[]{\includegraphics[width=0.48\linewidth]{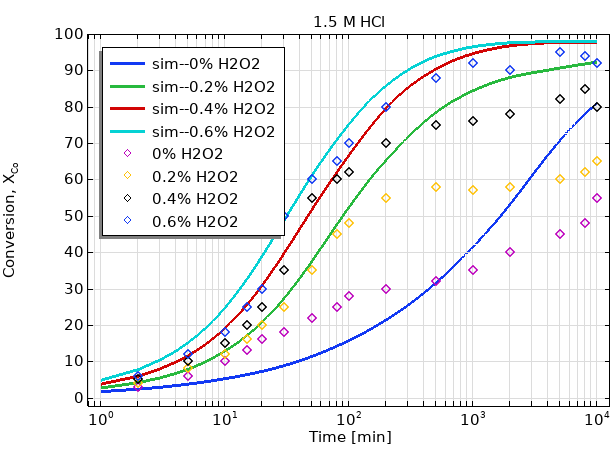}}
    \caption{At 1.5 M Acid: The model without the film passivation. The model over-predicts the data of \citet{cerrillo2022acid} at 0\% H$_2$O$_2$ or low concentrations of H$_2$O$_2$ and couldn't match with the data.}
    \label{fig:1p5M_X_nofilm}
\end{figure}

To demonstrate the need for the film passivation in the model, we simulated it by keeping $f_{acc}$ to 1, which indicates no film passivation, for the 1.5 M HCl. The predictions are shown in Figure~\ref{fig:1p5M_X_nofilm}. The results show the model's over-prediction and inability to capture the trend of recovery of Li and Co at the 0\% H$_2$O$_2$, which are non-sigmoid in shape. On the other hand the model is doing well at 0.6\% H$_2$O$_2$, where the film thickness effects are minimal. 

These observations motivated us to introduce an additional physics, beyond the diffusion limitations through the film, which inhibits the dissolution reactions further with film growing; so, we introduced the film passivation through accessibility factor which effectively reduces reaction surface for the dissolution reactions, dependent on the film volume. Rest of the simulations in the work are with the film passivation.

\subsection{At 0.5 M acid}
\begin{figure}
    \centering
    \subfloat[]{\includegraphics[width=0.48\linewidth]{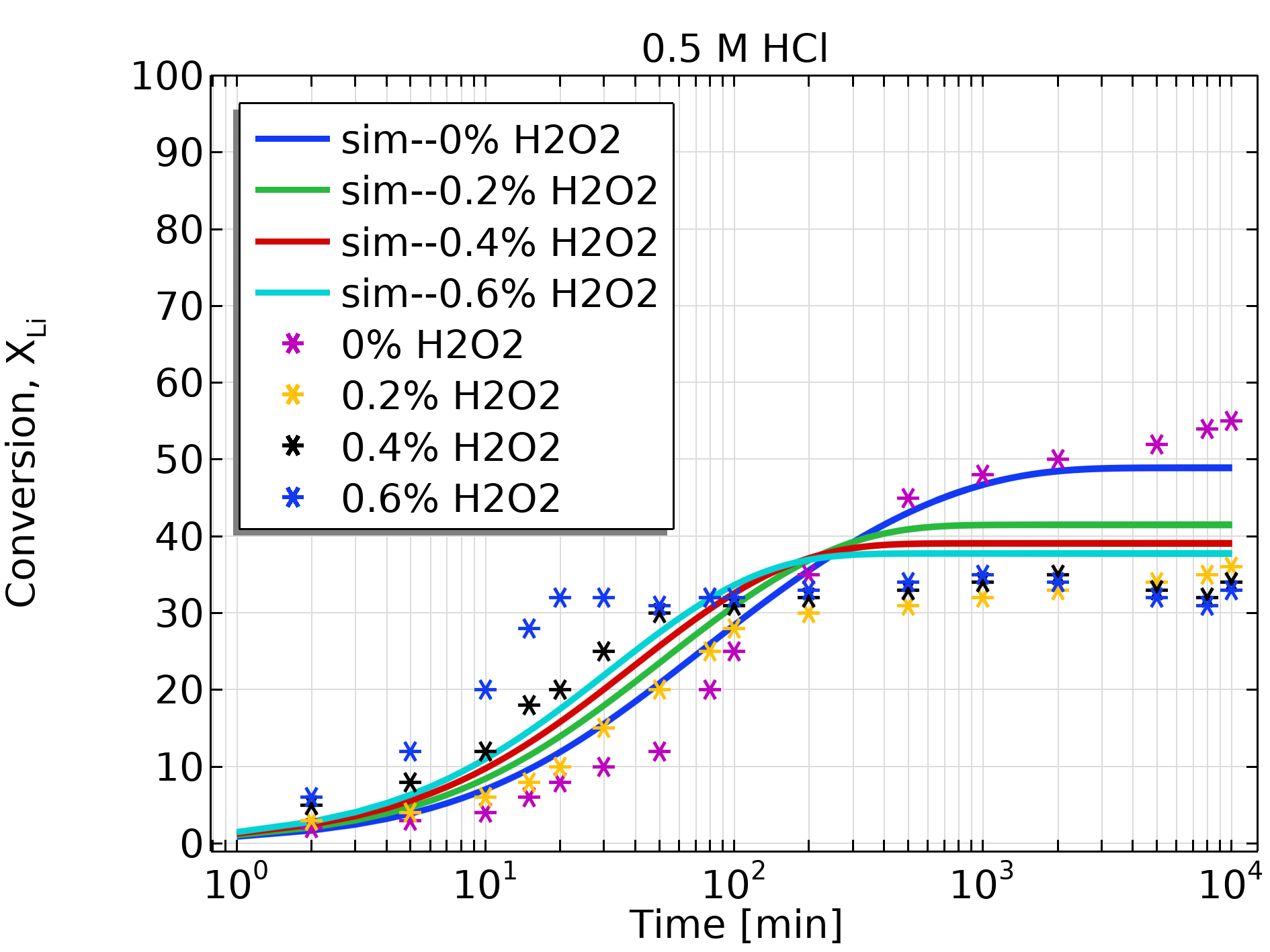}}
    \subfloat[]{\includegraphics[width=0.48\linewidth]{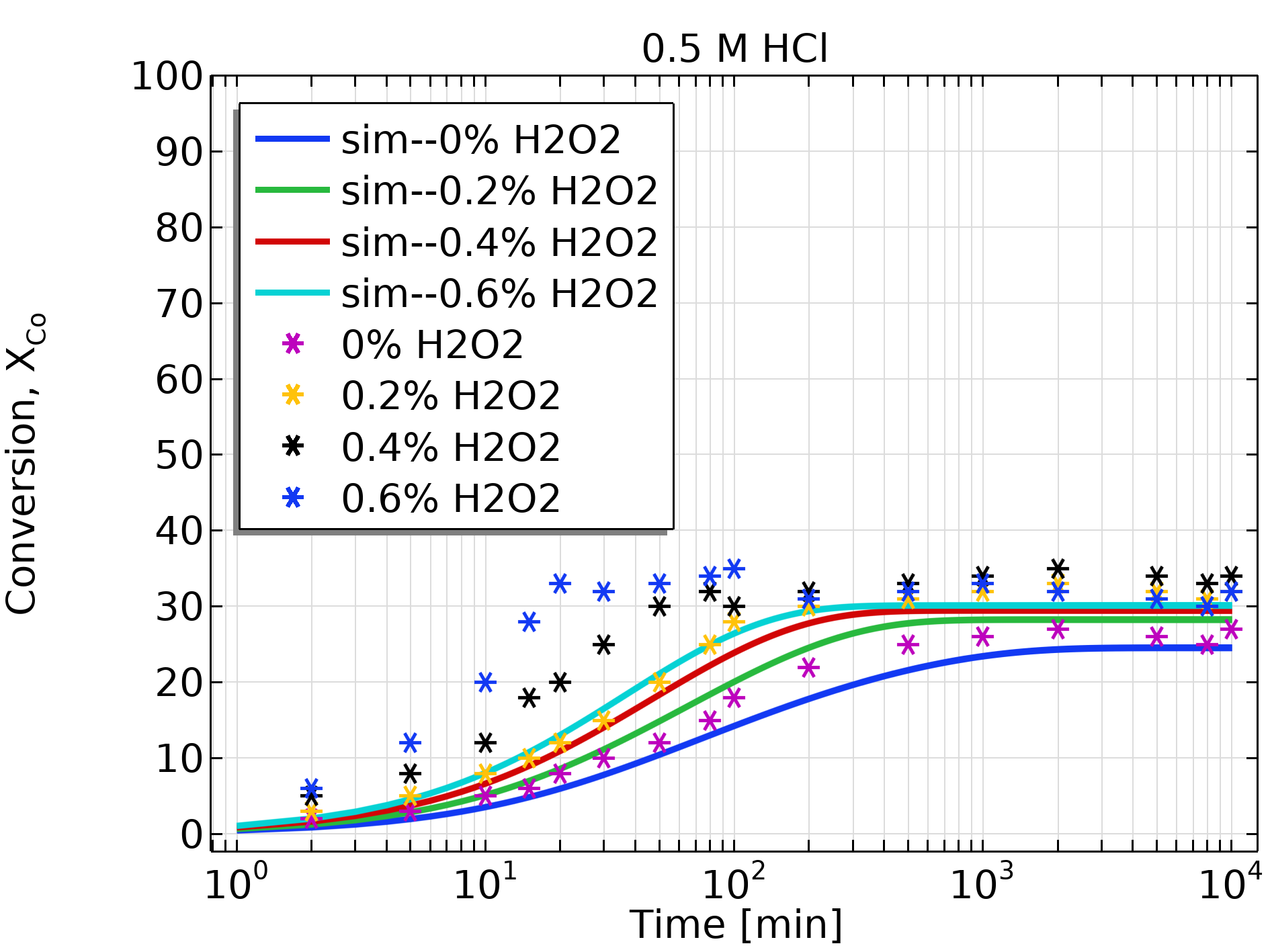}}
    \caption{At 0.5 M Acid: The model predictions with the film passivation compared against the experimental data of \citet{cerrillo2022acid} at four H$_2$O$_2$ concentrations. a) Conversion of Li and b) conversion of Co. }
    \label{fig:0p5M_X}
\end{figure}

We now extend the model simulations for 0.5 M Acid with the four H$_2$O$_2$ concentrations. The predictions are shown in Figure~\ref{fig:0p5M_X}. The model is able to predict the trajectories of Li and Co conversion data reasonably well at the four H$_2$O$_2$ concentrations. For Li, there is a cross-over of the profiles at around 2E2 min, leading to higher recovery of Li around 50\% at 0\% H$_2$O$_2$. For the rest of the H$_2$O$_2$ concentrations, the maximum recovery is around 40\%. The model predicted these trajectories well. For Co, there is no cross-over of trajectories and the maximum recovery stayed below 35\%; the model predicted these trajectories as well.  

\begin{figure}
    \centering
    \subfloat[]{\includegraphics[width=0.48\linewidth]{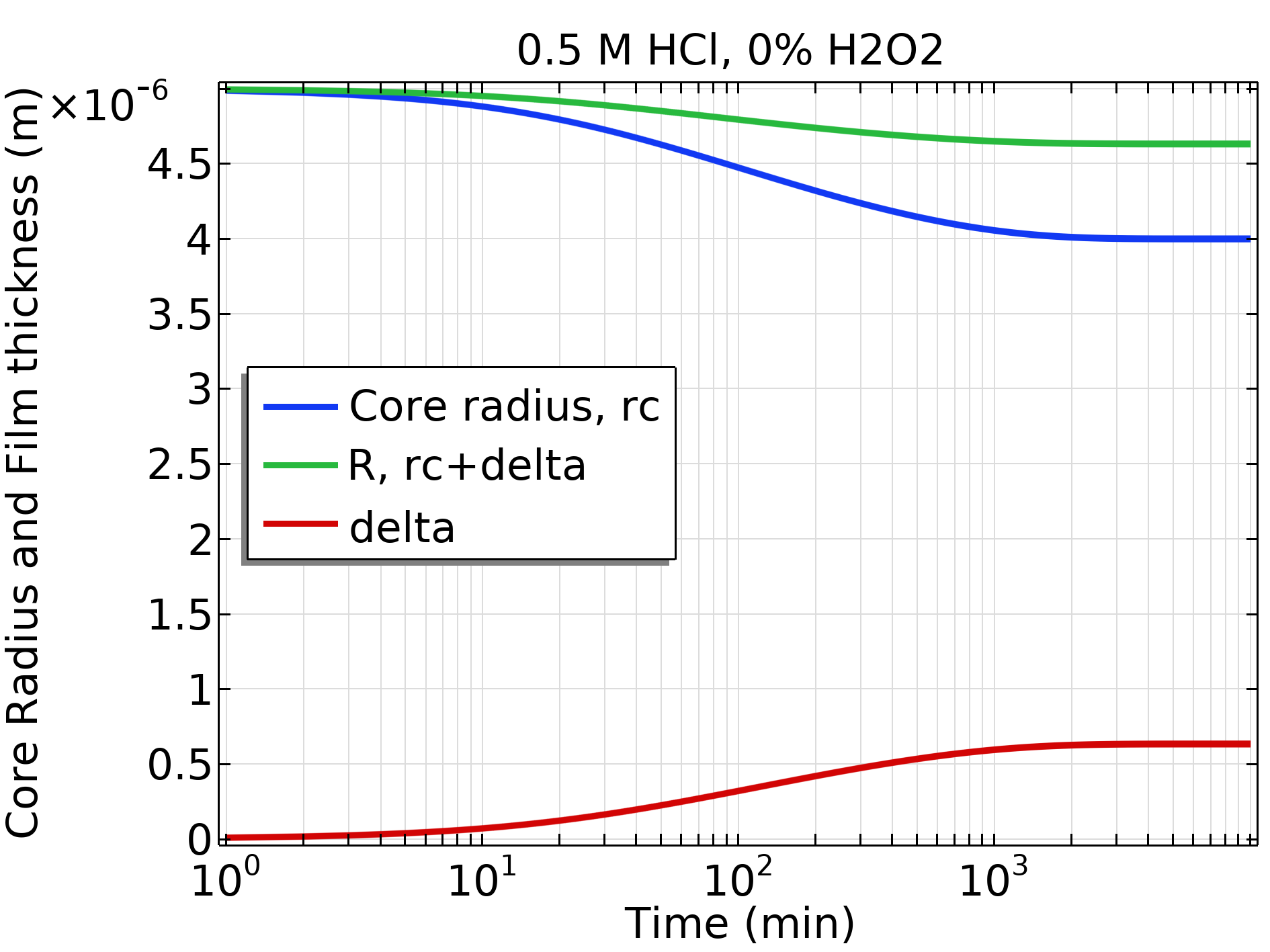}}
    \subfloat[]{\includegraphics[width=0.48\linewidth]{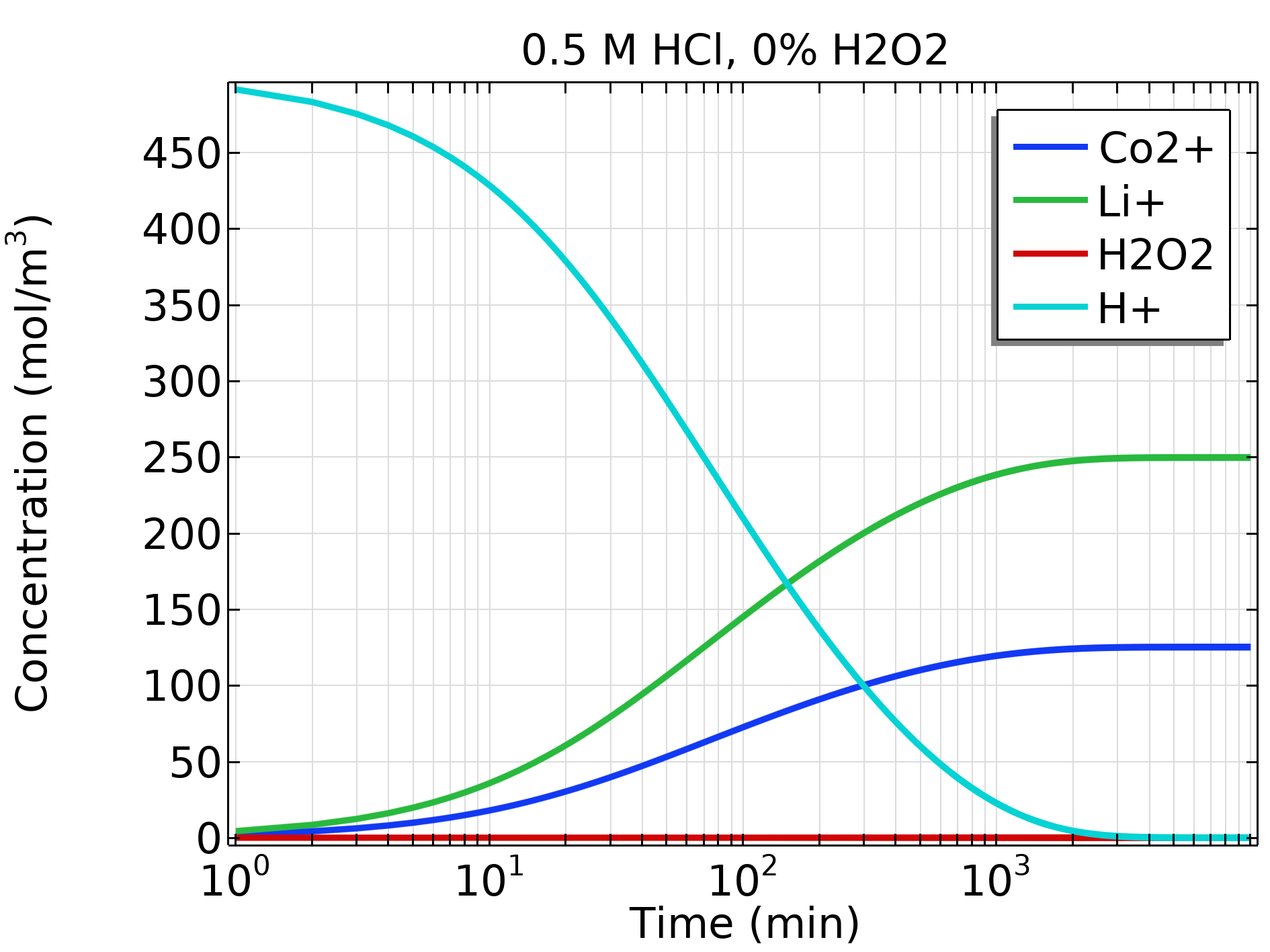}}
    \hfill
    \subfloat[]{\includegraphics[width=0.48\linewidth]{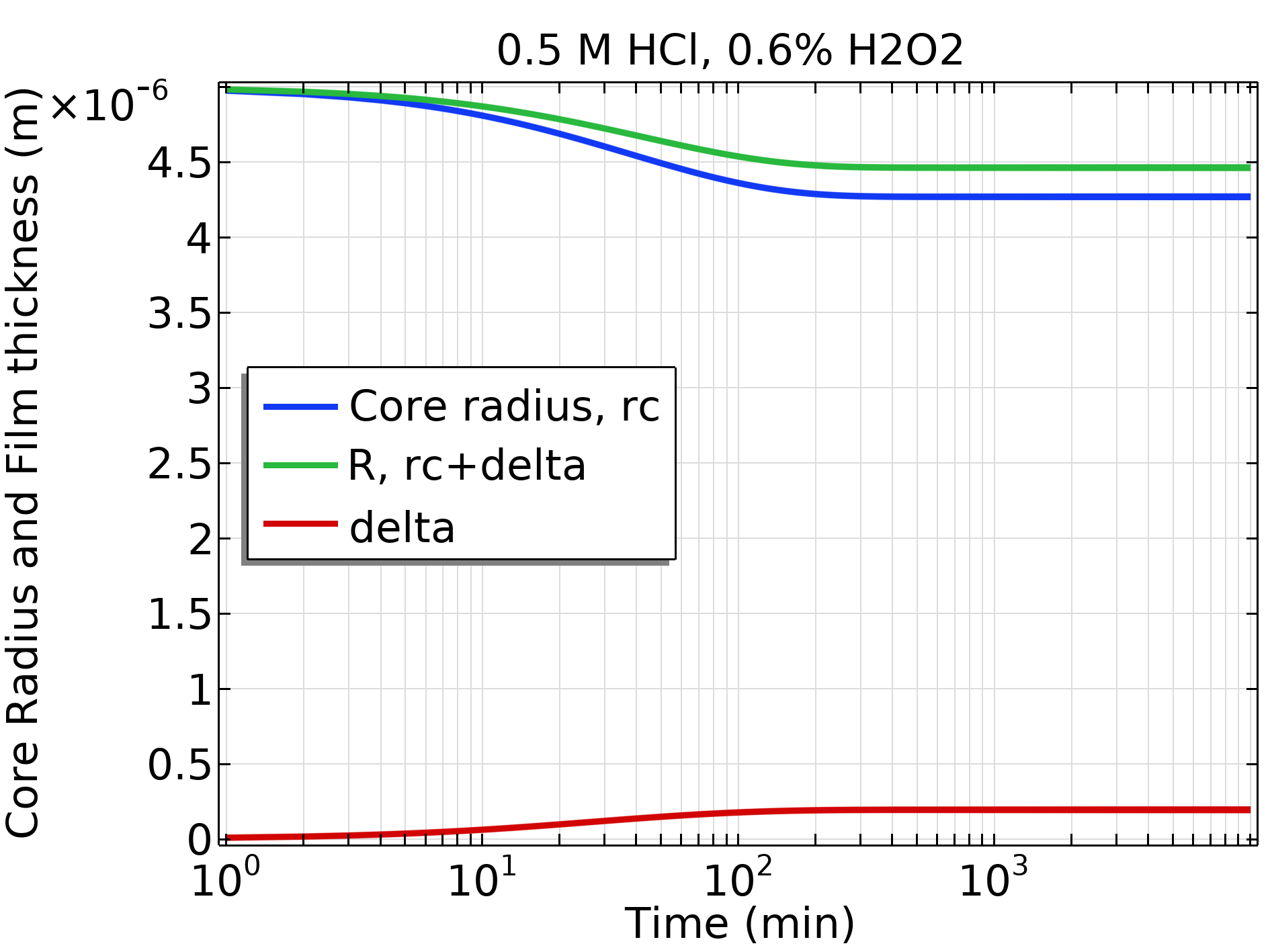}}
    \subfloat[]{\includegraphics[width=0.48\linewidth]{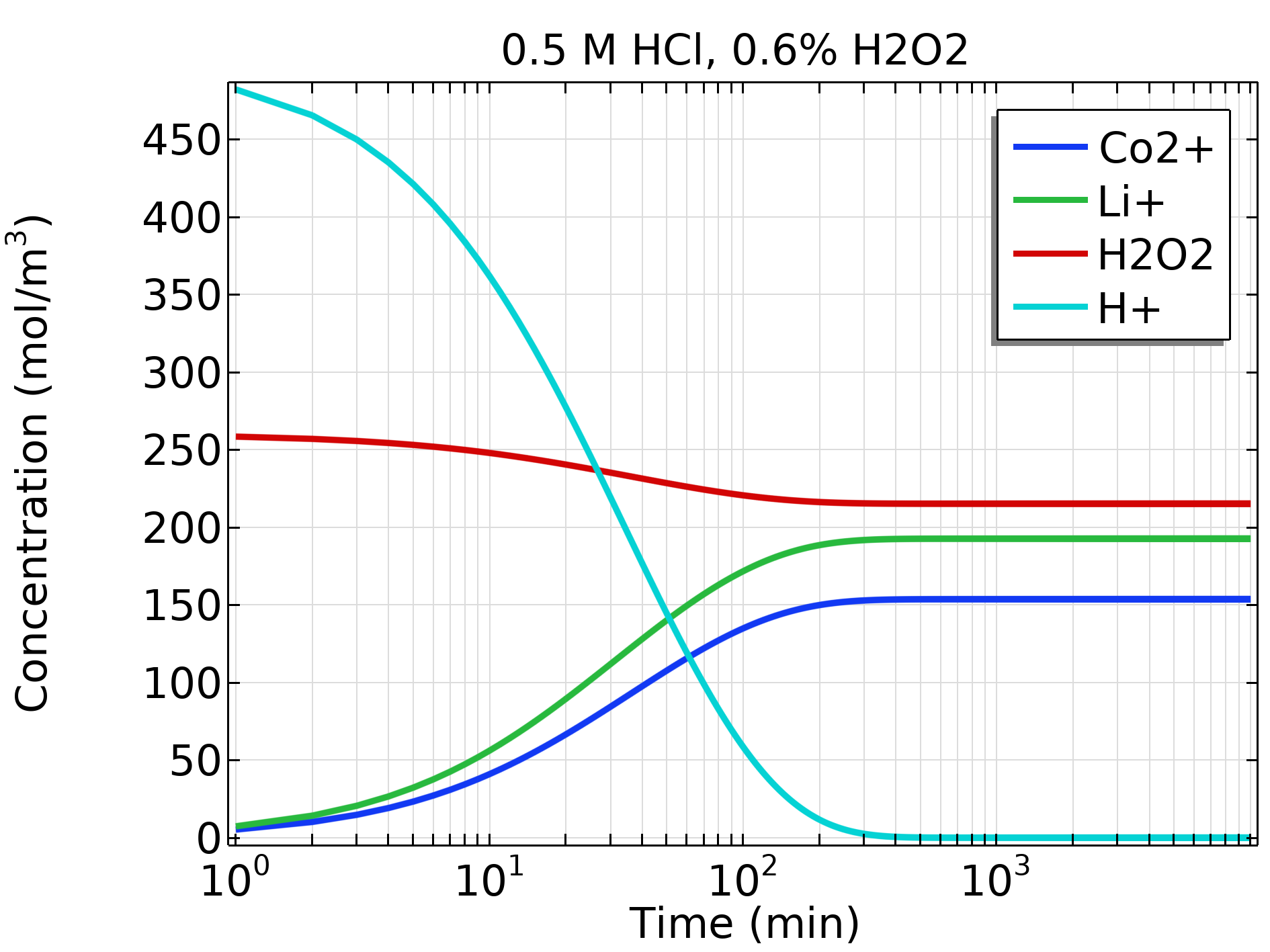}}
    \caption{At 0.5 M Acid: the model predictions with the film passivation. The profiles of LCO core particle radius (rc), the film thickness (delta), and total particle radius (rc + delta); and concentrations of Co$^{2+}$, Li$^{+}$, the acid and H$_2$O$_2$. a \& b) at 0\% H$_2$O$_2$ b). c \& d) 0.6\% H$_2$O$_2$.}
    \label{fig:0p5M_rc_conc}
\end{figure}

The reason for less recovery is evident through Figure~\ref{fig:0p5M_rc_conc}. The figure shows the concentrations of the four variables, the core particle radius and the film thickness, for the leaching with the 0.5\% acid at 0\% H$_2$O$_2$ and 0.6\% H$_2$O$_2$. It shows that, after 2E3 min at the 0\% H$_2$O$_2$ and 3E2 min at the 0.6\% H$_2$O$_2$, the acid is fully consumed and the recovery is then limited by its unavailability for further leaching at all the four H$_2$O$_2$ concentrations. The corresponding profiles for the rest of the variables look consistent with the acid profile and the conversion profiles. 

The cross-over observed earlier in the Li profiles is an artifact of the limited availability dynamics of the acid with time at different concentrations of H$_2$O$_2$ for the competitive R1 - R4 recovery reactions. In the absence of H$_2$O$_2$, the acid is mostly consumed by R1, leading to higher Li$^+$ recovery and lower Co$^{2+}$. 

The concentration and the film thickness related profiles for the other two H$_2$O$_2$ concentrations, the rate profiles, and the oxygen profiles for the 0.5 M HCl are shown in \ref{appnd:0p5M}.

\subsection{At 2.5 M acid}
\begin{figure}
    \centering
     
    \subfloat[]{\includegraphics[width=0.48\linewidth]{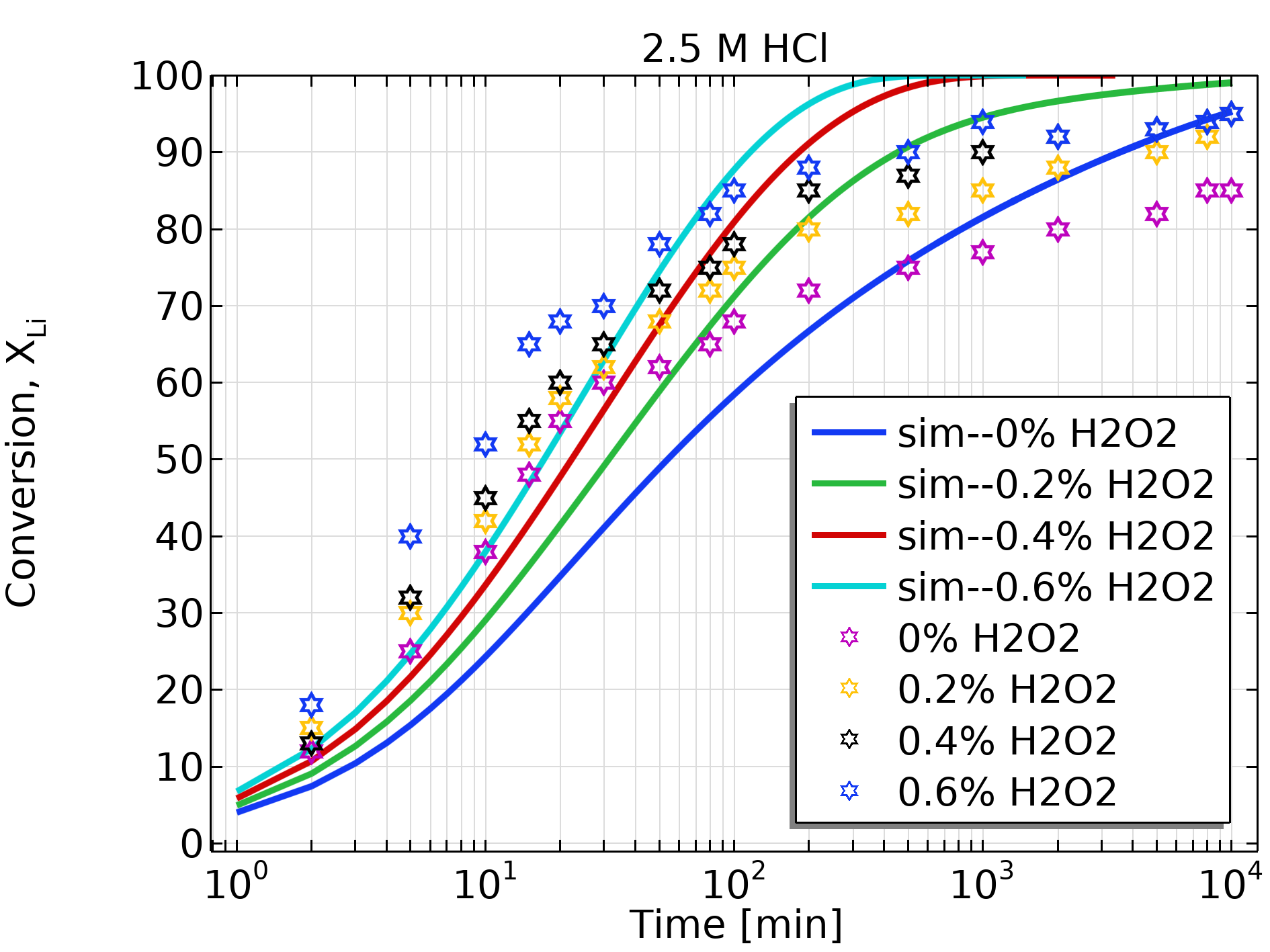}}
    \subfloat[]{\includegraphics[width=0.48\linewidth]{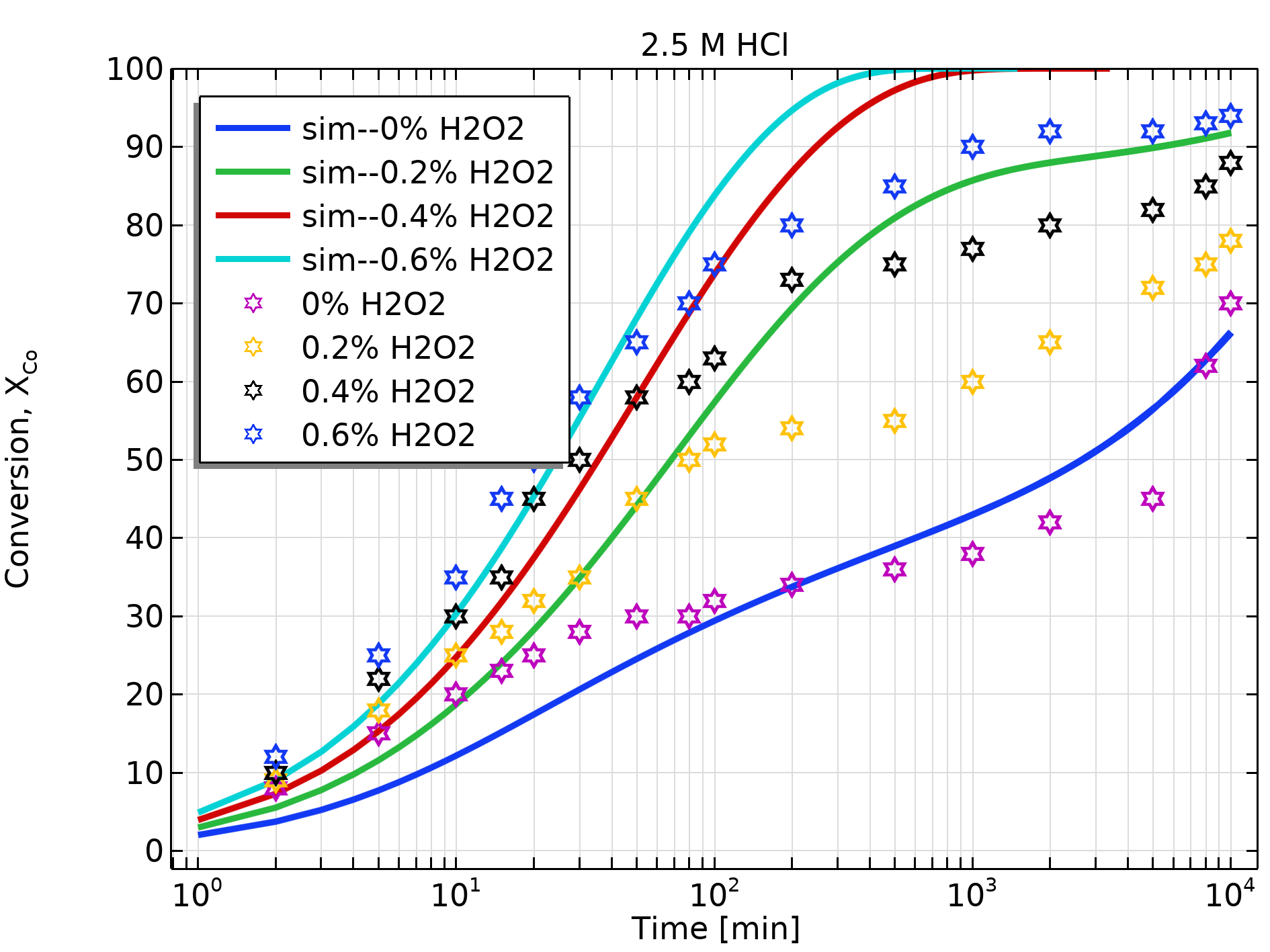}}
    \caption{At 2.5 M Acid: The model predictions with the film passivation compared against the experimental data of \citet{cerrillo2022acid} at four H$_2$O$_2$ concentrations. a) Conversion of Li and b) conversion of Co.}

    \label{fig:2p5M_X}
\end{figure}

\begin{figure}
    \centering
     
    \subfloat[]{\includegraphics[width=0.48\linewidth]{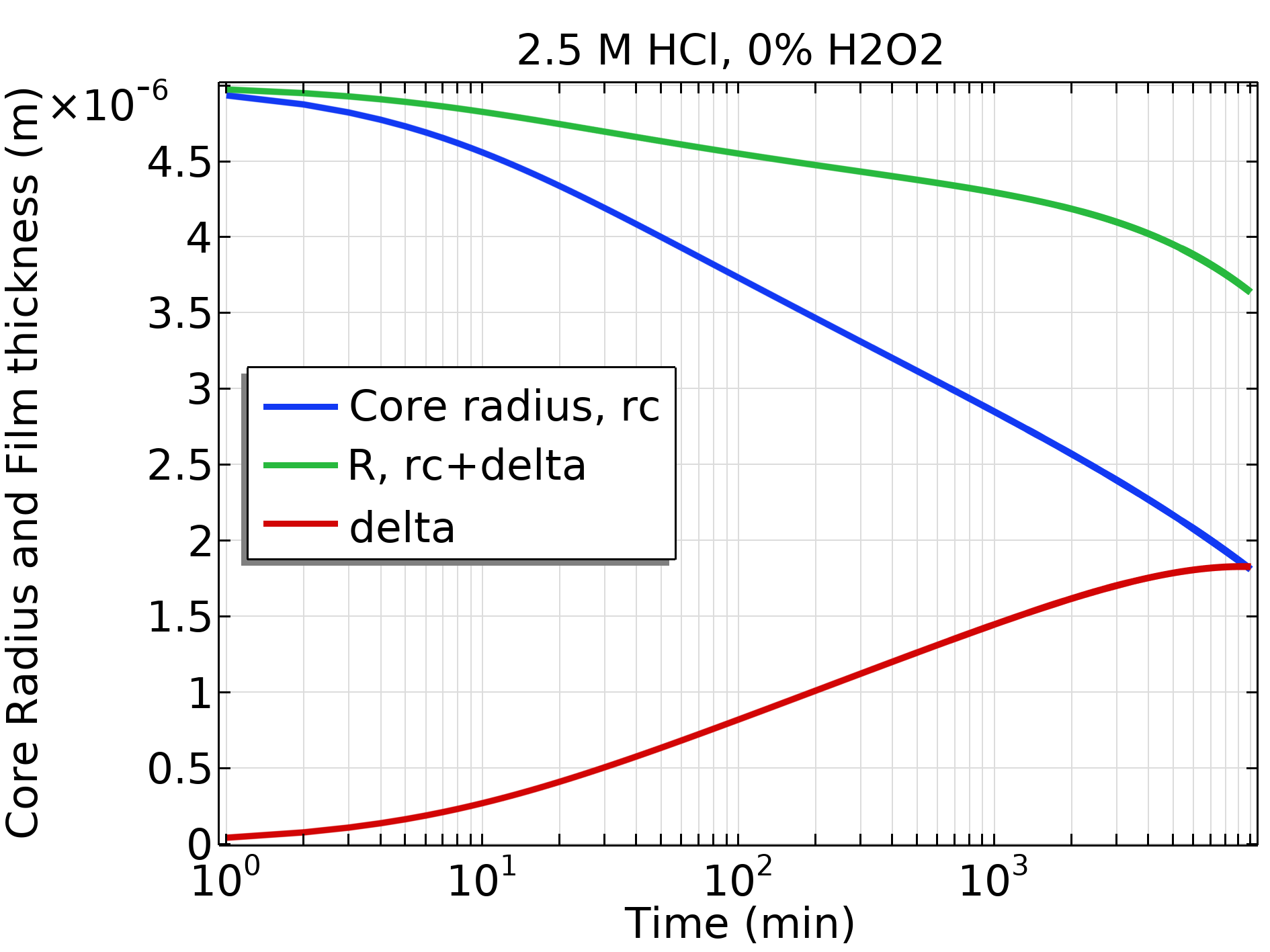}}
    \subfloat[]{\includegraphics[width=0.48\linewidth]{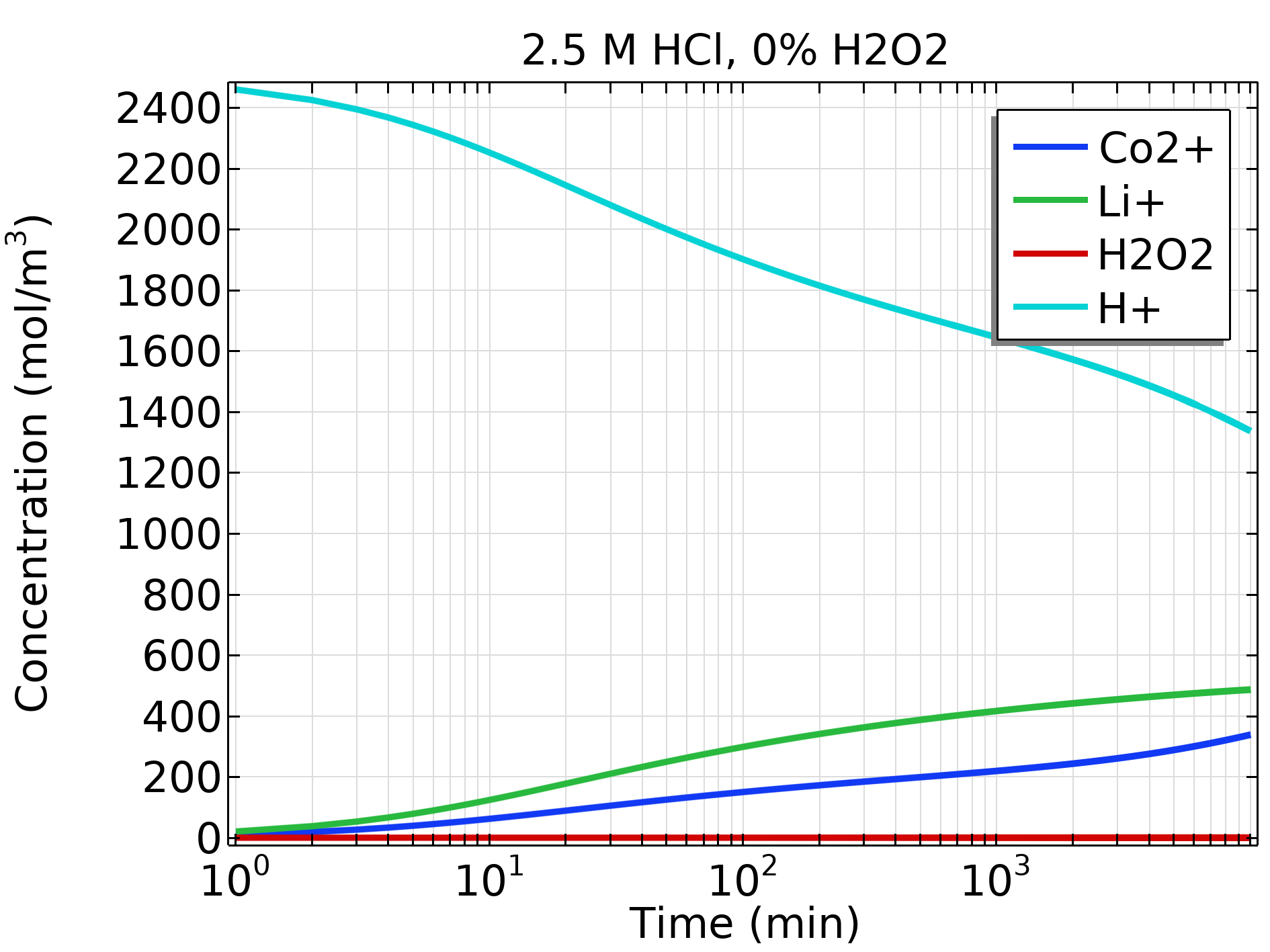}}
    \hfill
    \subfloat[]{\includegraphics[width=0.48\linewidth]{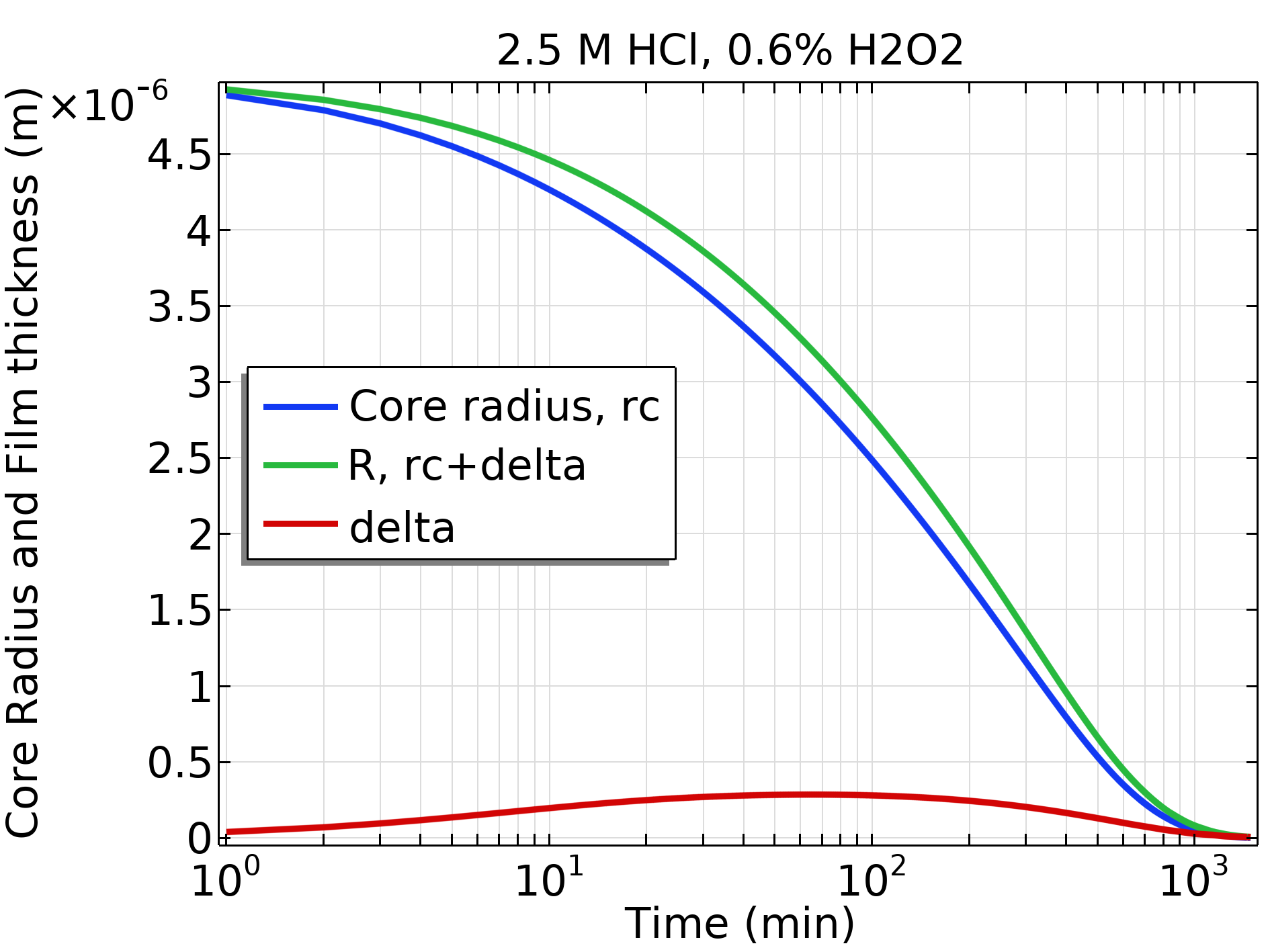}}
    \subfloat[]{\includegraphics[width=0.48\linewidth]{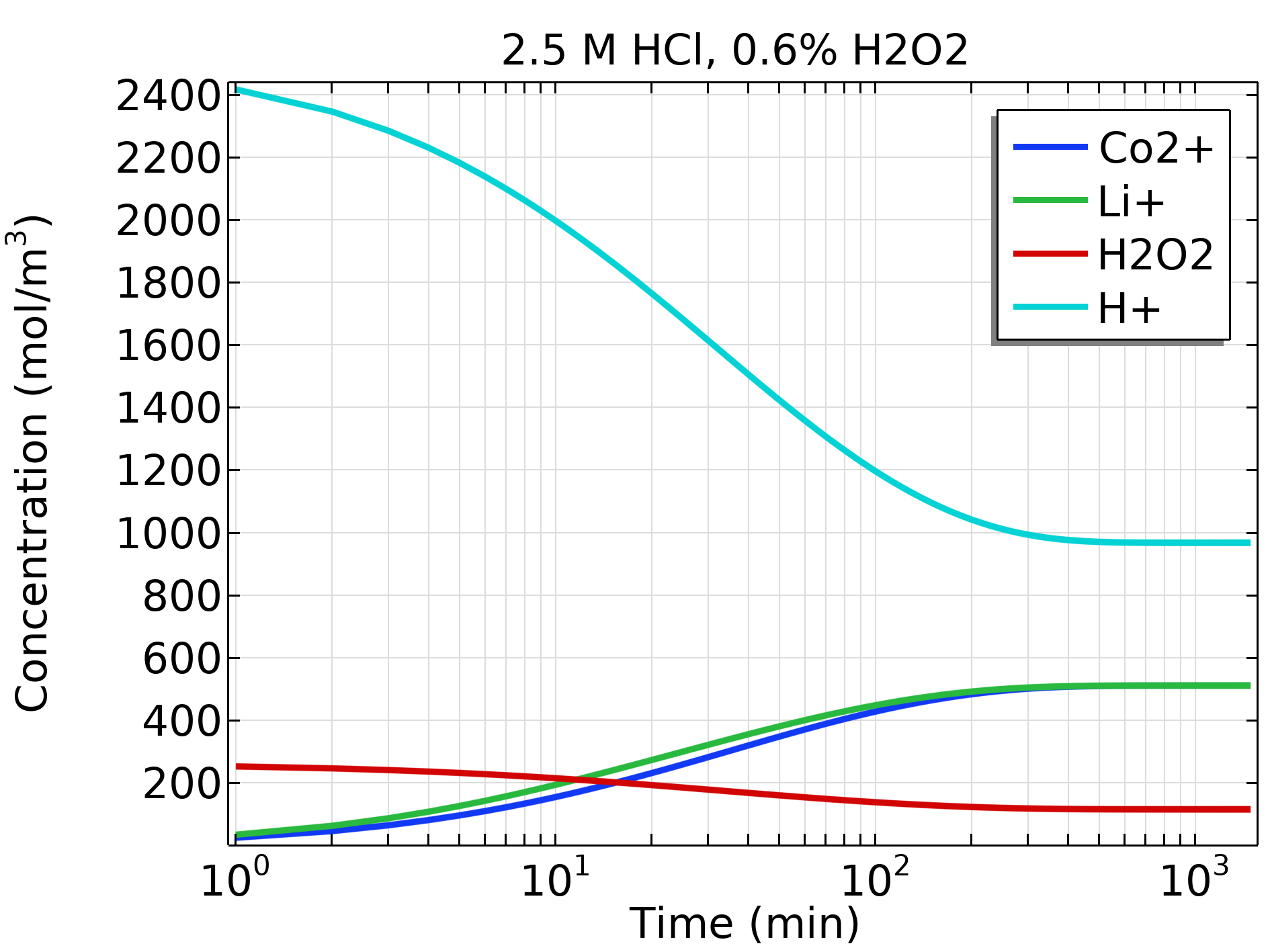}}
    \caption{At 2.5 M Acid: the model predictions with the film passivation. The profiles of LCO core particle radius (rc), the film thickness (delta), and total particle radius (rc + delta); and concentrations of Co$^{2+}$, Li$^{+}$, the acid and H$_2$O$_2$. a \& b) at 0\% H$_2$O$_2$ b). c \& d) 0.6\% H$_2$O$_2$.}

    \label{fig:2p5M_rc_conc}
\end{figure}

The simulations of Li and Co recovery for the 2.5 M acid concentration with the four concentrations of H$_2$O$_2$ are shown in Figure~\ref{fig:2p5M_X}. The corresponding change in the radius and the four concentration variables are presented in Figure~\ref{fig:2p5M_rc_conc} for 0\% and 0.6\% H$_2$O$_2$. We can clearly see that the particle radius and the film thickness approaching to zero, at 0.6\%, with the conversions approaching to 100\%, given no limitation due to H$_2$O$_2$ or acid concentrations. The concentrations of Li and Co have evolved to their theoretical maximum values ($\sim$ 510 mol/m$^3$) to be consistent. 

The results show that the model is over-predicting the conversion of Co and its rate, in presence of H$_2$O$_2$, while it predicted the recovery well at 0\% H$_2$O$_2$. Similar observations of over-prediction of the conversion for Li at higher concentration of H$_2$O$_2$ can be made from the figure. These observations indicate that the model rate expressions for R3 and R4, in presence of H$_2$O$_2$, are over-sensitive to the acid concentration at higher values. 

While it works reasonable well at 0.5 M and 1.5 M, at the 2.5 M, the order of the acid concentration needs to be revisited in a physically consistent manner. Our efforts to tune the rate constants, the diffusivities, accessibility factor, tortuosities etc didn't result a positive outcome and negatively affected the earlier fittings at the other acid concentrations, which were otherwise fine. So, we didn't change further the reasonably working set of parameters.

Using fractional orders for acid (less < 1) as \citet{cerrillo2022acid} followed could be one approach. We didn't prefer this approach due to lack of physical interpretation in our understanding. Another approach could be using activities of the acid  and introducing activity coefficient correction to the concentration at higher values, instead of just concentration. Also, the role of oxygen bubbles interference at higher concentrations, as suggested by \citet{cerrillo2022acid}, in presence of H$_2$O$_2$, could  be investigated to capture the sensitivity to the acid. The research community can further look into these directions for the improved fitting.

For completeness, the concentration and the film thickness related profiles for the other two H$_2$O$_2$ concentrations, the rate profiles, and the oxygen profiles for the 2.5 M HCl are shown in \ref{appnd:2p5M}.

\section{Conclusion}
Increased volumes of end-of-life lithium-ion batteries for battery recycling demand efficient recycling technologies. Physics-based models for the battery recycling models can help increase through-put,  enhance operating efficiencies for improved sustainability and increase profitability. However, the literature largely focussed on empirical modelling approaches for scale-up from the laboratories. 

Built on the foundation of \citet{cerrillo2022acid}'s work, the current work developed and tested a new reaction modelling framework for the leaching of Li and Co from the LCO cathode material using HCl with/without H$_2$O$_2$. The model includes reaction-diffusion limited dynamics of the acid and reducing agent through the Co$_3$O$_4$ film during leaching of LCO particles and the film passivation effects. The film passivation captures the reduced recovery in the absence of H$_2$O$_2$. 

The modelling framework features a full list of operating parameters, variables and clear set of modelling equations for reproducibility and for easier adoption to other chemistries. 

The mechanistic model tracked the conversion of LCO, the LCO core and total radiuses, the Co$_3$O$_4$ film thickness, concentrations of the acid and H$_2$O$_2$, formation-dissolution dynamics of the film under different H$_2$O$_2$ concentrations, conversions of lithium and cobalt, influence of the film passivation on the conversions and moles of O$_2$. It also compared and quantified the rates of the four reactions during leaching in response to the operating conditions. The model predictions of the variables discussed above have been demonstrated. 

The model was validated against the experimental data reported in the literature\cite{cerrillo2022acid} reasonably well for the recovery of Li and Co, with and without H$_2$O$_2$, at 0.5 M and 1.5 M of the acid. It captured the dynamics of acidic-reductive leaching of LCO under the given operating conditions. 

At 2.5 M acid, the model over-predicted the recovery of Li and Co, in presence of H$_2$O$_2$, while it worked well in the absence of H$_2$O$_2$. We suspect that the acid concentration terms in the rate of R3 and R4 were over-sensitive, at higher concentrations. We suggested the following approaches to improve the model functionality: fractional orders for acid (less < 1) in R3 and R4 with some physical interpretation; using activities of the acid  instead of the concentration; the role of the oxygen bubbles interference with the leaching process.

The framework can be extended and adopted for other cathode chemistries and acidic-reductive environments, given that the reaction pathways are known. The work can help optimise the operating conditions and enable scale-up for handling large volumes of the cells efficiently.

\section*{Funding}
G. Madabattula acknowledges the seed grant and the COMSOL Multiphysics software license from IIT (BHU) for enabling this work. 

\section*{Declaration of generative AI and AI-assisted technologies}
During the preparation of this work, the authors used ChatGPT (Go) in order to help with the basics of the development of the heterogenous reaction rates model. After using this tool, the authors reviewed and edited the model as needed and take full responsibility for the content of the published article.

\section*{Declaration of competing interest}
The authors declare that they have no known competing financial interests or personal relationships that could have appeared to influence the work reported in this paper. 

\section*{CRediT statement}
\textbf{U. Sarma}: Conceptualisation, methodology, model development, formal analysis, investigation, visualisation, writing-original draft preparation. 
\textbf{G. Madabattula}: Conceptualisation, methodology, model development, investigation, writing-original draft preparation and review and editing, funding, project administration, supervision.

\section*{ORCID ID}
Uddipta Sarma: https://orcid.org/0009-0003-3374-4794\\
Ganesh Madabattula: https://orcid.org/0000-0001-7915-0770  \\

\appendix
\section{Additional simulations for 0.5 M HCl}
\label{appnd:0p5M}
\begin{figure}
    \centering
    \subfloat[]{\includegraphics[width=0.48\linewidth]{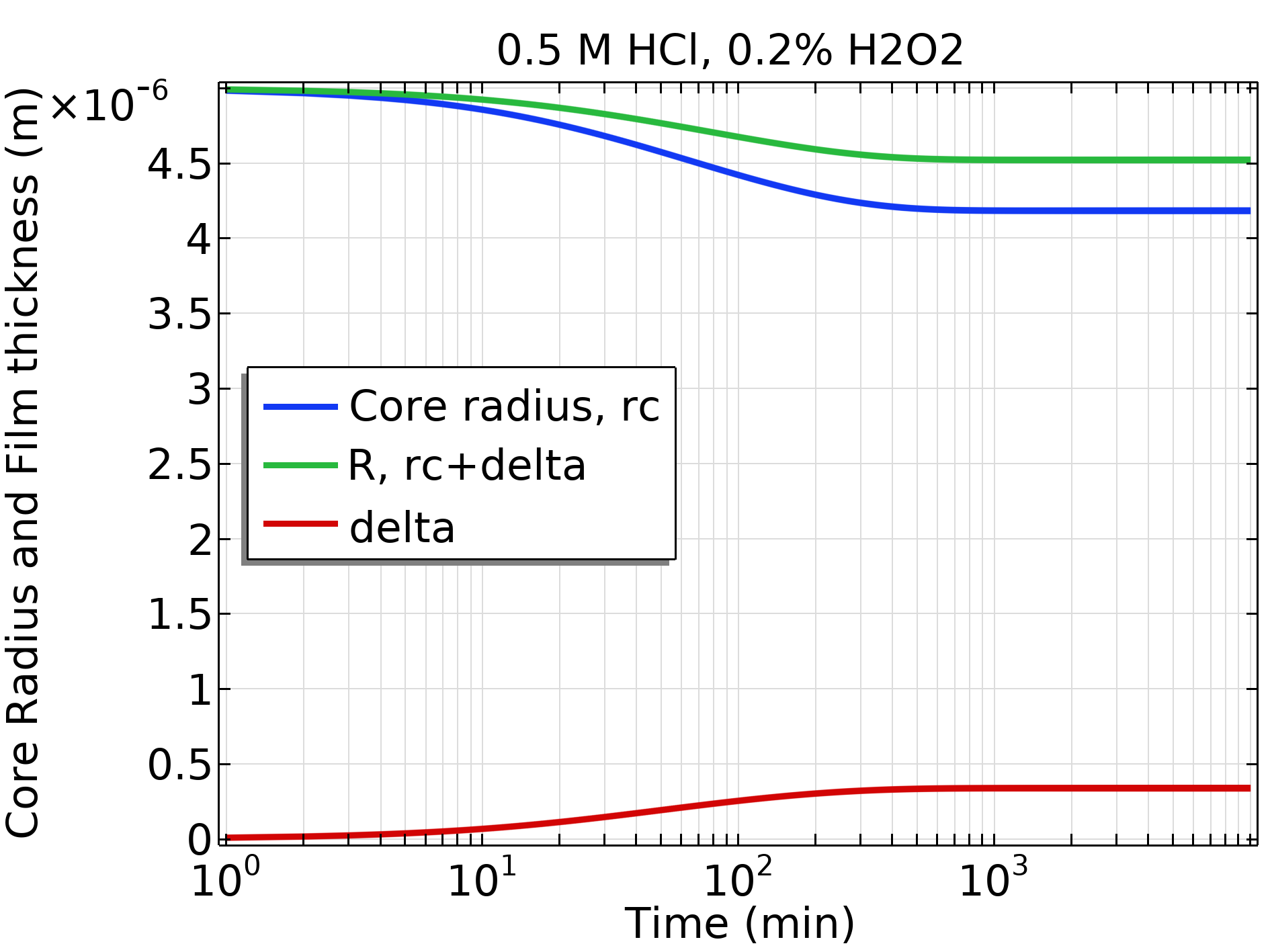}}
    \subfloat[]{\includegraphics[width=0.48\linewidth]{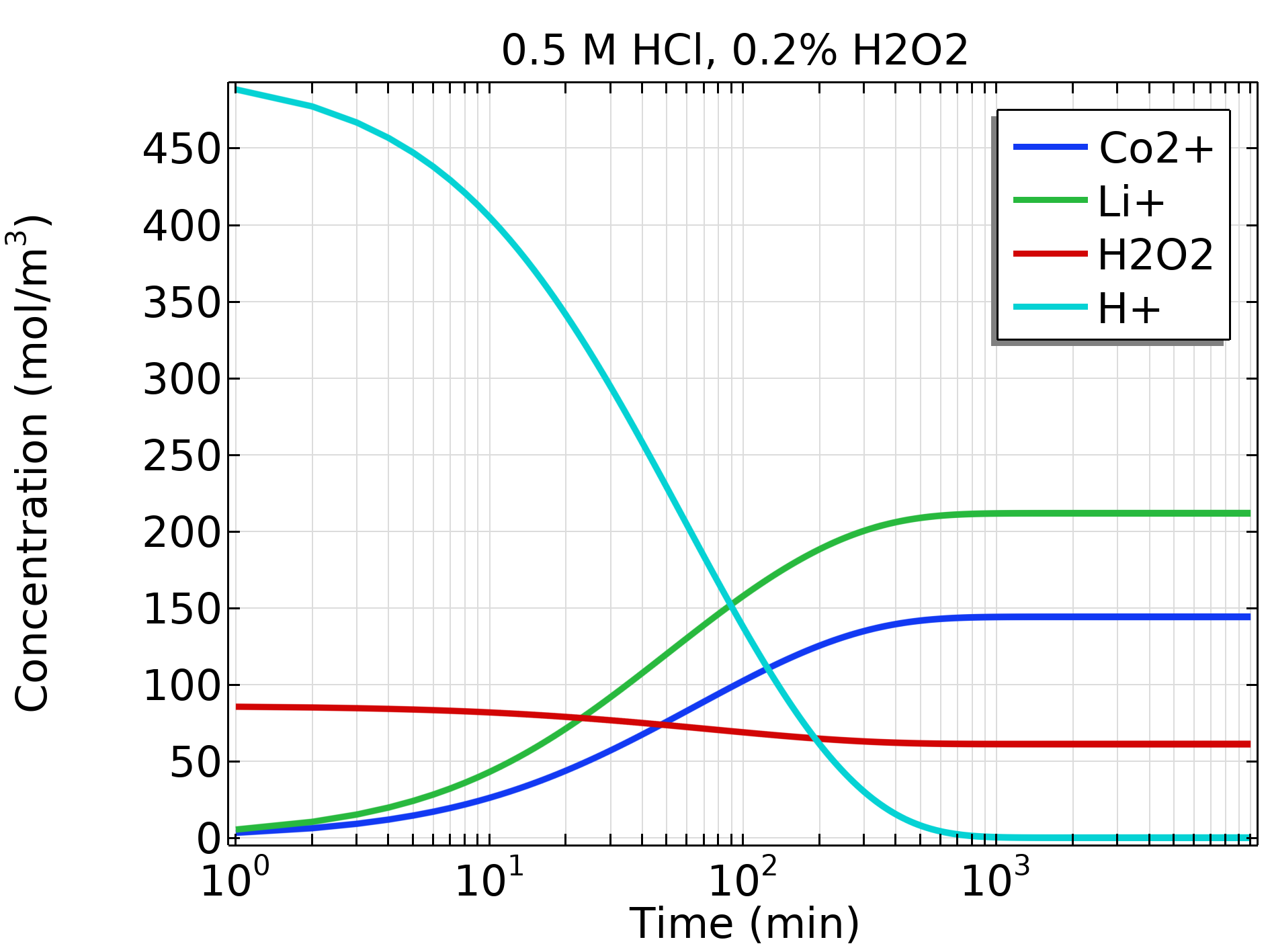}}
    \hfill
    \subfloat[]{\includegraphics[width=0.48\linewidth]{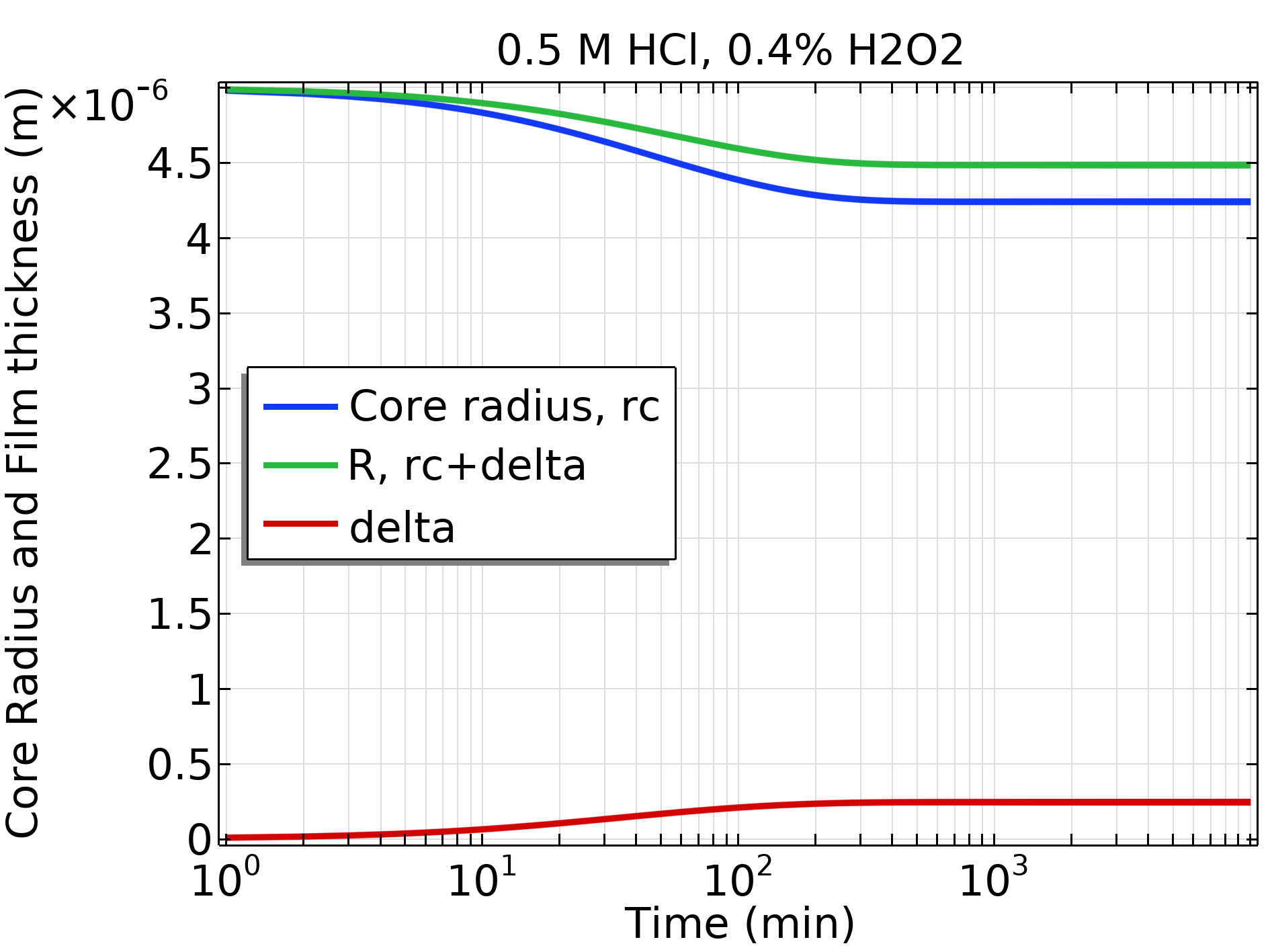}}
    \subfloat[]{\includegraphics[width=0.48\linewidth]{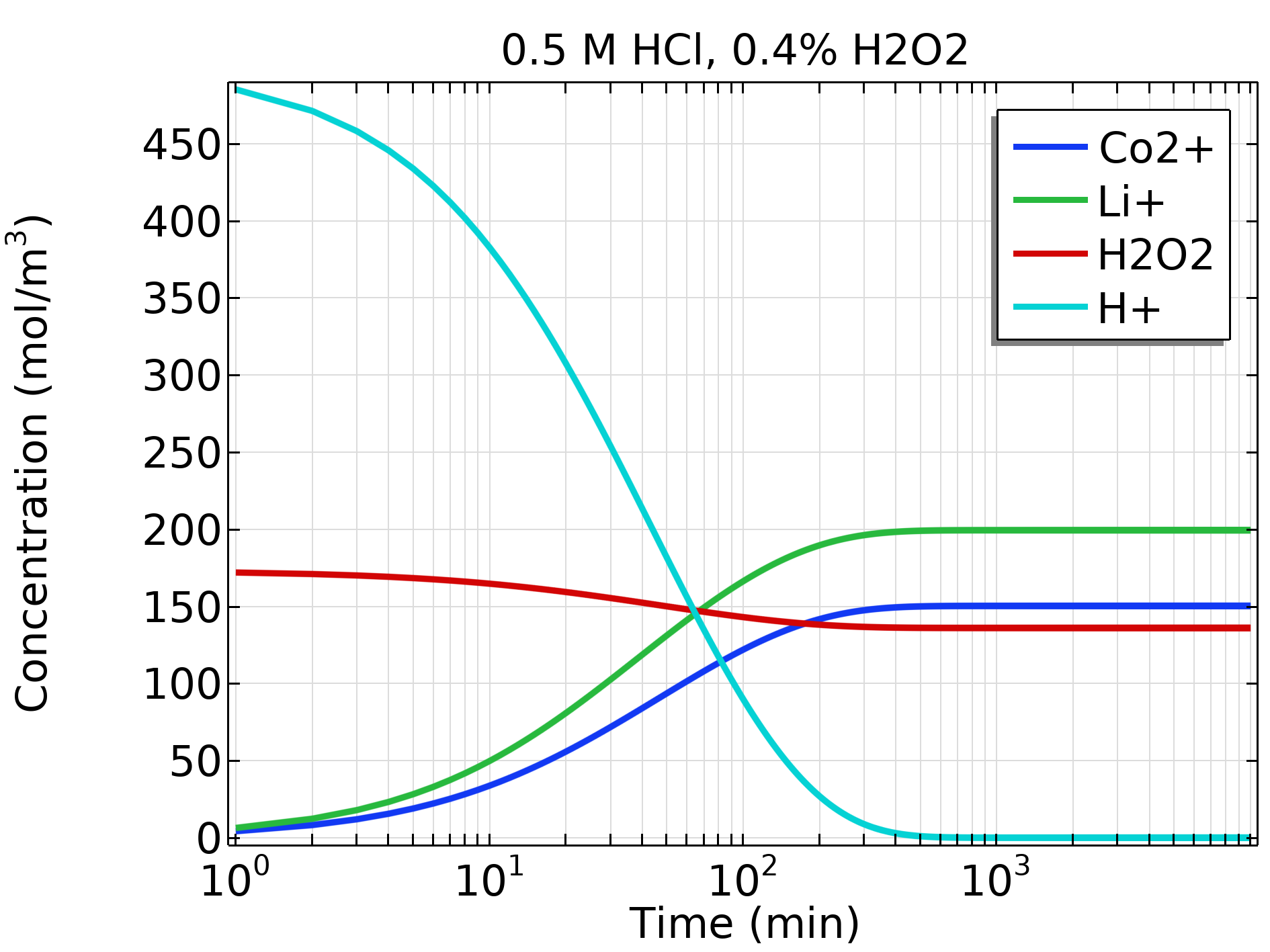}}
    \caption{At 0.5 M Acid: the model predictions with the film passivation. The profiles of LCO core particle radius (rc), the film thickness (delta), and total particle radius (rc + delta); and concentrations of Co$^{2+}$, Li$^{+}$, the acid and H$_2$O$_2$. a \& b) at 0.2\% H$_2$O$_2$ b). c \& d) 0.4\% H$_2$O$_2$.}
    \label{fig:0p5M_rc_concAppend}
\end{figure}
\begin{figure}
    \centering
    \subfloat[]{\includegraphics[width=0.48\linewidth]{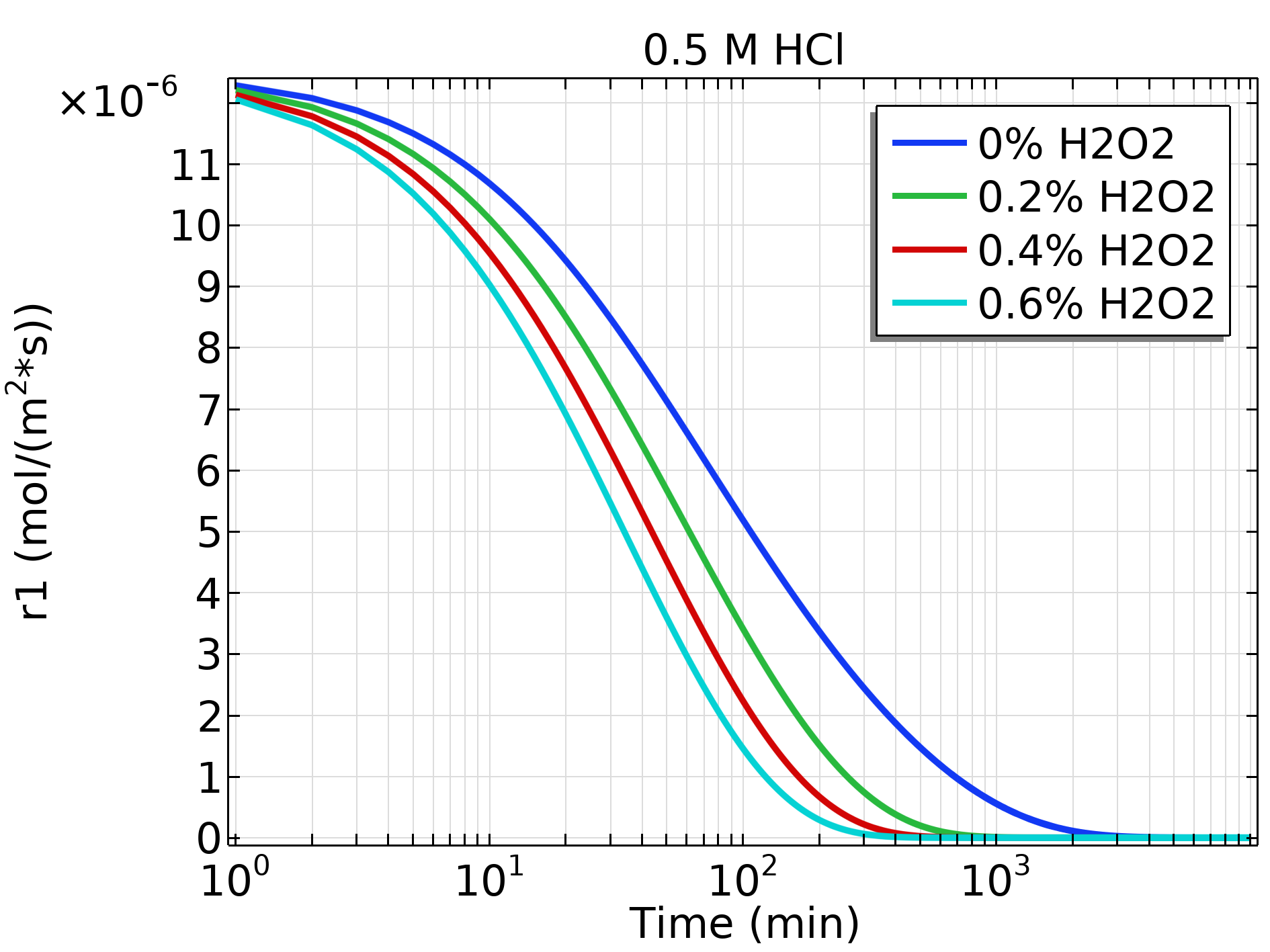}}
    \subfloat[]{\includegraphics[width=0.48\linewidth]{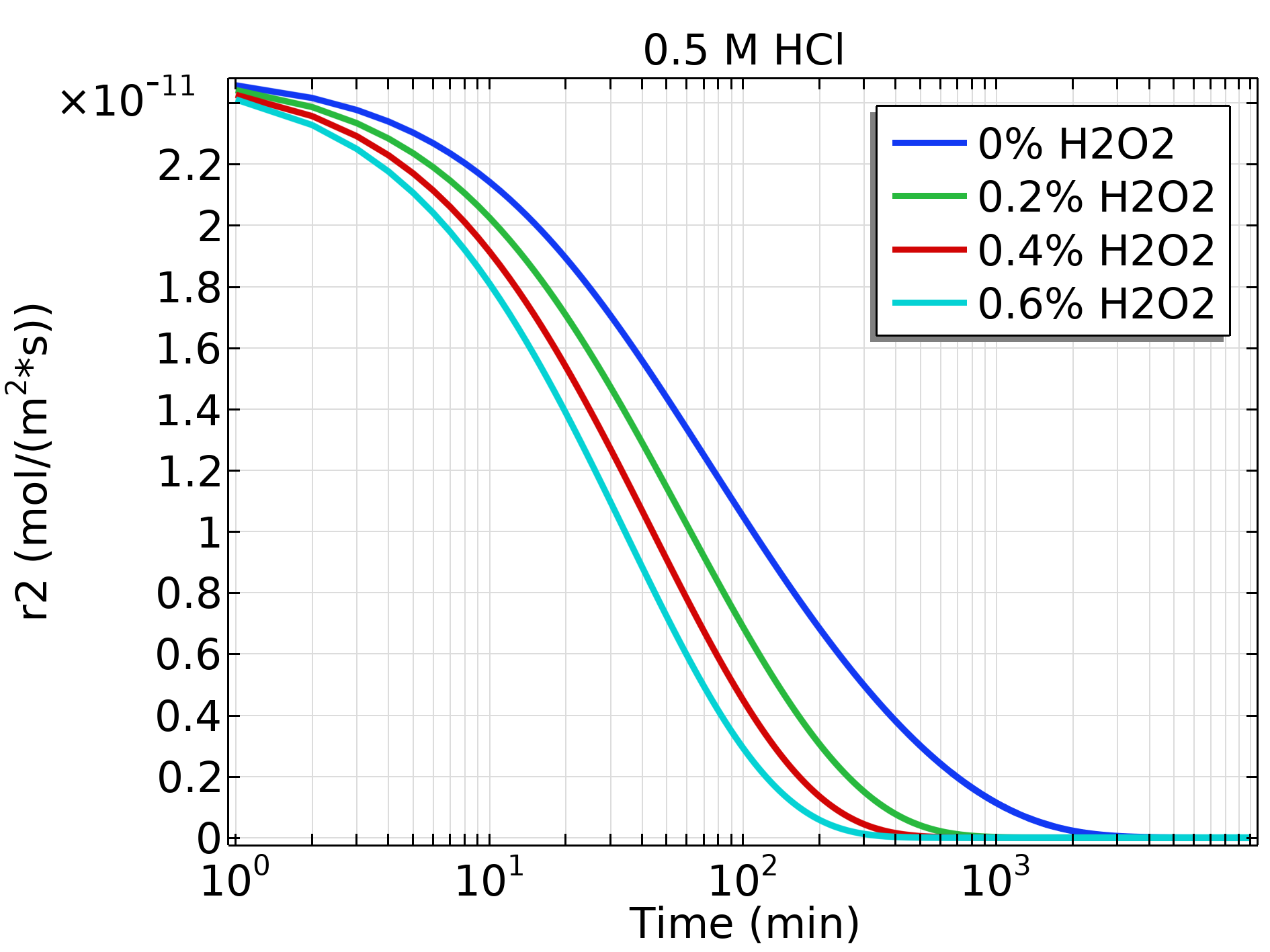}}
    \hfill
    \subfloat[]{\includegraphics[width=0.48\linewidth]{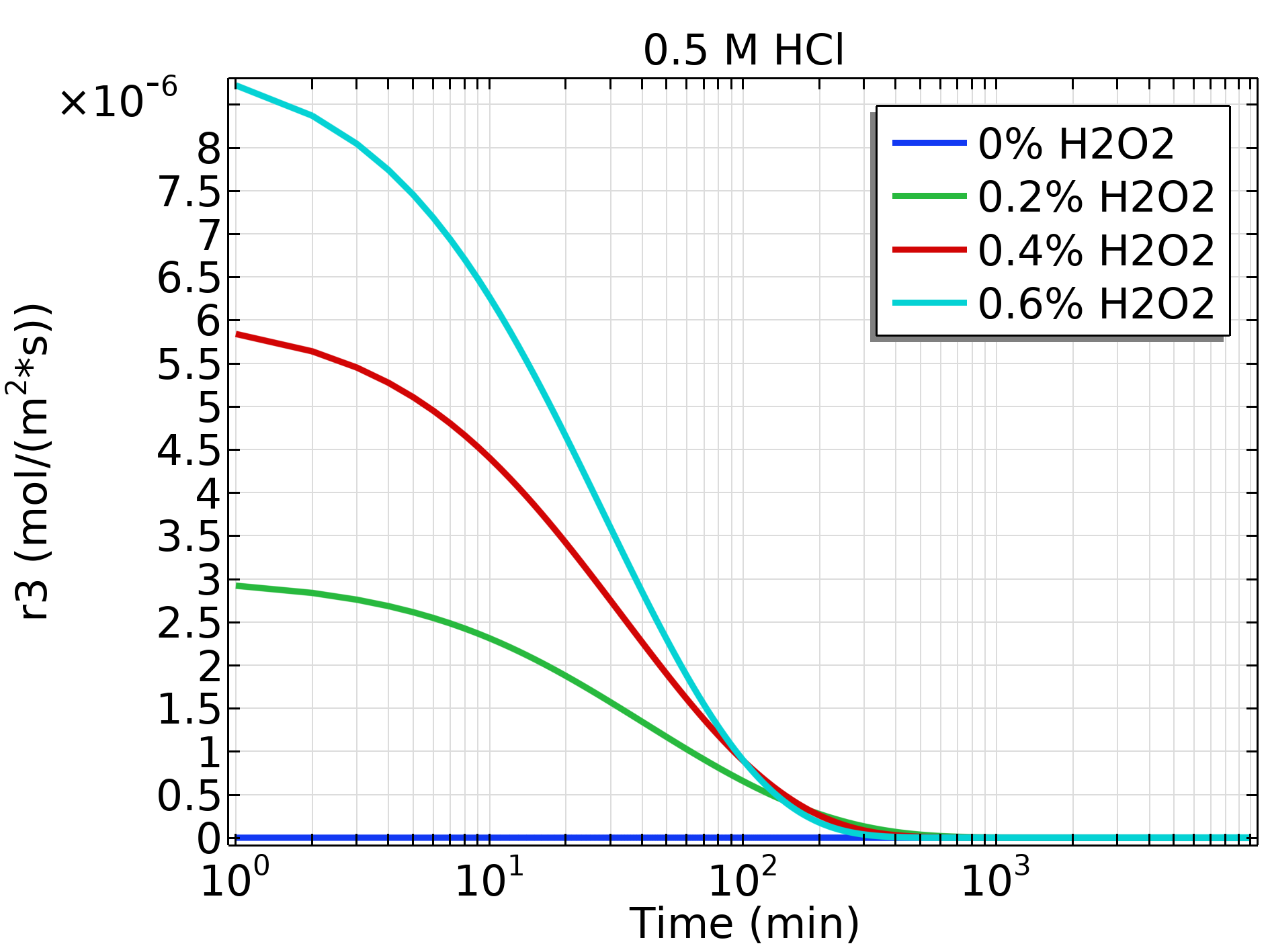}}
    \subfloat[]{\includegraphics[width=0.48\linewidth]{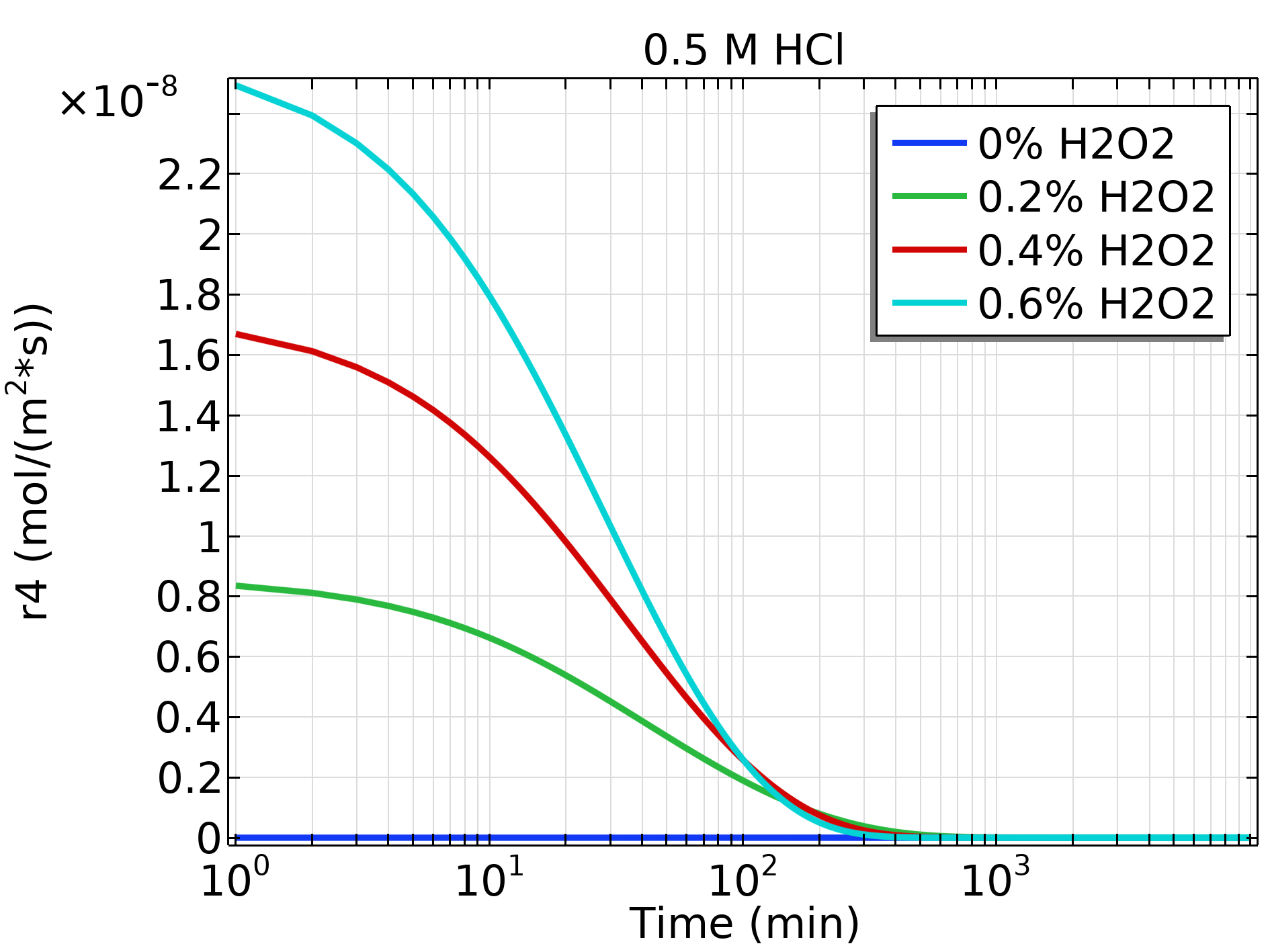}}
    \caption{At 0.5 M Acid: The model predictions of the rates (mol/m$^2$.s) of the four reactions at the four H$_2$O$_2$ concentrations   a) r$_1$, b) r$_2$, c) r$_3$ and d) r$_4$.}
    \label{fig:0p5M_ratesAppend}
\end{figure}

\begin{figure}
    \centering
    \includegraphics[width=0.5\linewidth]{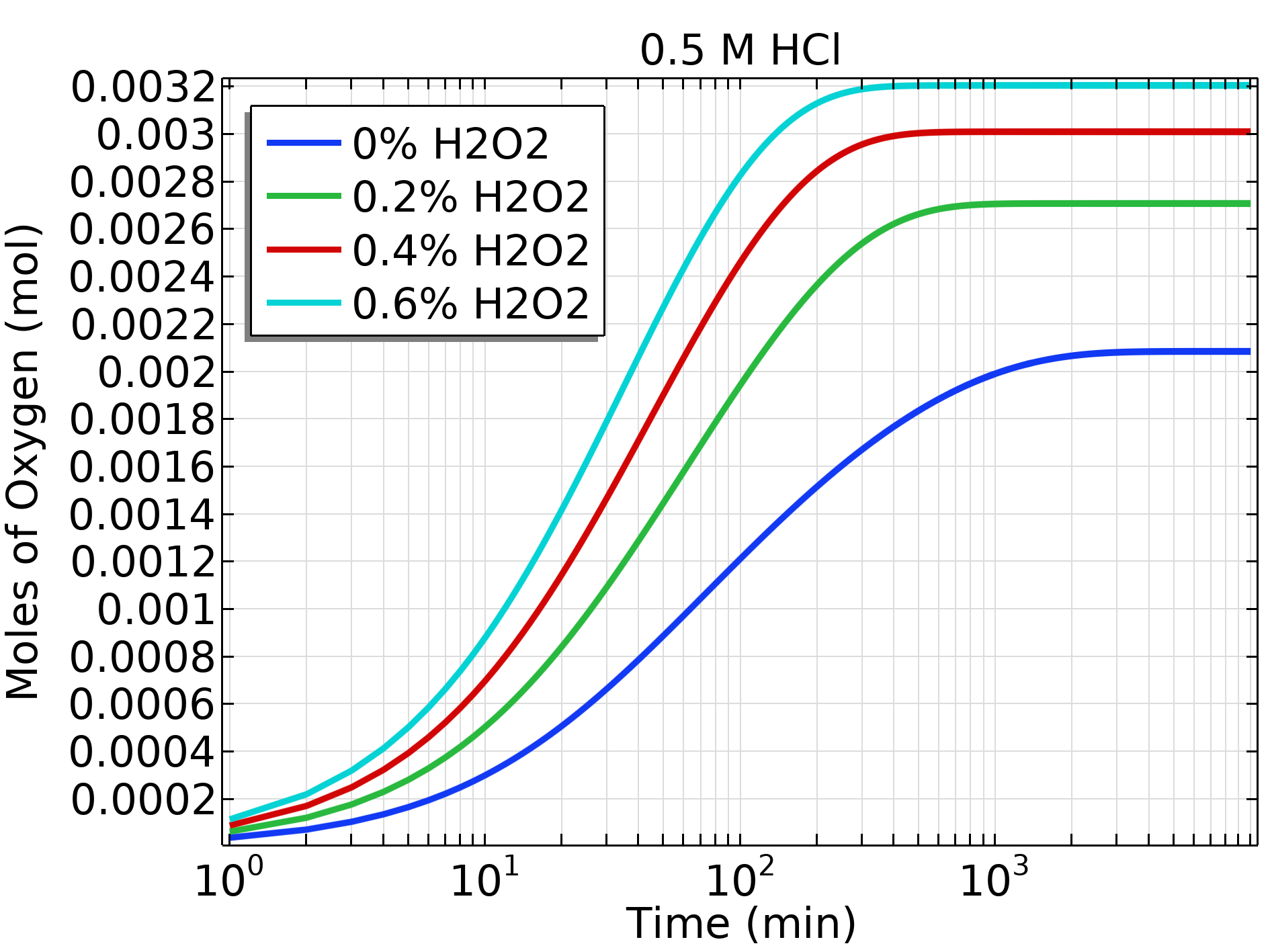}
    \caption{At 0.5 M HCl: The model predictions of the moles of O$_2$ released during leaching at different H$_2$O$_2$ concentrations. Higher the conversion, higher the moles of O$_2$ released.}
    \label{fig:molesO205Append}
\end{figure}

\clearpage
\section{Additional simulations for 2.5 M HCl}
\label{appnd:2p5M}
\begin{figure}
    \centering
    \subfloat[]{\includegraphics[width=0.48\linewidth]{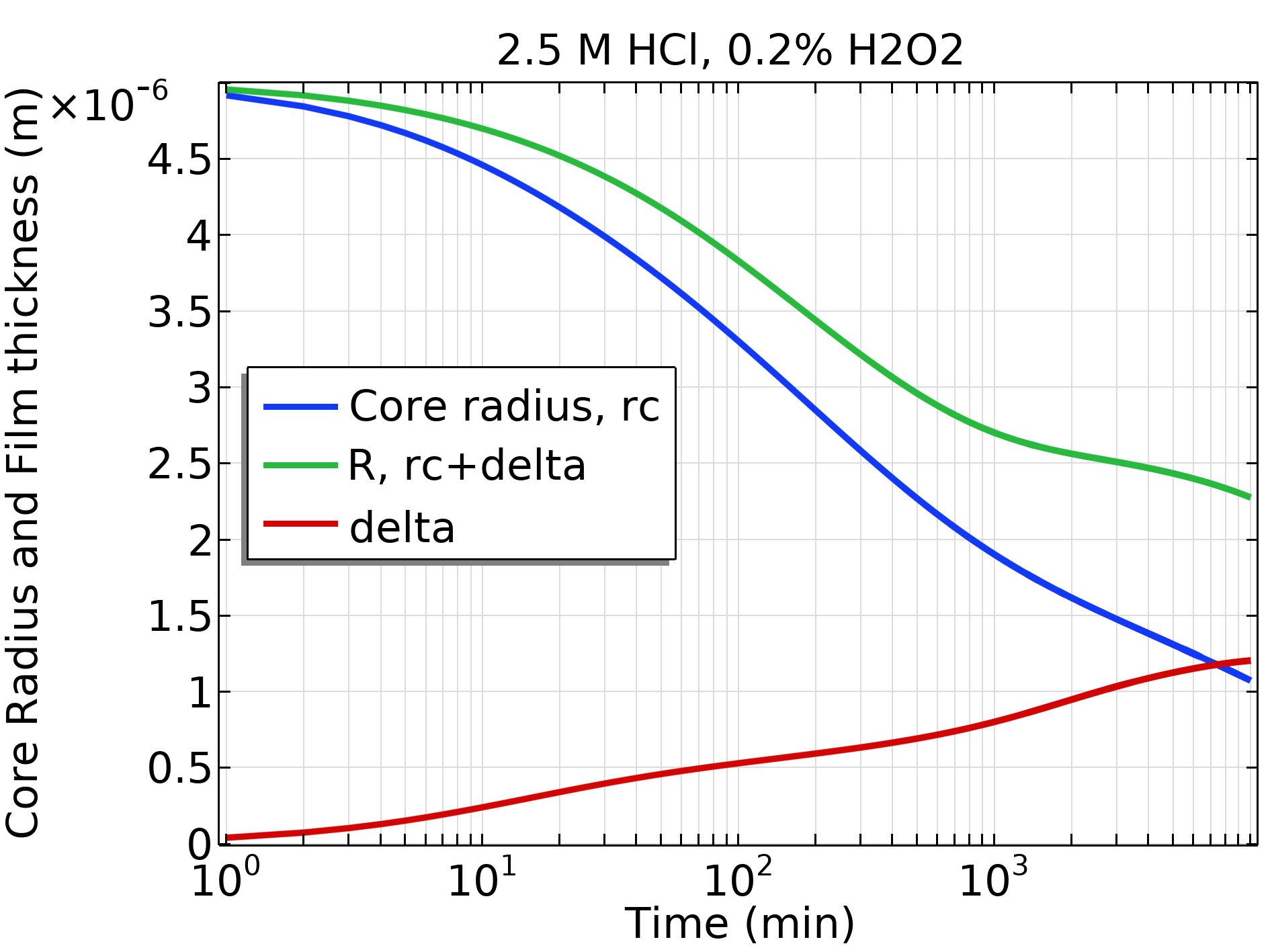}}
    \subfloat[]{\includegraphics[width=0.48\linewidth]{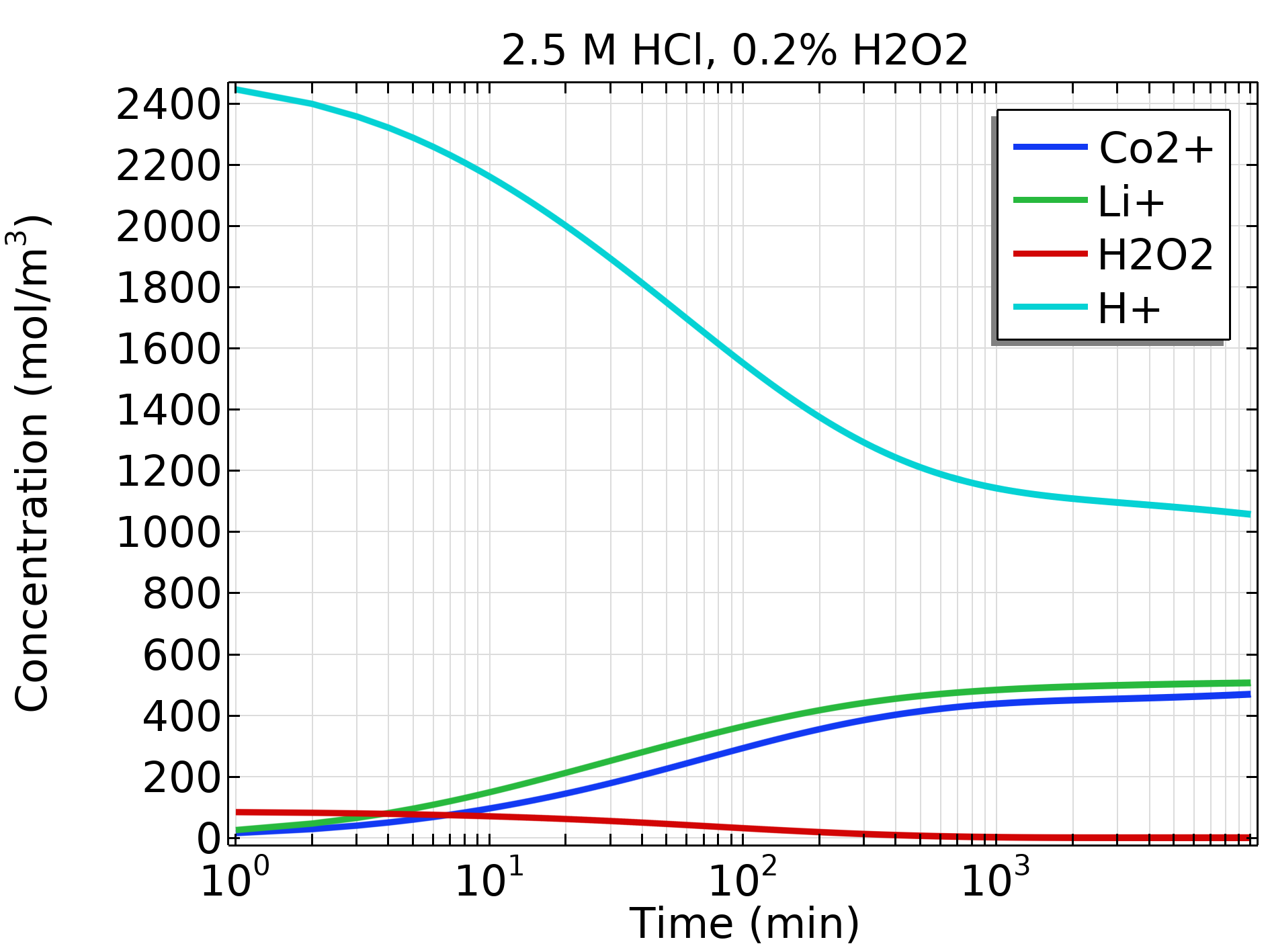}}
    \hfill
    \subfloat[]{\includegraphics[width=0.48\linewidth]{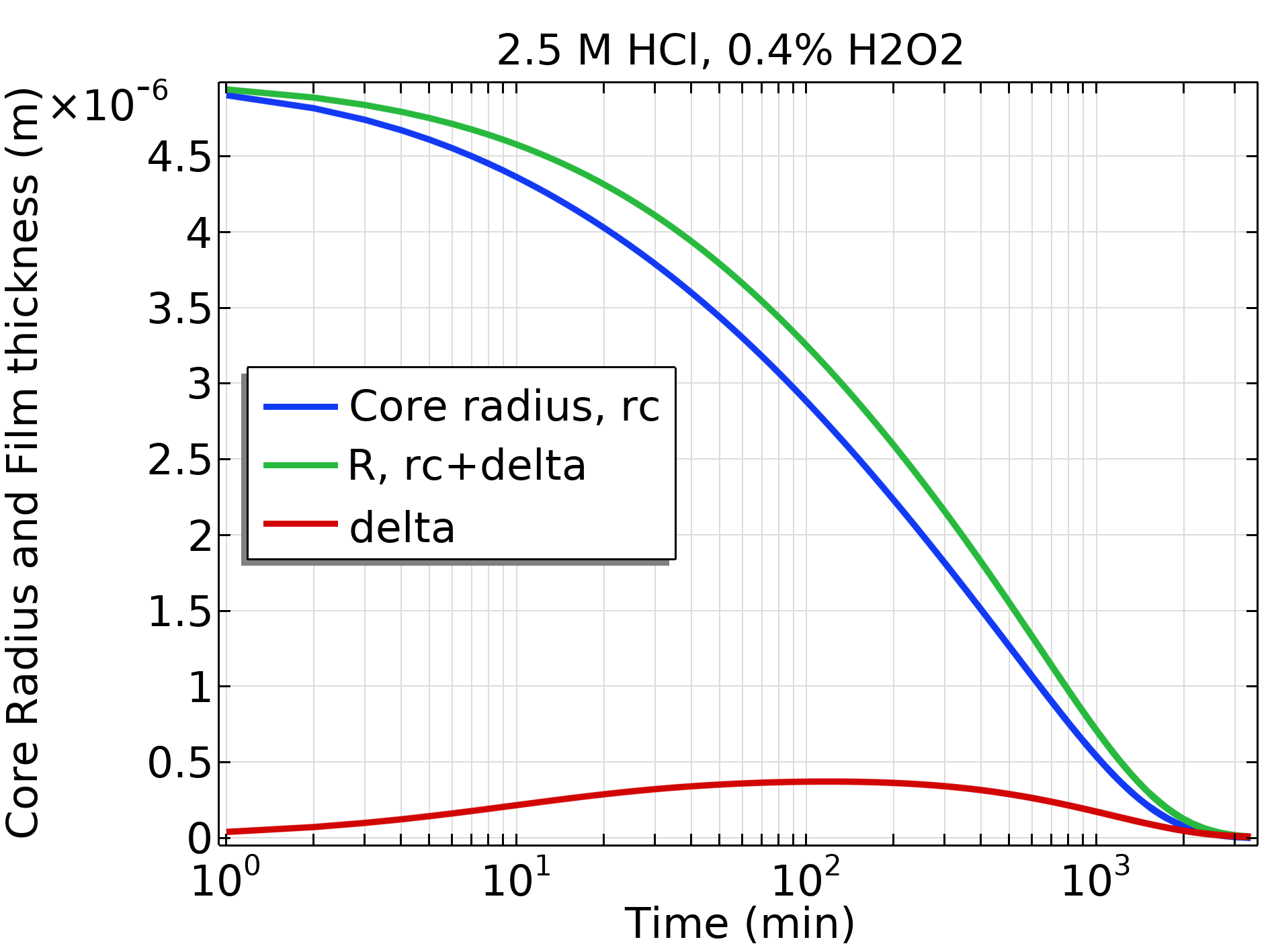}}
    \subfloat[]{\includegraphics[width=0.48\linewidth]{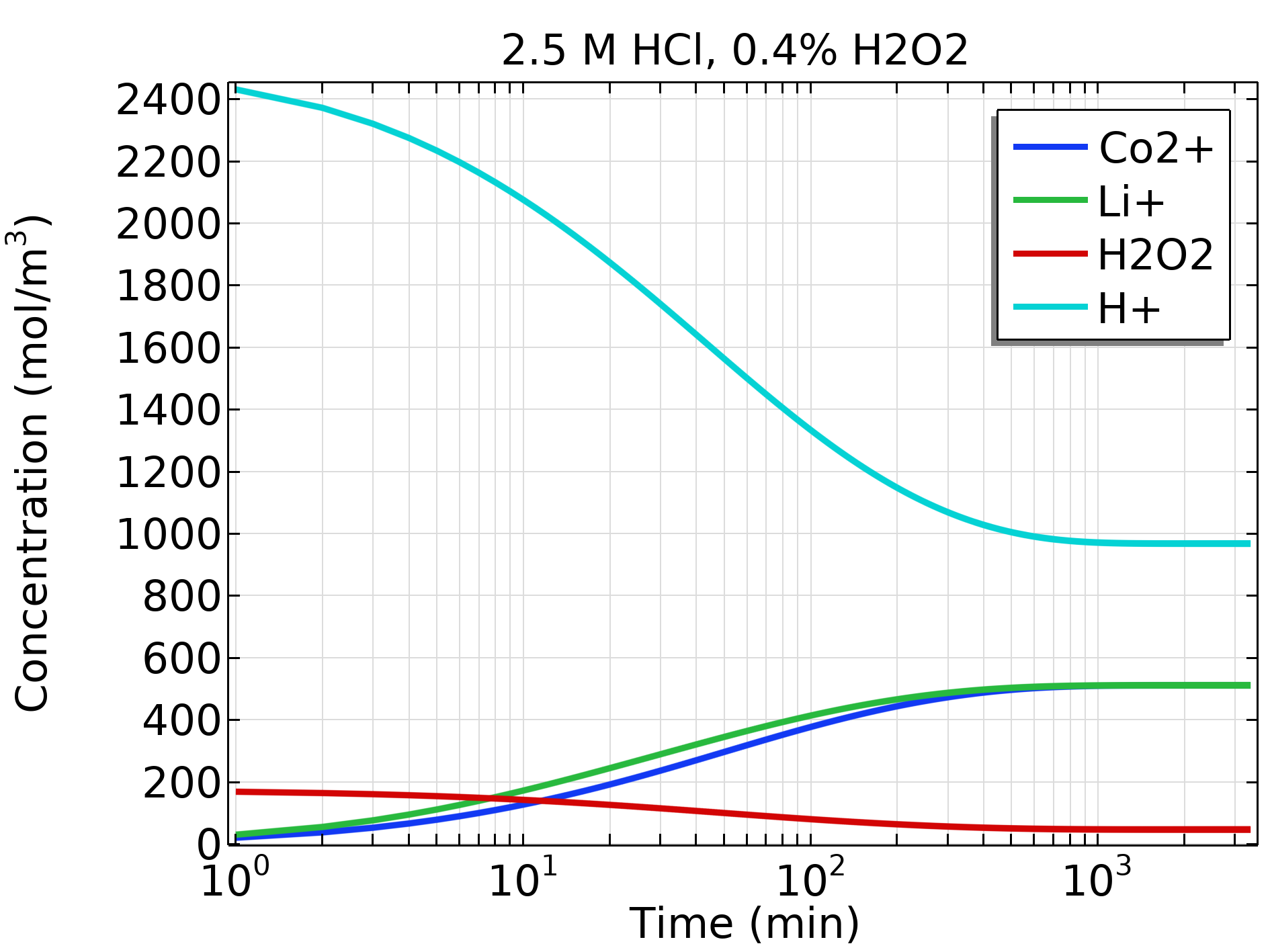}}
    \caption{At 2.5 M Acid: the model predictions with the film passivation. The profiles of LCO core particle radius (rc), the film thickness (delta), and total particle radius (rc + delta); and concentrations of Co$^{2+}$, Li$^{+}$, the acid and H$_2$O$_2$. a \& b) at 0.2\% H$_2$O$_2$ b). c \& d) 0.4\% H$_2$O$_2$.}
    \label{fig:2p5M_rc_concAppend}
\end{figure}

\begin{figure}
    \centering
    \subfloat[]{\includegraphics[width=0.48\linewidth]{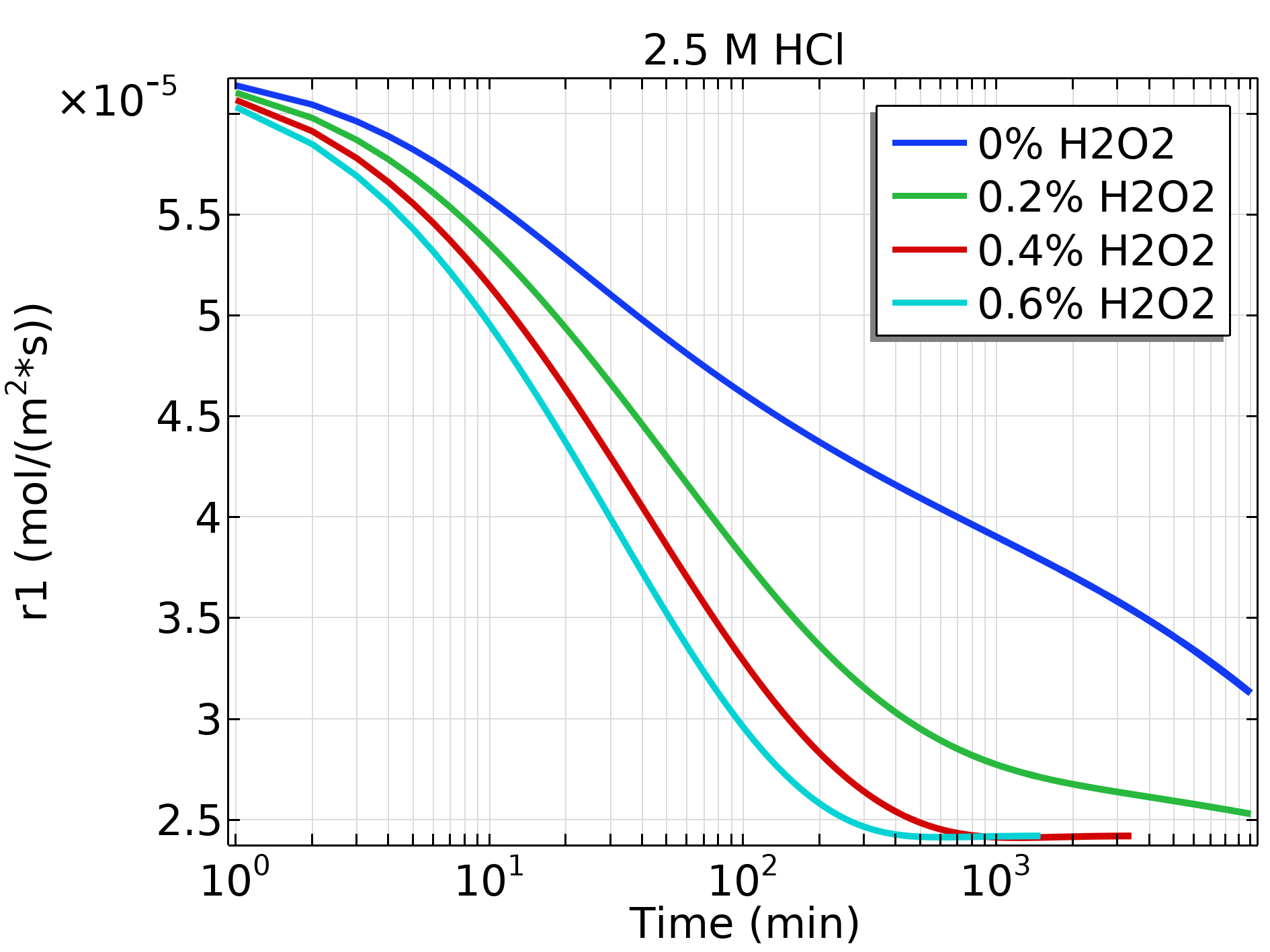}}
    \subfloat[]{\includegraphics[width=0.48\linewidth]{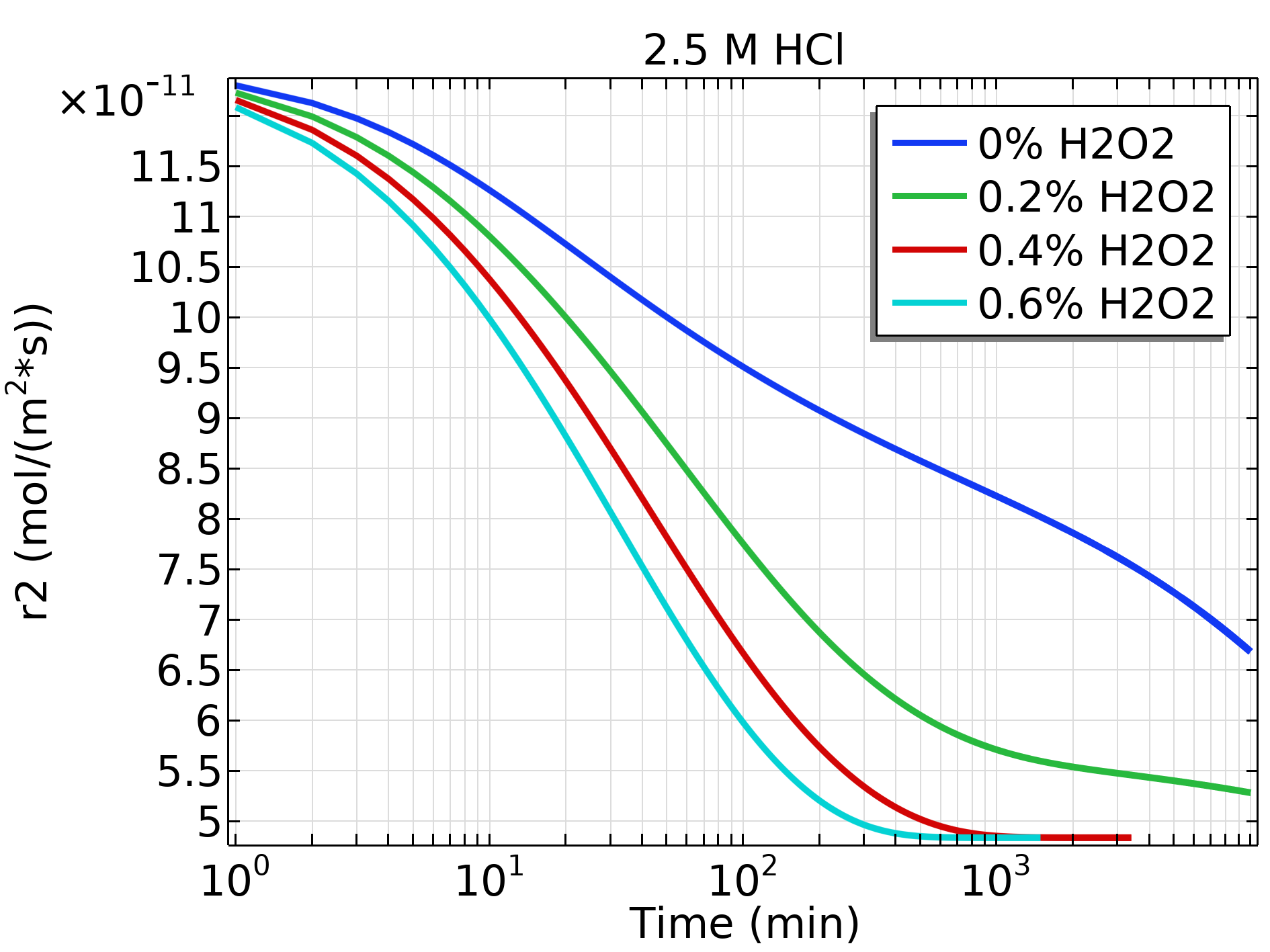}}
    \hfill
    \subfloat[]{\includegraphics[width=0.48\linewidth]{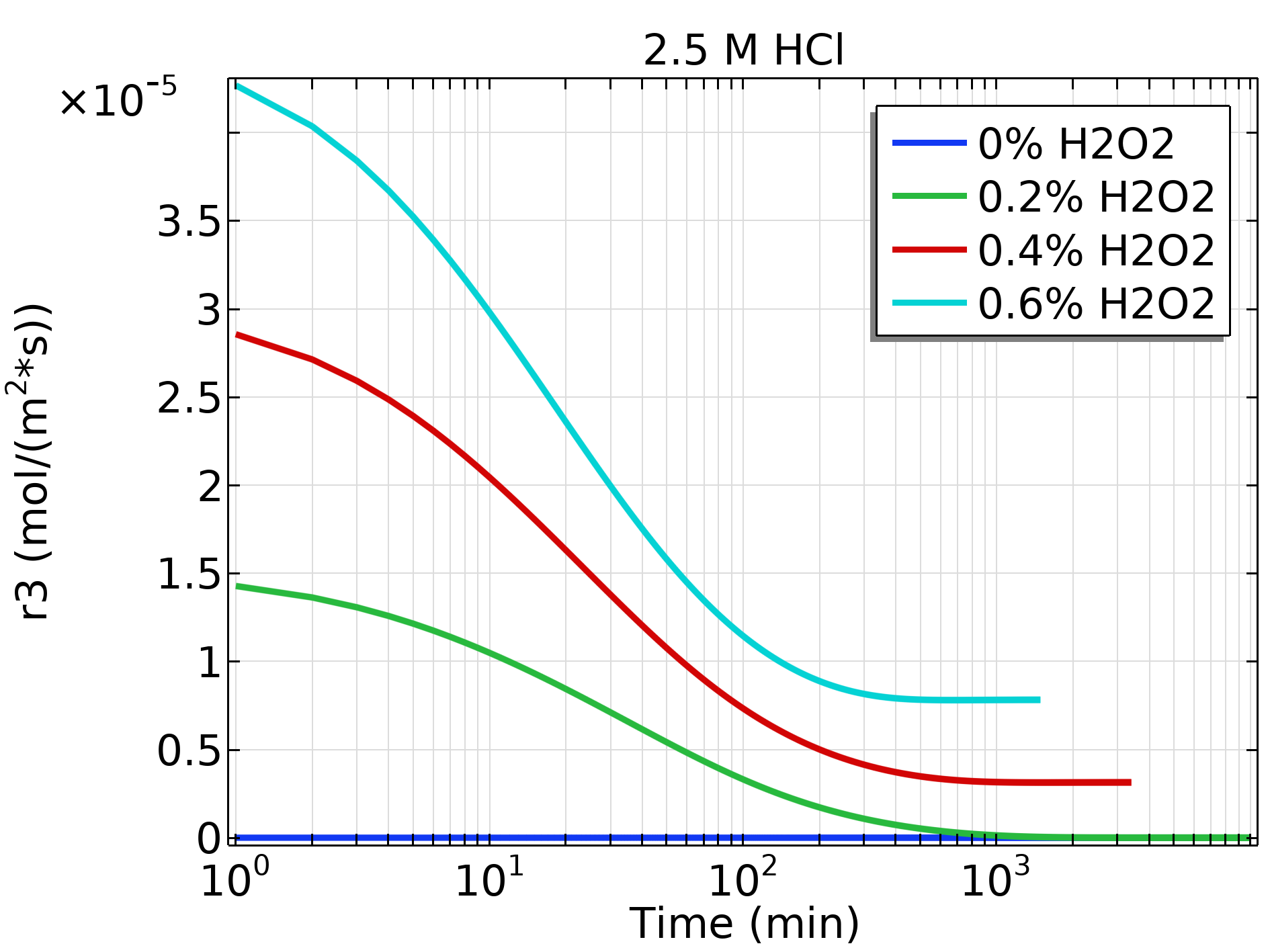}}
    \subfloat[]{\includegraphics[width=0.48\linewidth]{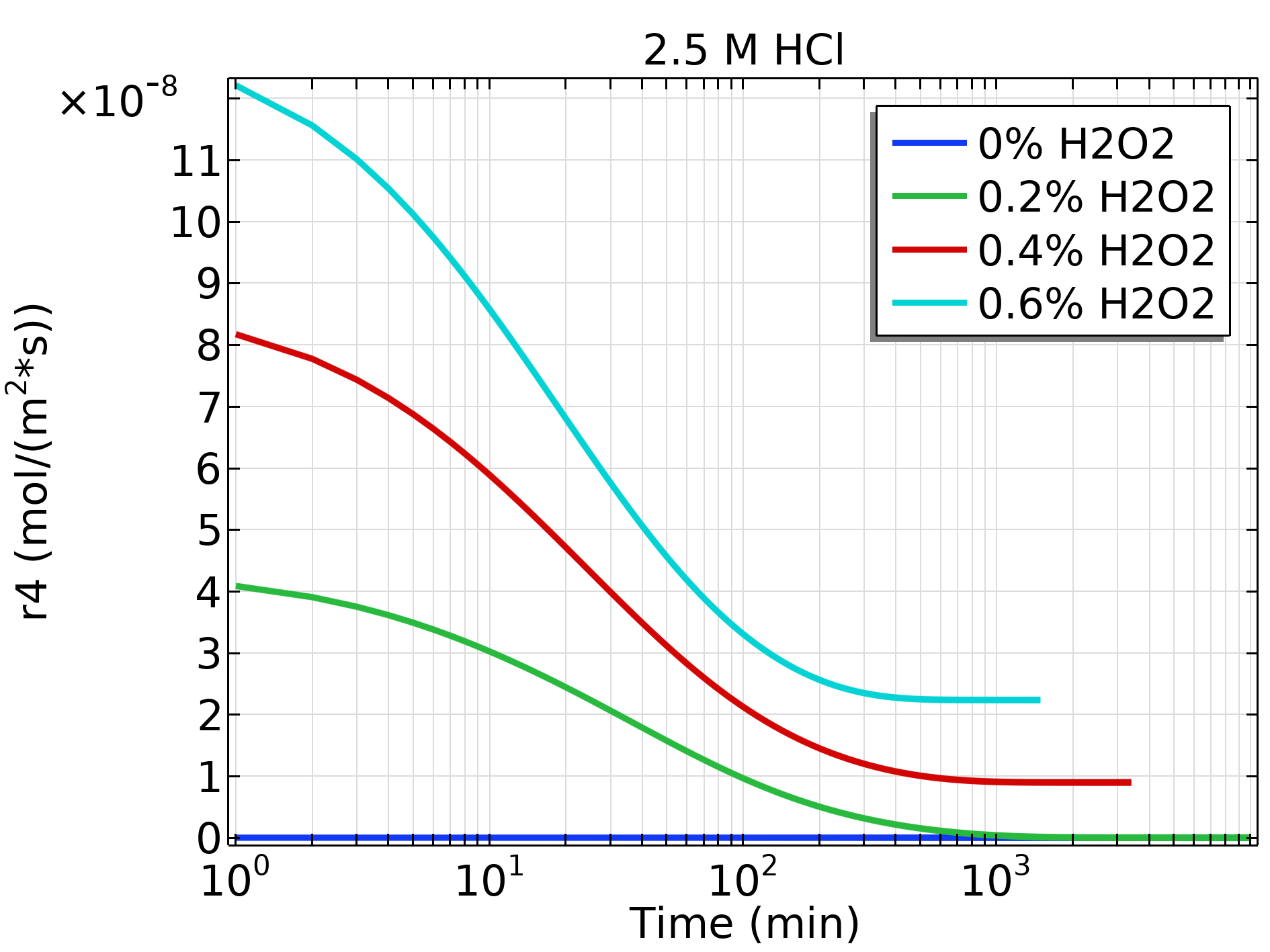}}
    \caption{At 2.5 M Acid: The model predictions of the rates (mol/m$^2$.s) of the four reactions at the four H$_2$O$_2$ concentrations.  a) r$_1$, b) r$_2$, c) r$_3$ and d) r$_4$.}
    \label{fig:2p5M_ratesAppend}
\end{figure}

\begin{figure}
    \centering
    \includegraphics[width=0.5\linewidth]{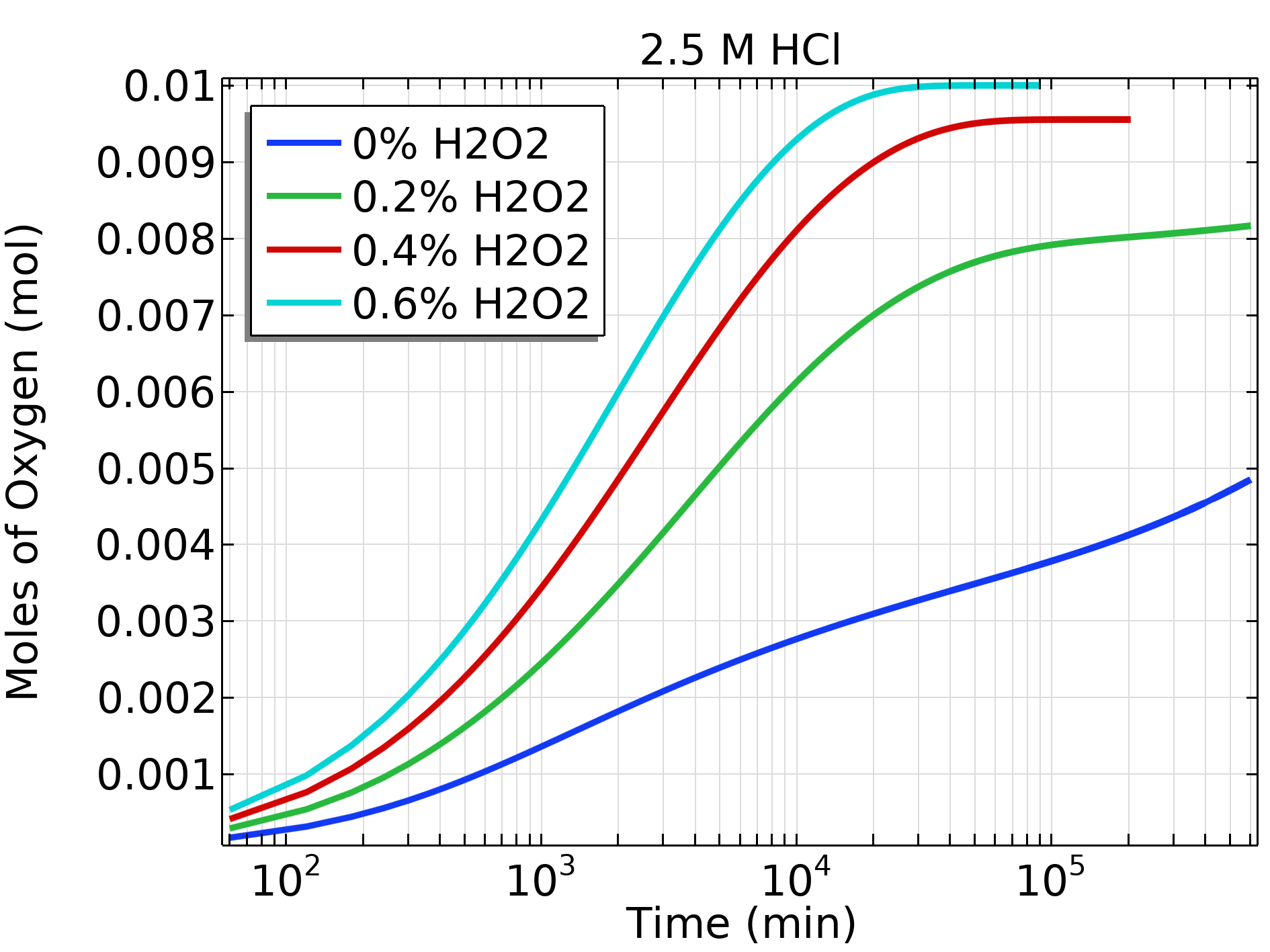}
    \caption{At 2.5 M HCl: The model predictions of the moles of O$_2$ released during leaching at different H$_2$O$_2$ concentrations. Higher the conversion, higher the moles of O$_2$ released.}
    \label{fig:molesO2-25Append}
\end{figure}
\clearpage
\section{Nomenclature}
\label{sec:appendix}
\begin{table}[h!]
\centering
\caption{Model Variables}
\begin{tabular}{lll}
\toprule
Symbol & Description & Units \\
\midrule
$V_{LCO}$ & Total core volume of LiCoO$_2$ particles & m$^3$ \\
$V_{s}$ & Total solid film volume of Co$_3$O$_4$ layer & m$^3$ \\
$C_H$ & Proton concentration & mol/m$^3$ \\
$C_{H,s}$ & Interfacial proton concentration & mol/m$^3$ \\
$C_{H_2O_2}$ & Hydrogen peroxide concentration & mol/m$^3$ \\
$C_{H_2O_2,s}$ & Interfacial hydrogen peroxide concentration & mol/m$^3$ \\
$C_{Li}$ & Lithium concentration & mol/m$^3$ \\
$C_{Co^{2+}}$ & Cobalt concentration & mol/m$^3$ \\
$n_{O_2}$ & Oxygen evolved & mol \\
$r_c$ & Core radius of a LiCoO$_2$ particle  & m \\
$R$ & Particle radius of a LiCoO$_2$ with Co$_3$O$_4$ layer & m \\
$\delta$ & Film thickness of Co$_3$O$_4$ layer on a LiCoO$_2$ particle & m \\
\bottomrule
\end{tabular}
\label{tab:variables}
\end{table}

\begin{table}[h!]
\centering
\caption{Model Parameters}
\begin{tabular}{lll}
\toprule
Symbol & Description & Units \\
\midrule
$R_0$ & Initial LCO particle radius & m \\
$N_p$ & Number of particles & -- \\
$V$ & Liquid volume in the reactor & m$^3$ \\
$\alpha$ & Effective area factor for reactions [0,1] & -- \\
$\varepsilon$ & Porosity of Co$_3$O$_4$ film & -- \\
$\tau$ & Tortuosity of Co$_3$O$_4$ film & -- \\
$a_s$ & Surface area per solid volume of porous Co$_3$O$_4$ layer & m$^2$/m$^3$ \\
$D_i$ & Diffusivity of species $i$ & m$^2$/s \\
$k_{s,i}$ & Interfacial surface rate constant for reactions \ref{eqn:R1} and \ref{eqn:R3} for species $i$& m/s \\
$k_2$ & First-order surface rate constant for reaction \ref{eqn:R2} & m/s \\
$k_3$ & Bimolecular interfacial surface rate constant for reaction \ref{eqn:R3} & m$^4$/(mol·s) \\
$k_4$ & Bimolecular surface rate constant for reaction \ref{eqn:R4}& m$^4$/(mol·s) \\
$M_i$ & Molar mass of species $i$ & kg/mol \\
$\rho_i$ & Density of species $i$ & kg/m$^3$ \\
$m_{LCO}$ & Mass of LCO sample & kg \\

$f_{\mathrm{acc}}$ & Accessibility factor & -- \\
$\lambda$ & Passivation coefficient & -- \\
\bottomrule
\end{tabular}
\label{tab:parameters}
\end{table}


\clearpage
\bibliography{reference_LCOkinetics}
\end{document}